\documentclass{aimsarxiv}

\usepackage{cite}
\usepackage{txfonts}
\usepackage{subfigure}
\def\typeofarticle{}

\def\ppages{}
\def\DOI{}

\numberwithin{equation}{section}

\begin{document}

\title{A lattice model for active--passive 
pedestrian dynamics: a quest for drafting effects}

\author{%
  Emilio N. M. Cirillo\affil{1},
  Matteo Colangeli\affil{2}
  Adrian Muntean \affil{3}
  and
  T. K. Thoa Thieu\affil{4}\corrauth
}

\shortauthors{the Author(s)}

\address{%
	\addr{\affilnum{1}}{Dipartimento di Scienze di Base e Applicate per l'Ingegneria, Sapienza Universit\`{a} di Roma}
  \addr{\affilnum{2}}{Dipartimento di Ingegneria e Scienze dell'Informazione e Matematica, Universit\`{a} degli Studi dell'Aquila, Italy}
    \addr{\affilnum{3}}{Department of Mathematics and Computer Science, Karlstad University, Sweden}
    \addr{\affilnum{4}}{Department of Mathematics, Gran Sasso Science Institute, L'Aquila, Italy}
}

\corraddr{emilio.cirillo@uniroma1.it, matteo.colangeli1@univaq.it, adrian.muntean@kau.se, thikimthoa.thieu@gssi.it
}

\begin{abstract}
	We study the pedestrian escape from an obscure corridor using a lattice gas model with two species of particles. One species, called passive, performs a symmetric random walk on the lattice, whereas the second species, called active, is subject to a drift guiding the particles towards the exit. The drift mimics the awareness of some pedestrians of the geometry of the corridor and of the location of the exit.
	We provide numerical evidence that, in spite of the hard core interaction between particles -- namely, there can be at most one particle of any species per site, -- 
	adding a fraction of active particles in the system enhances the evacuation rate of all particles from the corridor. A similar effect is also observed when looking at the outgoing particle flux, when the system is in contact with an external particle reservoir that induces the onset of a steady state. We interpret this phenomenon as a discrete space counterpart of the drafting effect typically observed in a continuum set--up as the aerodynamic drag experienced by pelotons of competing cyclists. 
\end{abstract}



\maketitle

\section{Introduction}
\label{s:int}

This work is part of a larger recent research initiative oriented towards investigating the evacuation behavior of large crowds of pedestrians where the geometry where the dynamics takes place is partly unknown and possibly also with limited visibility. Such scenarios are encountered for instance when catastrophic situations occur in urban environments (e.g., in large office spaces), in tunnels, within underground spaces and/or in forests in fire.  We refer the reader, for instance,  to \cite{Albi,XJ+2016,FRNF2013,MWS2014,WS2014} 
and references cited therein, 
as well as to our previous results \cite{cm2012, cm2013, CCMmms2016, CCMcrm2016, CCCM2018,playing}.

In the current framework, our hypothesis is that the crowd under consideration is always heterogeneous in the sense that some part of the population is well informed about the details of the geometry of the location and corresponding exits as well as of the  optimal escape routes, while the rest of the population is ignorant. This is precisely the standing assumption we have investigated  in \cite{FT,bookchapter} for a dynamics in smoke scenario.  In turns out that in situations where the information can only difficultly be transmitted from pedestrian to pedestrian  (like when large crowds are present and/or if the geometry of the evacuation is largely unknown or invisible and/or groups are not able to act rationally \cite{curseu2014}), the use of leaders to guide  crowds towards the exists might not always be possible, or it works inefficiently. In such cases, leadership is not essential to speed up evacuation\footnote{This is contrary to situations like those described on \cite{couzin},  where leadership is an efficient crowd management mechanism. }. So, what can then be still done to improve evacuations for such unfavorable conditions, i.e. to decrease the evacuation time of the overall crowd?   One of the main points we want to make here is the following: Even if information cannot be transmitted within the crowd, simply having a suitable fraction of informed pedestrians speeds up the overall evacuation time. 

To defend our point of view, we study the typical time needed by a bunch of pedestrian 
to escape a dark corridor. We consider two pedestrian species, the \emph{active} and 
the \emph{passive} one, i.e., those who know the 
location of the exit and those who do not.
Using a lattice--type model, we show that the presence of pedestrians aware of the location of 
the exit helps  the unaware companions to find the exit of the corridor even 
in absence of any information exchange among them. 
This effect will be called \emph{drafting} and it has a twofold 
interpretation:
i)
active particles, quickly moving 
towards the exit, will leave a wake of empty sites in which 
the unaware particles can jump in, so that they 
are guided to the exit;
ii)
active pedestrians, 
in their rational motion towards the exit, push their passive companions 
to the exit as well. 

Our attention is focused here on the flow of mixed pedestrian populations and on the modelling  of their interactions with obstacles in the presence of regions with lack of visibility (see \cite{ACK2015, Cristiani, BCGTV2016}, e.g. for more on crowd dynamics modelling and \cite{CKMSpre2016, CP2017, CCS2018, CC2018} for relevant works handling the presence of the obstacles and the way they affect the pedestrian motion).  It is worth however mentioning that remotely resembling situations appear frequently in soft matter physics; see e.g. \cite{abc2011, abc2014}.

The reminder of the paper is organized as follows. 
In Section~\ref{s:mod} we define the model. 
Section~\ref{s:eva} is devoted to the study of the evacuation time. 
In Section~\ref{s:flu} a slightly modified version of the model 
is considered, so that a not trivial stationary state is reached. 
For such a model the stationary exit flux is thus studied. 
In Section~\ref{s:con}, we summarize our conclusions and give a glimpse of possible further research.

\section{The model}
\label{s:mod}
\begin{figure}
	\centering
	\includegraphics[width=0.4\textwidth]{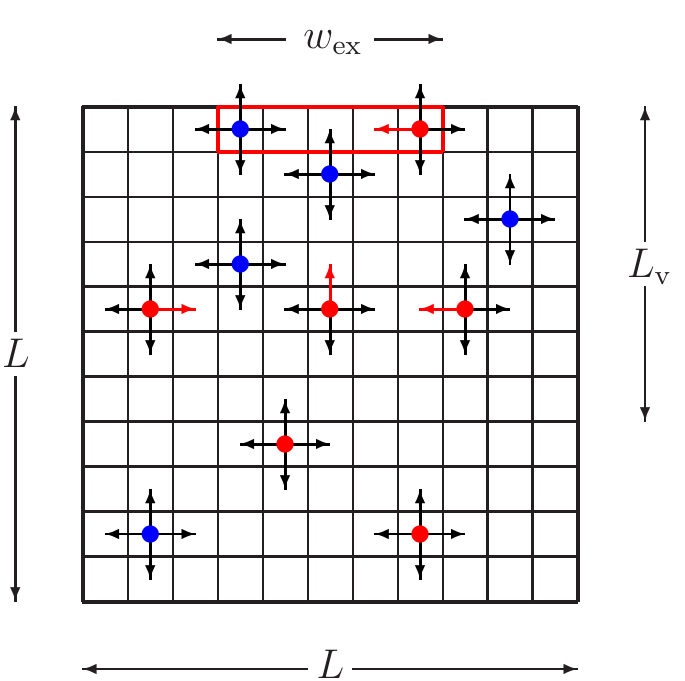}
	\caption{Schematic representation of our lattice model. Blue and red 
disks denote passive and active particles, respectively. 
The rectangle of sites delimited by the red contour denotes the exit.
Black and red arrows (color online) denote
transitions performed with rates $1$ and $1+\varepsilon$, respectively.}
	\label{fig:fig0}
\end{figure}

The \emph{corridor} is the square 
lattice $\Lambda=\{1,\dots,L\}\times\{1,\dots,L\}\subset\mathbb{Z}^2$ of side 
length $L$, with $L$ an odd positive integer number,
see Figure~\ref{fig:fig0}.
An element $x=(x_1,x_2)$ of the corridor $\Lambda$ is called \emph{site} 
or \emph{cell}.
Two sites $x,y\in\Lambda$ are said \emph{nearest neighbor} if and only 
if $|x-y|=1$.
We conventionally call \emph{horizontal} the first coordinate 
axis and \emph{vertical} the second one. The words 
\emph{left}, 
\emph{right}, 
\emph{up}, 
\emph{down}, 
\emph{top}, 
\emph{bottom}, 
\emph{above}, 
\emph{below}, 
\emph{row}, 
and 
\emph{column} 
will be used accordingly.
We call \emph{exit} a set of 
$w_{\mathrm{ex}}$ pairwise adjacent
sites, with $w_{\mathrm{ex}}$ an odd positive integer smaller than $L$,
of the top row of the corridor $\Lambda$ symmetric with respect to its median 
column.
In other words, 
the exit is 
a centered slice of the top row of the corridor mimicking the presence of an 
exit door. 
The number $w_{\mathrm{ex}}$ will be called \emph{width} of the exit.
Inside the top part of the 
corridor we define a rectangular interaction zone, namely, 
the \emph{visibility region} $V$, made of the 
first $L_{\text{v}}$ top rows of $\Lambda$, with the positive integer 
$L_{\text{v}}\le L$ called \emph{depth} of the visibility region.
By writing $L_{\text{v}}=0$, we refer to the case in which no 
visibility region is considered. 

We  consider two different species of particles, i.e.,
\emph{active} and \emph{passive}, moving inside $\Lambda$
(we shall sometimes use in the notation the symbols A and P to 
respectively refer to them).
Note that the sites 
of the external boundary of the corridor, that is to say the sites  
$x\in\mathbb{Z}^2\setminus\Lambda$ such that there exists 
$y\in\Lambda$ nearest neighbor of $x$,
cannot be accessed by the particles.
The state of the system will be a 
\emph{configuration} $\eta\in\Omega=\{-1,0,1\}^\Lambda$ 
and 
we shall say that the site $x$ is 
\emph{empty} if $\eta_x=0$,
\emph{occupied by an active particle} if $\eta_x=1$,
and
\emph{occupied by a passive particle} if $\eta_x=-1$.
The number of active (respectively, passive) 
particles in the configuration $\eta$ 
is given by 
$n_{\text{A}}(\eta)=\sum_{x\in\Lambda}\delta_{1,\eta_x}$
(resp.\ $n_{\text{P}}(\eta)=\sum_{x\in\Lambda}\delta_{-1,\eta_x}$), 
where $\delta_{\cdot,\cdot}$ is Kronecker's symbol.
Their sum is the total number of particles in the configuration $\eta$.

The dynamics in the corridor is modeled via a 
simple exclusion random walk with
the two species of particles undergoing 
two different microscopic dynamics: the passive particles perform a 
symmetric simple exclusion dynamics on the whole lattice, while the active particles, 
on the other hand, perform a symmetric simple exclusion walk 
outside the visibility region, whereas inside such a region they 
experience a drift pushing them towards the exit. 
In other words, the whole corridor is obscure for the passive 
particles, while, for the active ones, only the region outside the 
visibility region is obscure. 

What concerns this precise setup, 
the dynamics is incorporated in the continuous time Markov chain $\eta(t)$ on 
$\Omega$ with rates $c(\eta,\eta')$ defined as follows:
Let $\varepsilon\ge0$ be the \emph{drift};
for any pair $x=(x_1,x_2),y=(y_1,y_2)$ of nearest neighbor sites in $\Lambda$ 
we set $\epsilon(x,y)=0$, excepting the following cases 
in which we set $\epsilon(x,y)=\varepsilon$:
\begin{itemize}
\item[--]
$x,y\in V$ and $y_2=x_2+1$, namely, $x$ and $y$ belong to the 
	visibility region and $x$ is below $y$;
\item[--]
$x,y\in V$ and $y_1=x_1+1<(L+1)/2$, namely, 
	$x$ and $y$ belong to the left part of the visibility region 
	and $x$ is to the left with respect to $y$;
\item[--]
$x,y\in V$ and $y_1=x_1-1>(L+1)/2$, namely, 
	$x$ and $y$ belong to the right part of the visibility region 
	and $x$ is to the right with respect to $y$.
\end{itemize}
Next, we let 
the rate $c(\eta,\eta')$ be equal 
\begin{itemize}
\item[--]
to
$1$
if $\eta'$ can be obtained by $\eta$ by 
replacing with $0$ a $-1$ or a $1$ at the exit
(particles leave the corridor);
\item[--]
to
$1$
if $\eta'$ can be obtained by $\eta$ by 
exchanging a $-1$ with a $0$ between two neighboring sites of $\Lambda$ 
(motion of passive particles inside $\Lambda$);
\item[--]
to
$1+\epsilon(x,y)$
if $\eta'$ can be obtained by $\eta$ by 
exchanging a $+1$ at site $x$ with a $0$ at site $y$, 
with $x$ and $y$ nearest neighbor sites of $\Lambda$ 
(motion of active particles inside $\Lambda$);
\item[--]
to
$0$ in all the other cases.
\end{itemize}

\begin{figure}
	\centering
	\begin{tabular}{lll}
	\includegraphics[width = 0.28\textwidth]{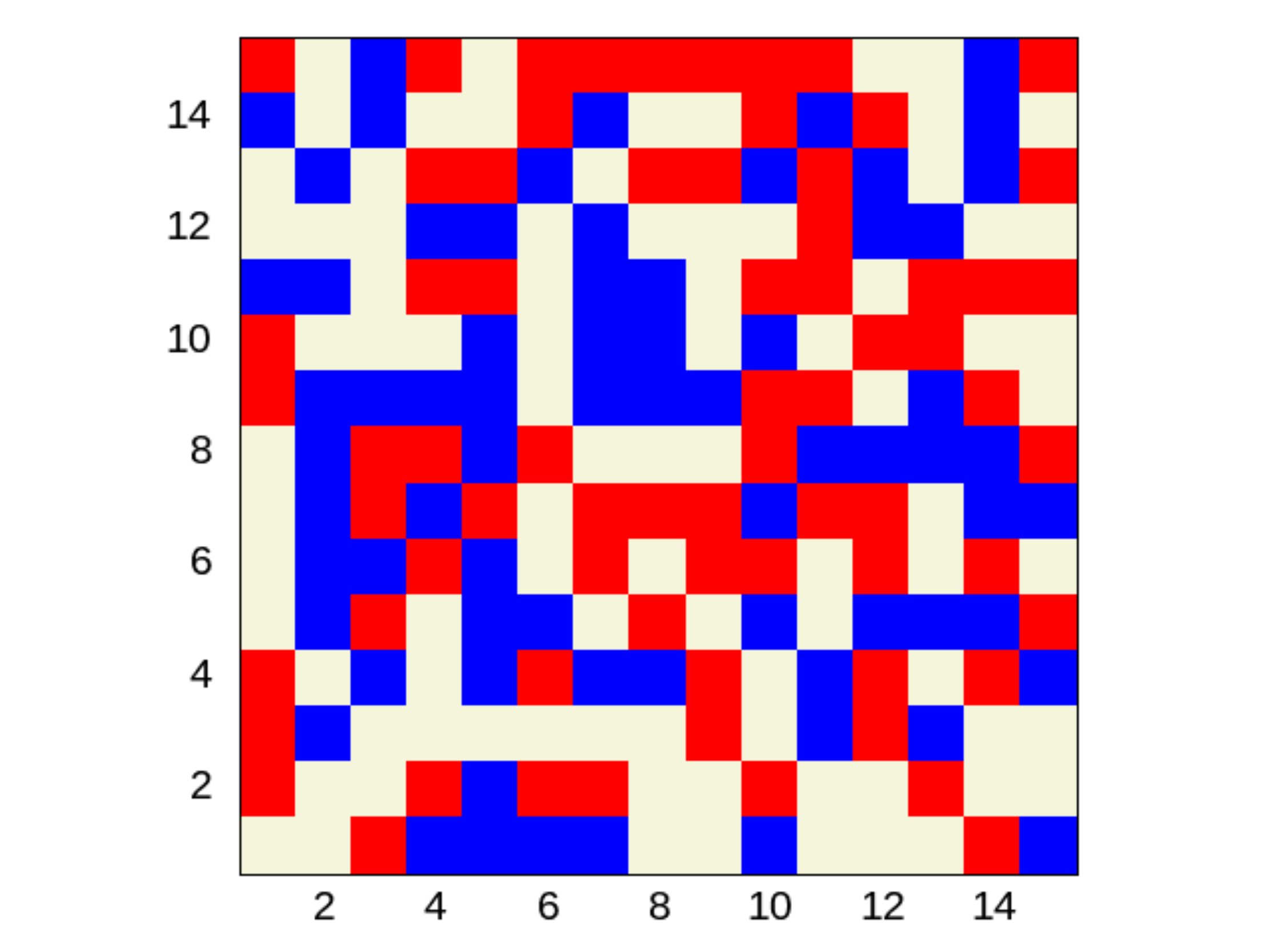} &
	\includegraphics[width = 0.28\textwidth]{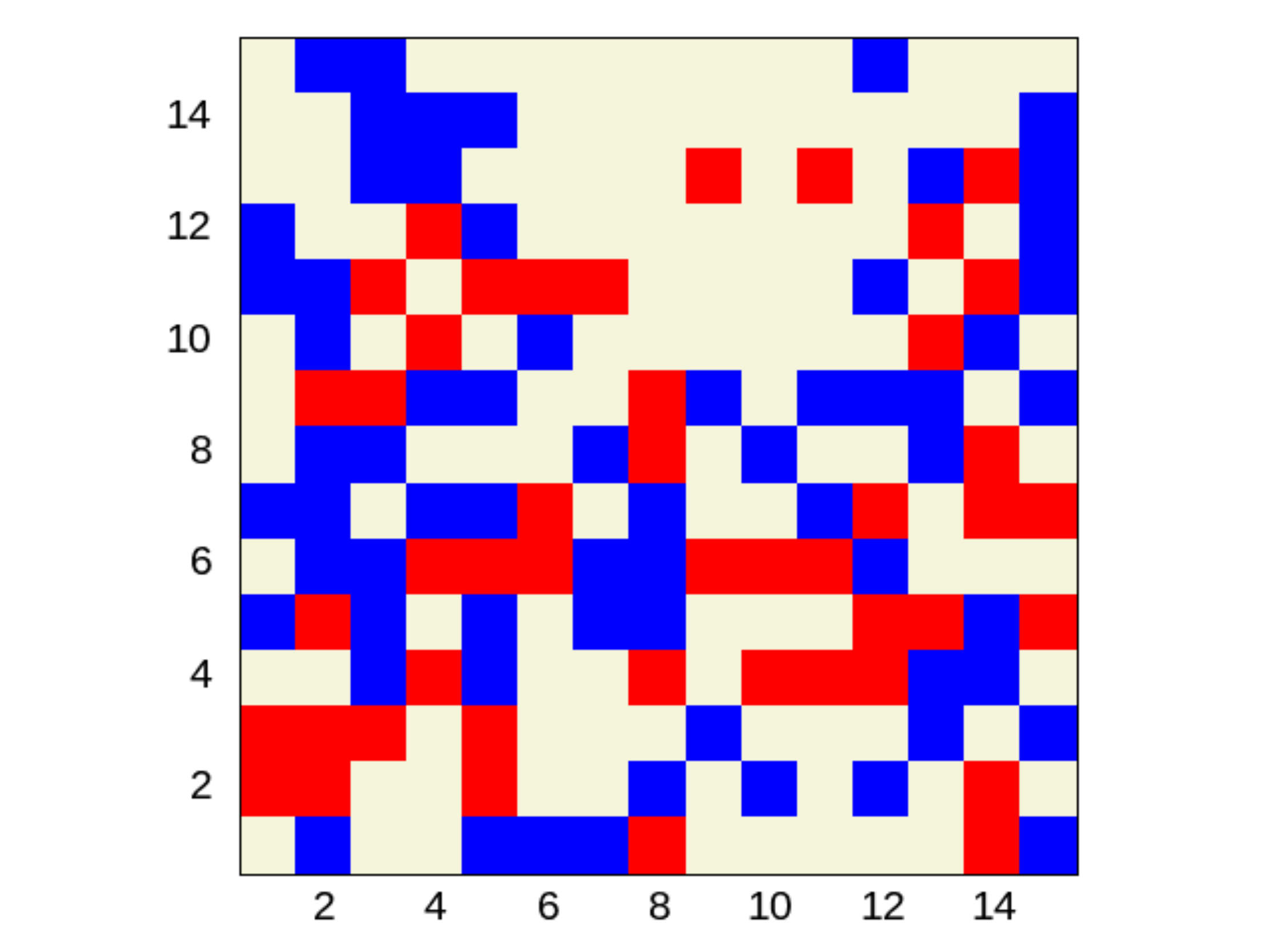} &
	\includegraphics[width = 0.28\textwidth]{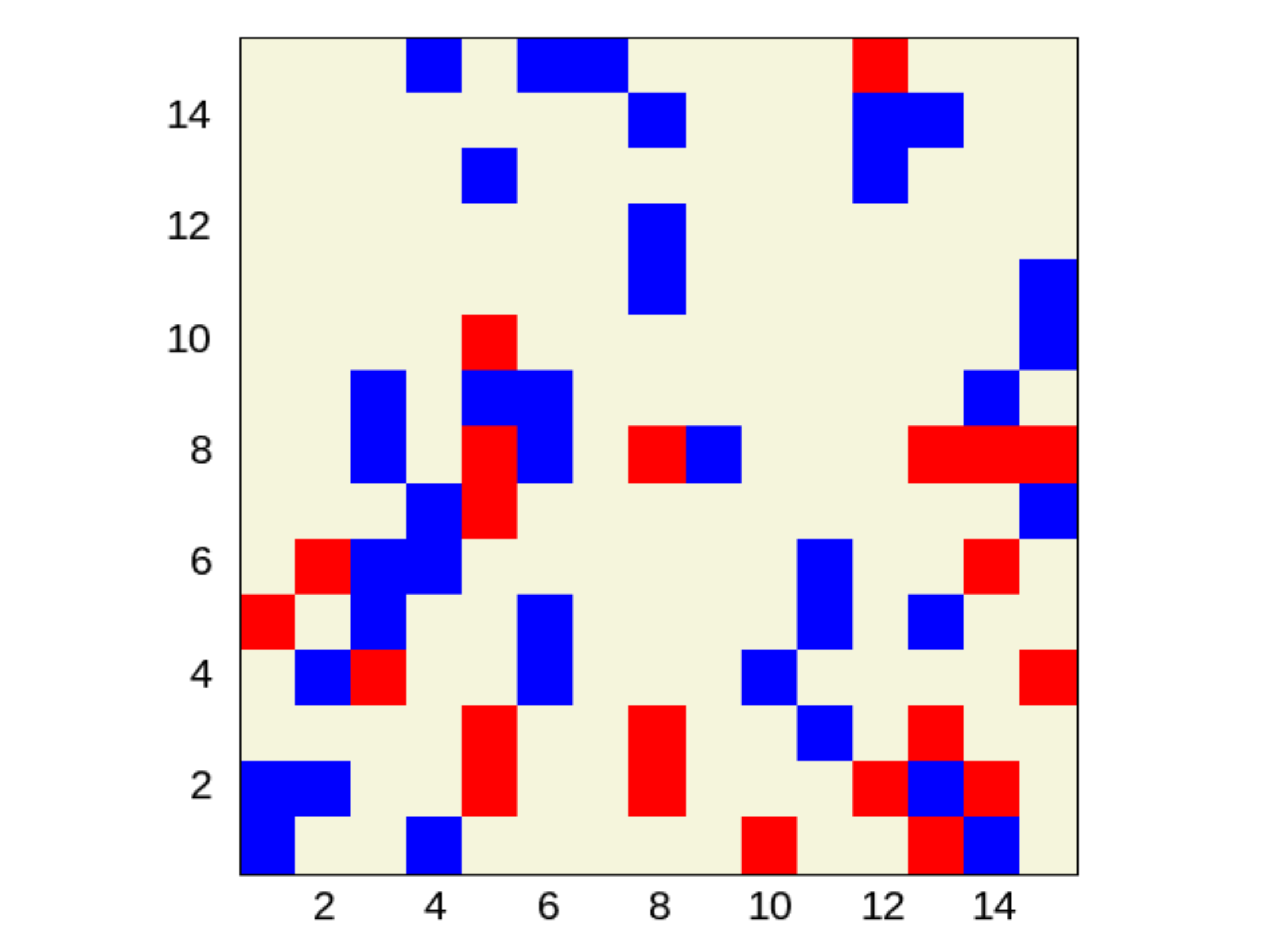}\\[0.1cm]
	\includegraphics[width = 0.28\textwidth]{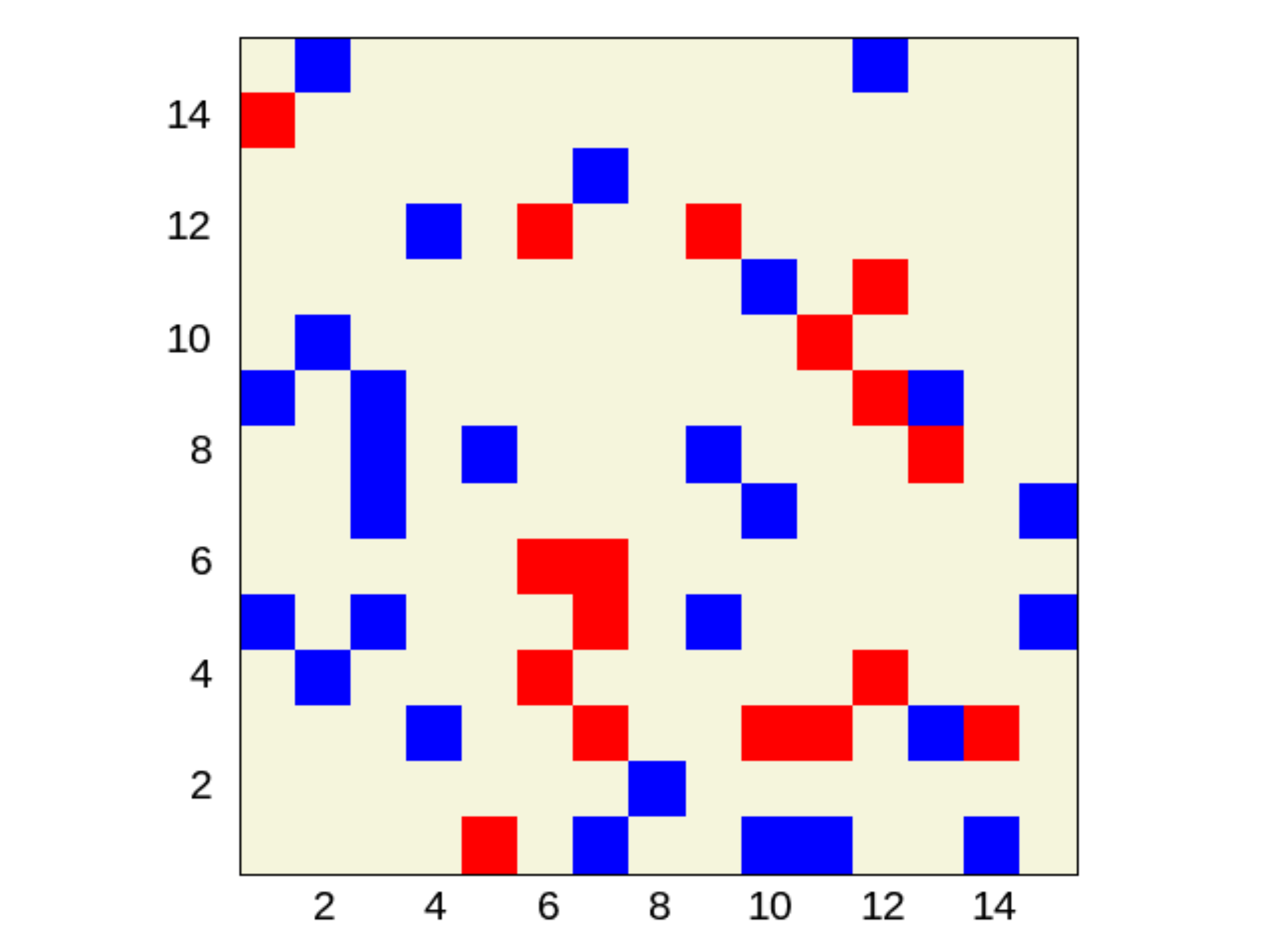} &
	\includegraphics[width = 0.28\textwidth]{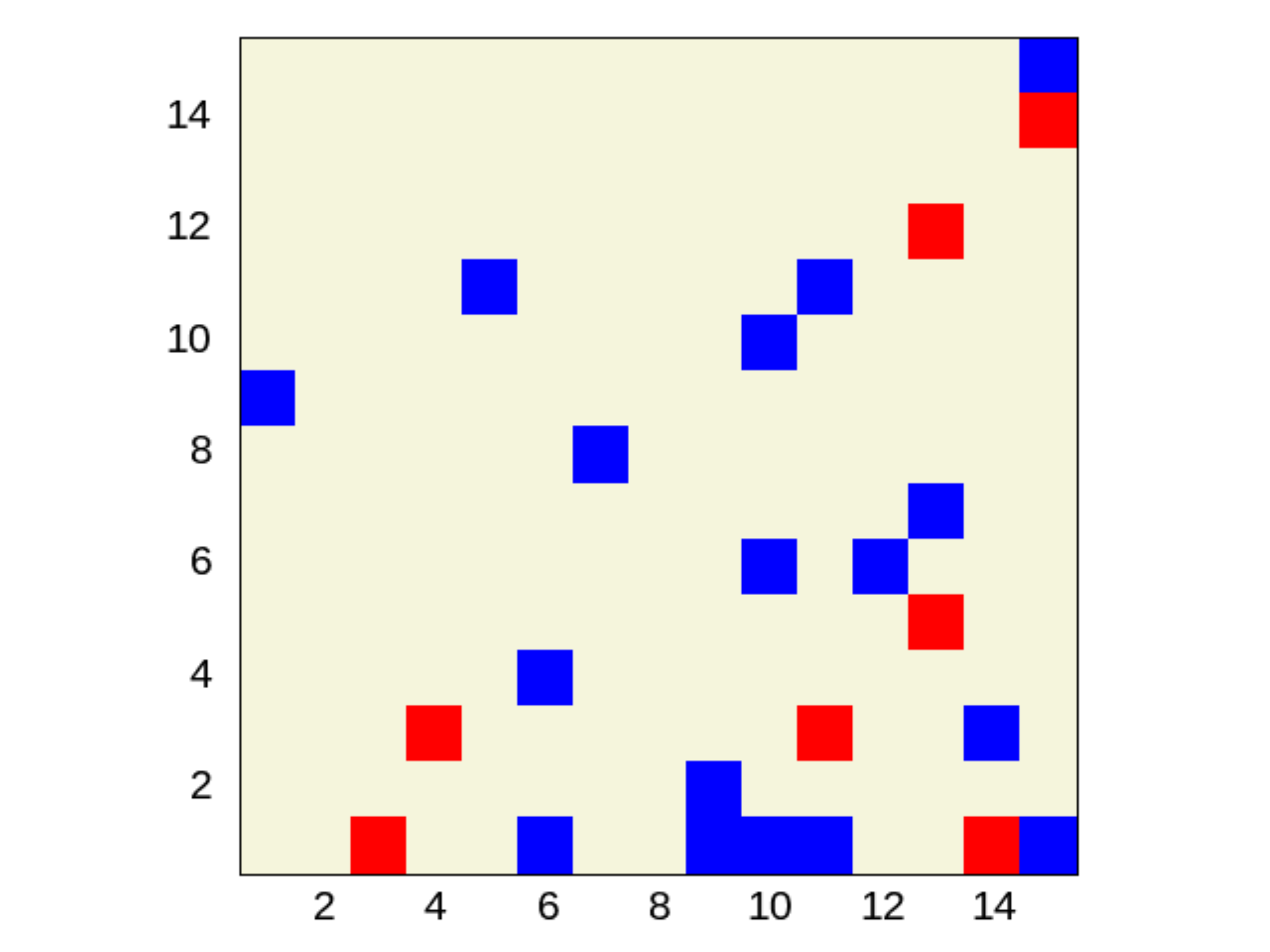}&
	\includegraphics[width = 0.28\textwidth]{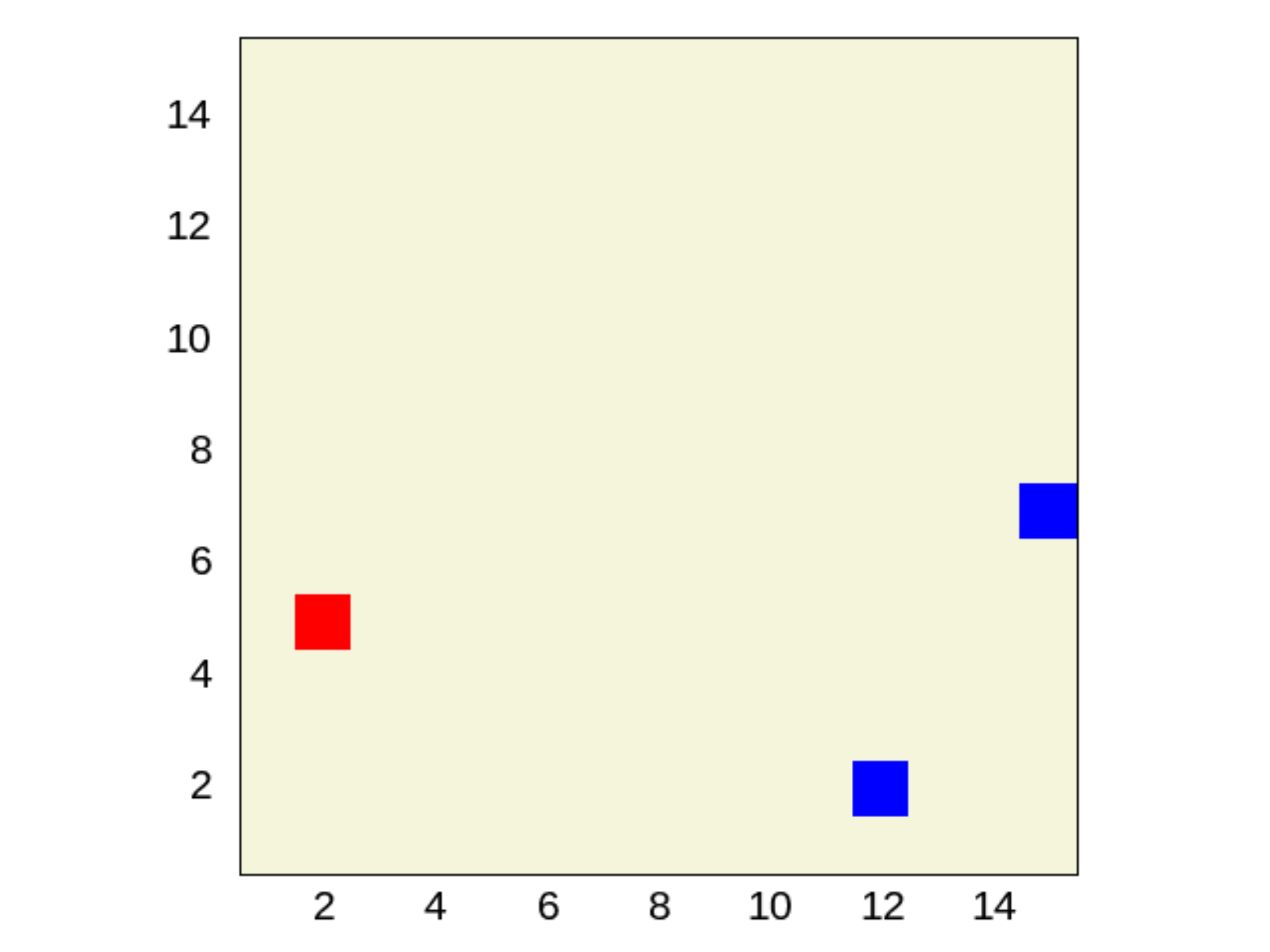}
	\end{tabular}
	\caption{\small Configurations of the model sampled at different 
times (increasing in lexicographic order).
Parameters: $L=15$, $w_{\mathrm{ex}}=7$, $L_{\text{v}}=5$, and 
$\varepsilon=0.3$. 
Red pixels represent active particles, blue pixels denote passive 
particles, and gray sites are empty. 
In the initial configuration (top left panel) there are $70$ active and 
$70$ passive particles.}
\label{fig1:conf}
\end{figure}

\begin{figure}
	\centering
	\begin{tabular}{lll}
	\includegraphics[width = 0.28\textwidth]{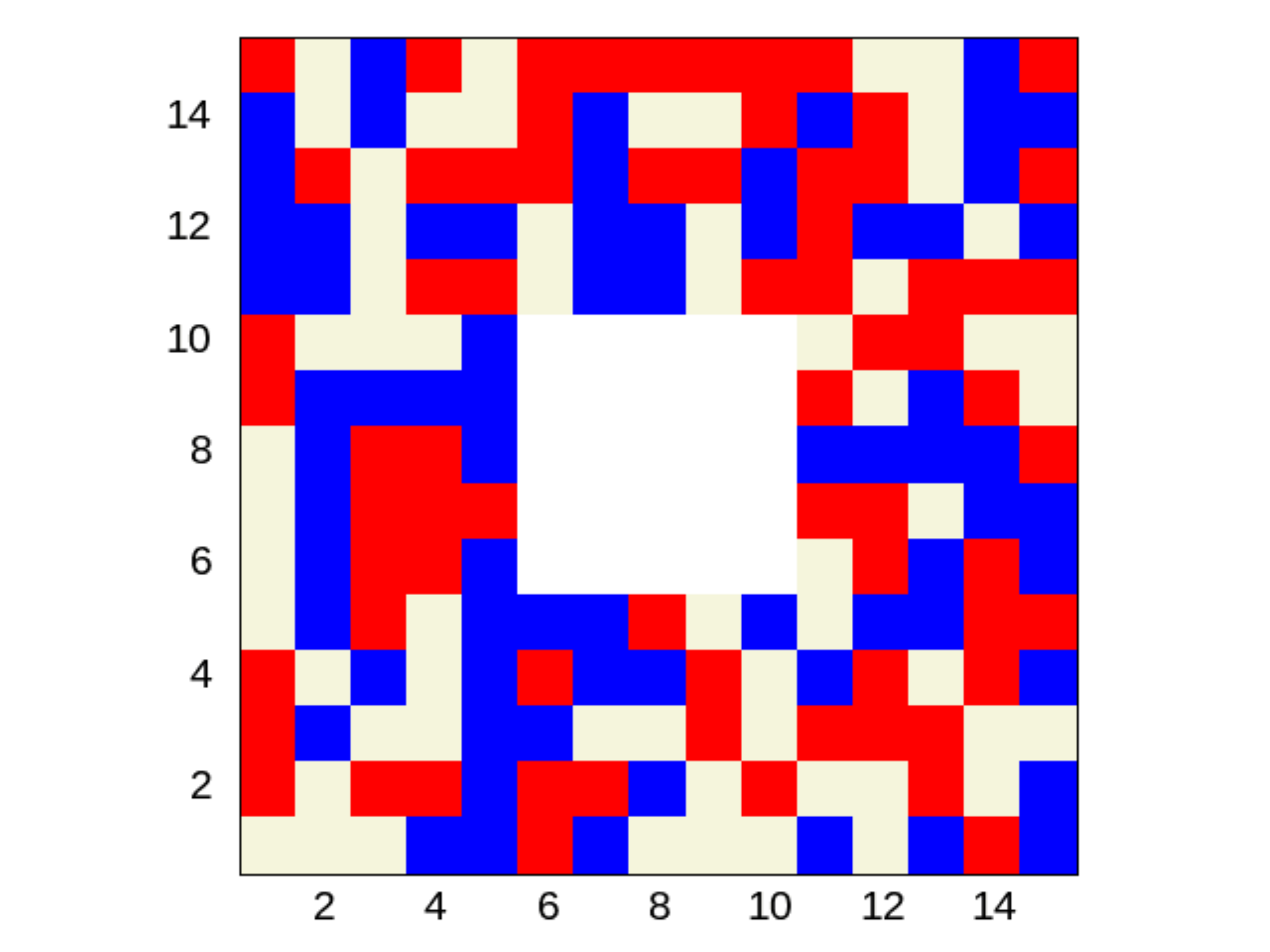} &
	\includegraphics[width = 0.28\textwidth]{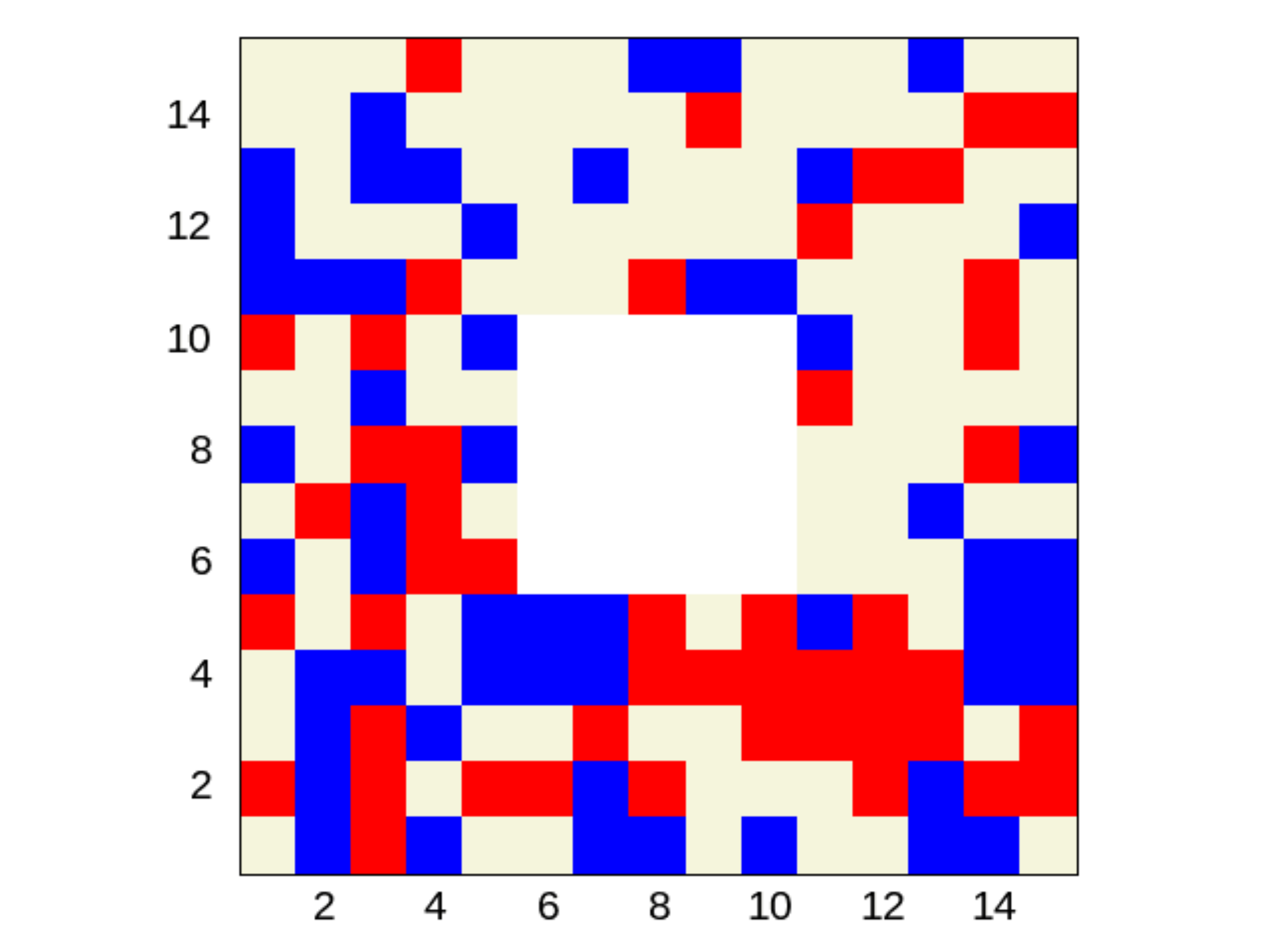} &
	\includegraphics[width = 0.28\textwidth]{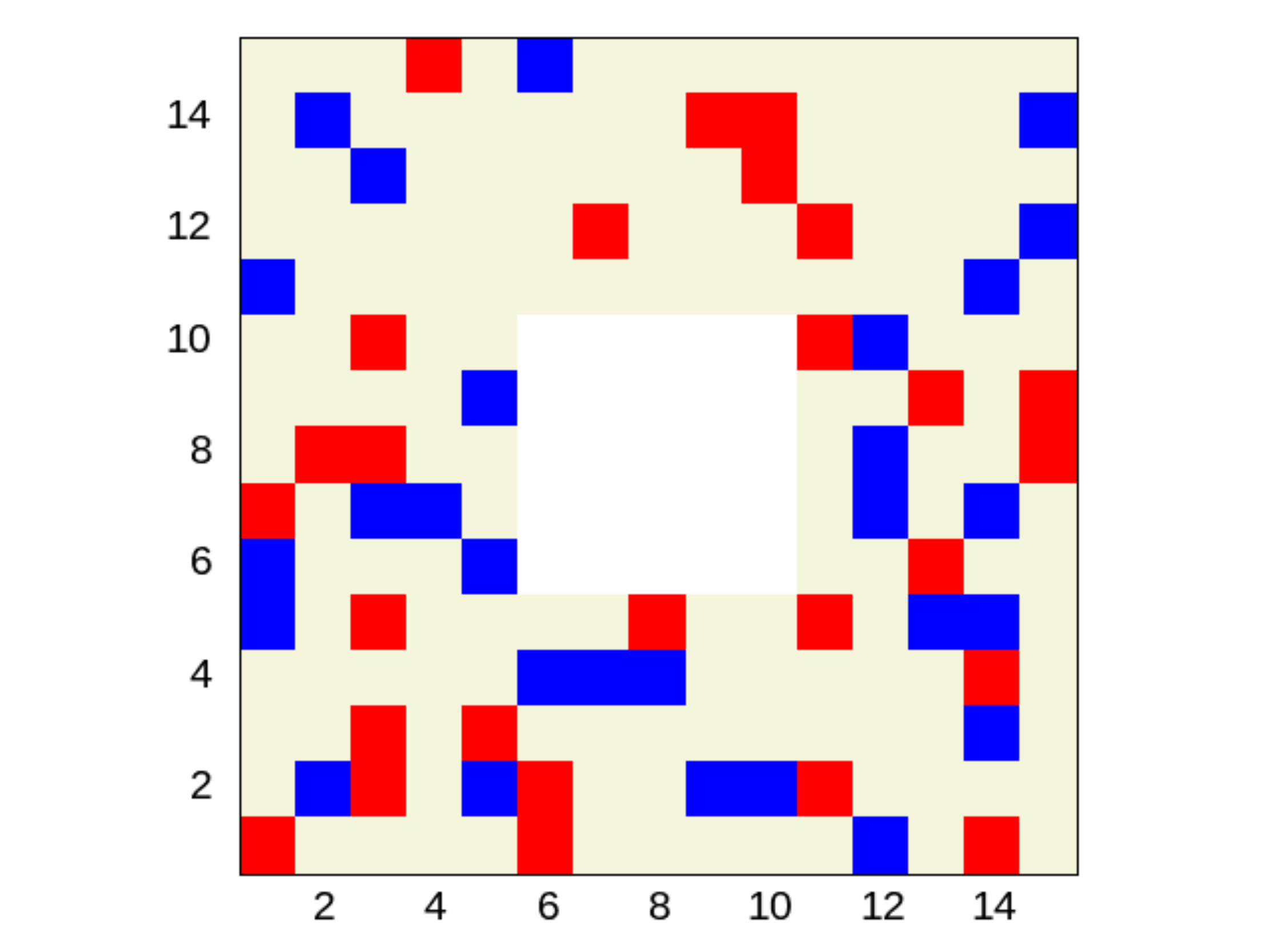}\\[0.1cm]
	\includegraphics[width = 0.28\textwidth]{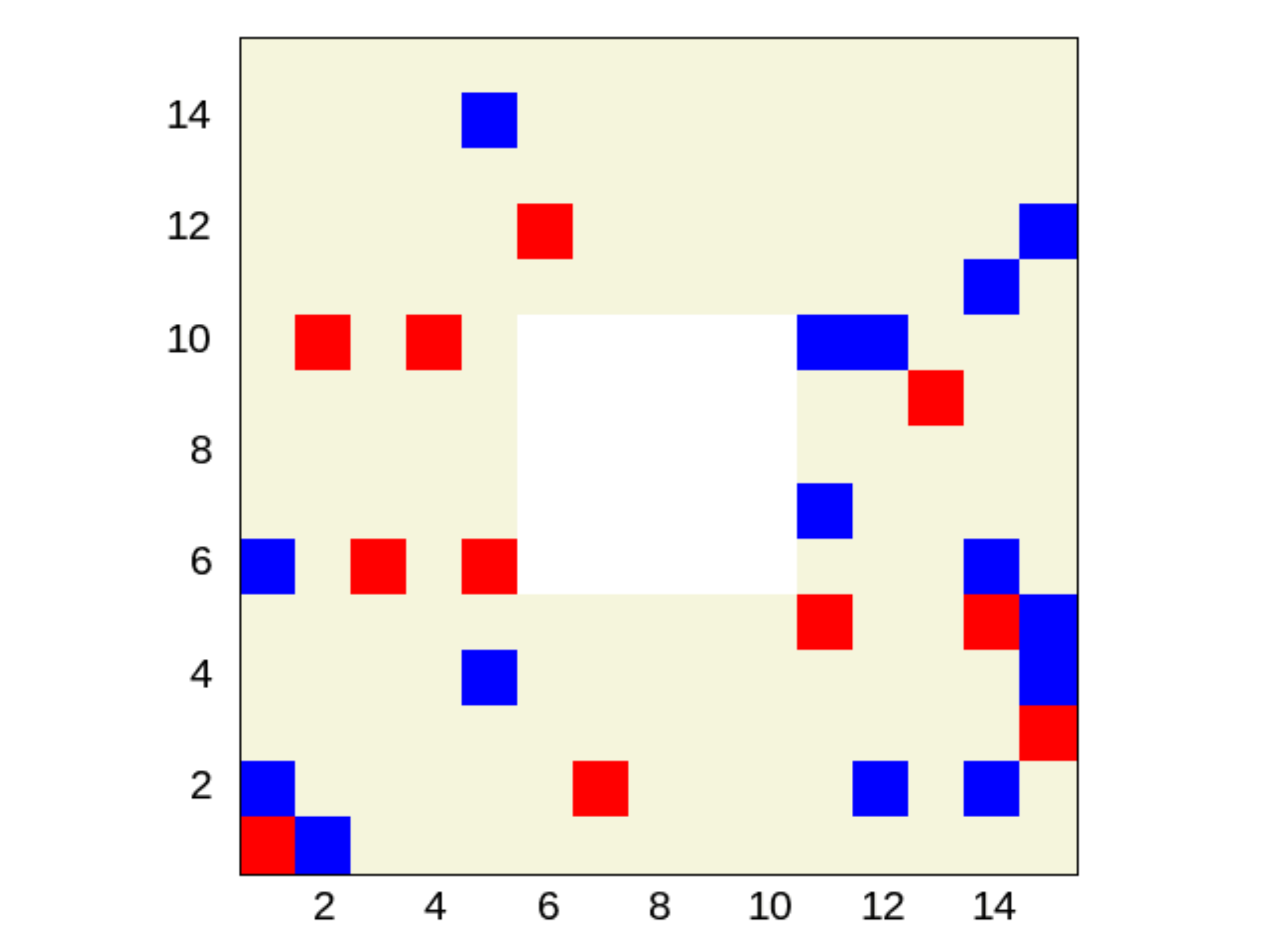} &
	\includegraphics[width = 0.28\textwidth]{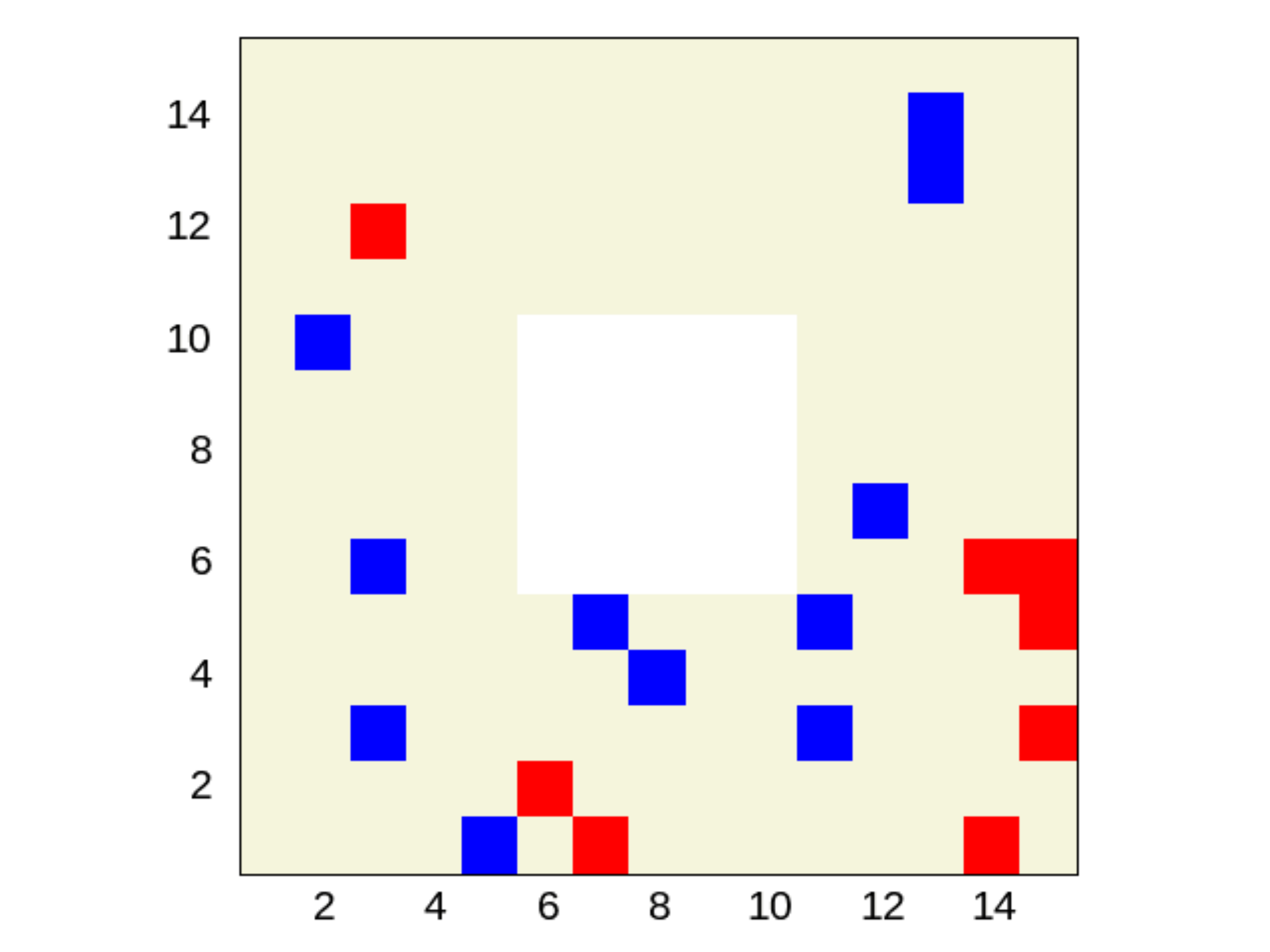}&
	\includegraphics[width = 0.28\textwidth]{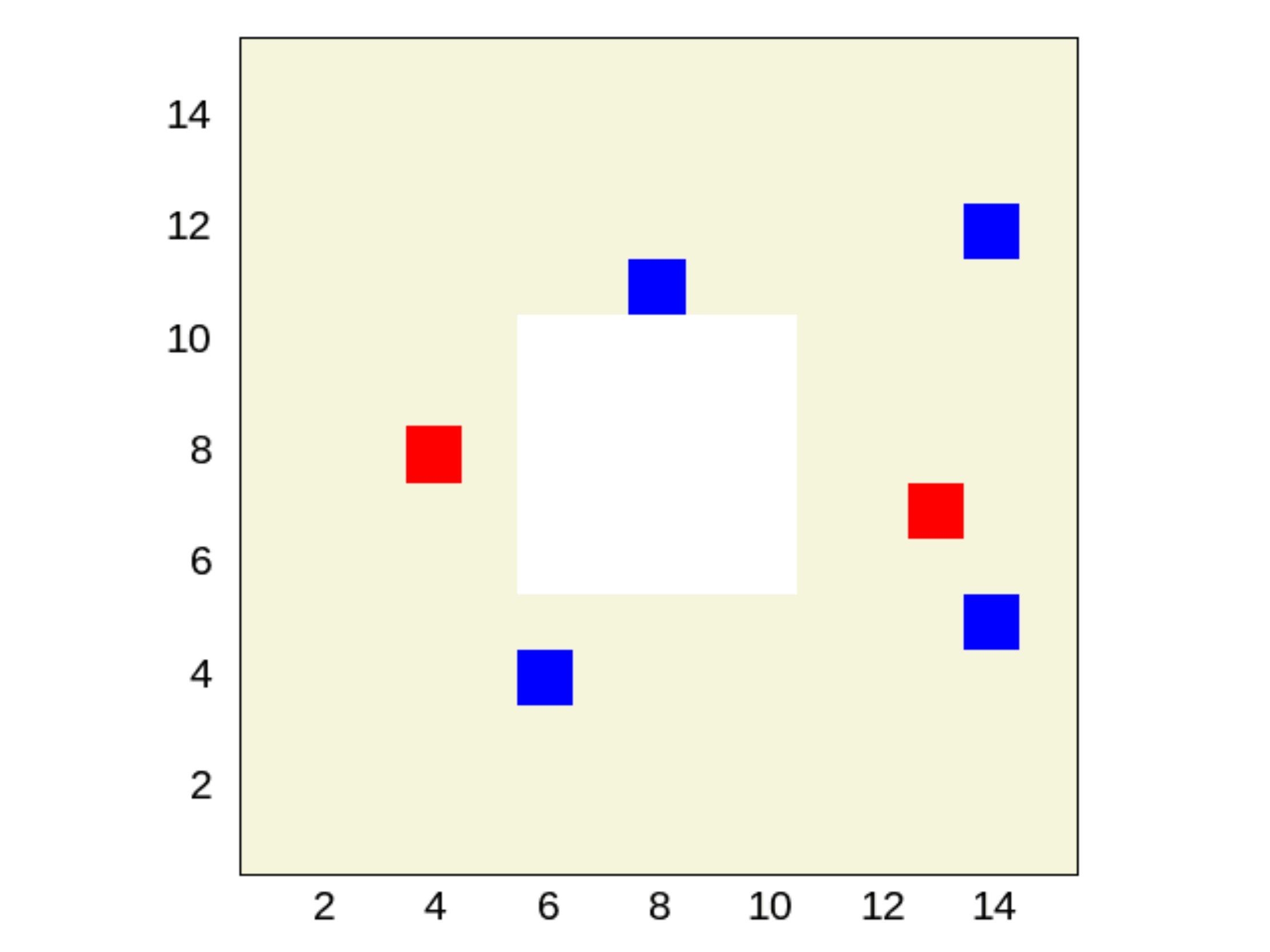}
	\end{tabular}
	\caption{\small As in Figure~\ref{fig1:conf}, the obstacle is a   centered $5\times5$ 
square. This fixed obstacle is depicted with white pixels. 
}
\label{fig1:conf2}
\end{figure}

The \emph{infinitesimal generator} $\mathcal{L}$ acts on continuous bounded functions 
$f:\Omega\to\mathbb{R}$ as 
\begin{equation}
\label{mod040}
\mathcal{L}(\eta)
=\sum_{\eta'\in\Omega}
c(\eta,\eta')[f(\eta')-f(\eta)].
\end{equation}
The probability measure induced by the Markov chain 
started at $\eta$ is denoted by 
$P_\eta$ and the related 
expectation is 
denoted by $E_\eta$. We refer to \cite{Sp91, Pavliotis} where 
similarly--in--spirit models are described mathematically in a rigorous fashion.

The initial number of active (respectively, passive) 
particles is denoted by 
$N_{\text{A}}=n_{\text{A}}(\eta_x(0))$
(respectively, $N_{\text{P}}=n_{\text{P}}(\eta_x(0))$).
We also let $N=N_{\text{A}}+N_{\text{P}}$ be the initial total 
number of particles. 

For any choice of the initial configuration $\eta(0)$ in $\Omega$, the 
process will eventually reach 
the \emph{empty configuration}
$\underline{0}$
corresponding to zero particles in the corridor which 
is an absorbing point of the chain.  

As alternative working scenario, we will study the dynamics described above also in 
the presence of a solid obstacle hindering the pedestrian flow in the 
corridor. The \emph{obstacle} is made of 
an array of sites permanently occupied by fictitious particles. 
In such a way these sites are never accessible to the actual interacting particles 
within  our system. Although, in principle, there is no restriction on the choice of the obstacle geometry, in this framework we  always 
consider centered squared obstacles. A  thorough investigation of the effect of the choice of obstacle geometry on the evacuation time for mixed pedestrian populations moving thorough partially obscure corridors deserves special attention and will be done in a forthcoming work.

We 
simulate this process using the 
following scheme:
at time $t$ we 
extract an exponential random time $\tau$ with parameter
the total rate
$\sum_{\zeta\in\Omega}c(\eta(t),\zeta)$
and set the time equal to $t+\tau$.
We then select a configuration using the probability 
distribution 
$c(\eta(t),\eta)/\sum_{\zeta\in\Omega}c(\eta(t),\zeta)$
and set $\eta(t+\tau)=\eta$.

As we have already pointed out in Section~\ref{s:int}, the main goal 
of the paper is to detect drafting in pedestrian flows, namely, 
identify situations when 
the evacuation of passive particles is favored by the presence 
of the active ones, even if no leadership or other kind of information exchange is allowed. 
We expect that this phenomenon will be effective, provided the 
active particles
will spend a sufficiently long time in the 
corridor. This seems to help efficiently passive particles to escape. This effect is illustrated in the Figures~\ref{fig1:conf} and \ref{fig1:conf2},
where we show the configuration of the system at different times 
both 
in the absence, and respectively,  in the presence of an obstacle.
Indeed, the sequences of configurations show that, though the 
evacuation of active particles is faster than that of the passive ones, 
even at late times the fraction of active particles is still 
reasonably high.

\section{The evacuation time}
\label{s:eva}
Consider the dynamics defined in Section~\ref{s:mod}, 
given a configuration $\eta\in\Omega$. We let $\tau_\eta$ be the first 
hitting time to the empty configuration, i.e.
\begin{equation}
\label{eva000}
\tau_\eta
=
\inf\{t>0:\,\eta(t)=\underline{0}\},
\end{equation}
for the chain started at $\eta$. 
Given a configuration $\eta\in\Omega$, we define the 
\emph{evacuation time} starting from $\eta$ as
\begin{equation}
\label{eva020}
T_\eta
=
\mathbb{E}_\eta[\tau_\eta]
\;.
\end{equation}

We have defined the evacuation time as the time 
needed to evacuate all the particles initially in the system, that is to 
say the evacuation time is the time at which the last particle leaves the 
corridor. 
In this section as well as in the next one, we study numerically 
the evacuation time for a fixed initial 
random condition and then produce various realizations of the process for specific  values of the initial drift $\varepsilon$ and  of the visibility depth 
$L_{\text{v}}$.
To this end, we consider two geometrically different situations: (i) the empty corridor and (ii) the 
corridor with a squared obstacle positioned at the center. 

\subsection{The empty corridor case}
\label{s:emp}
We consider the system defined in Section~\ref{s:mod} 
for $L=15$ (side length of the corridor), 
$w_{\text{ex}}=7$ (exit width), 
$N_{\text{P}}=70$  (initial number of passive particles)
$N_{\text{A}}=0,70$  (initial number of active particles)
$L_{\text{v}}=2,5,7,15$ (visibility depth), 
and 
$\varepsilon = 0.1,0.3,0.5$ (drift).
More details are provided in the figure captions. 

\begin{figure}[tb]
    \centering
    \begin{subfigure}[]
        {\includegraphics[width=0.4\textwidth]{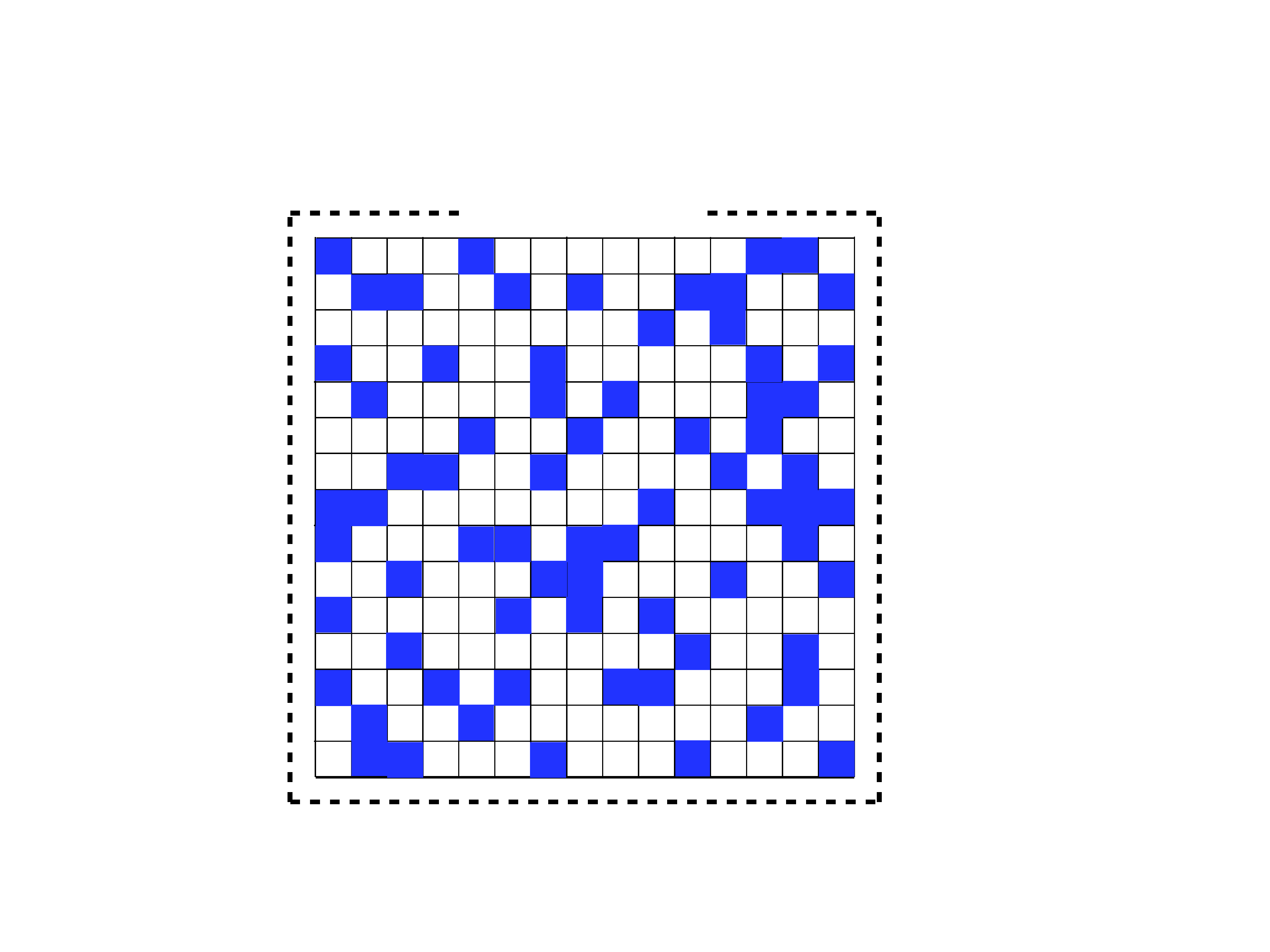}}
    \end{subfigure}
    \hskip -2 cm
    \begin{subfigure}[]
        {\includegraphics[width=0.4\textwidth]{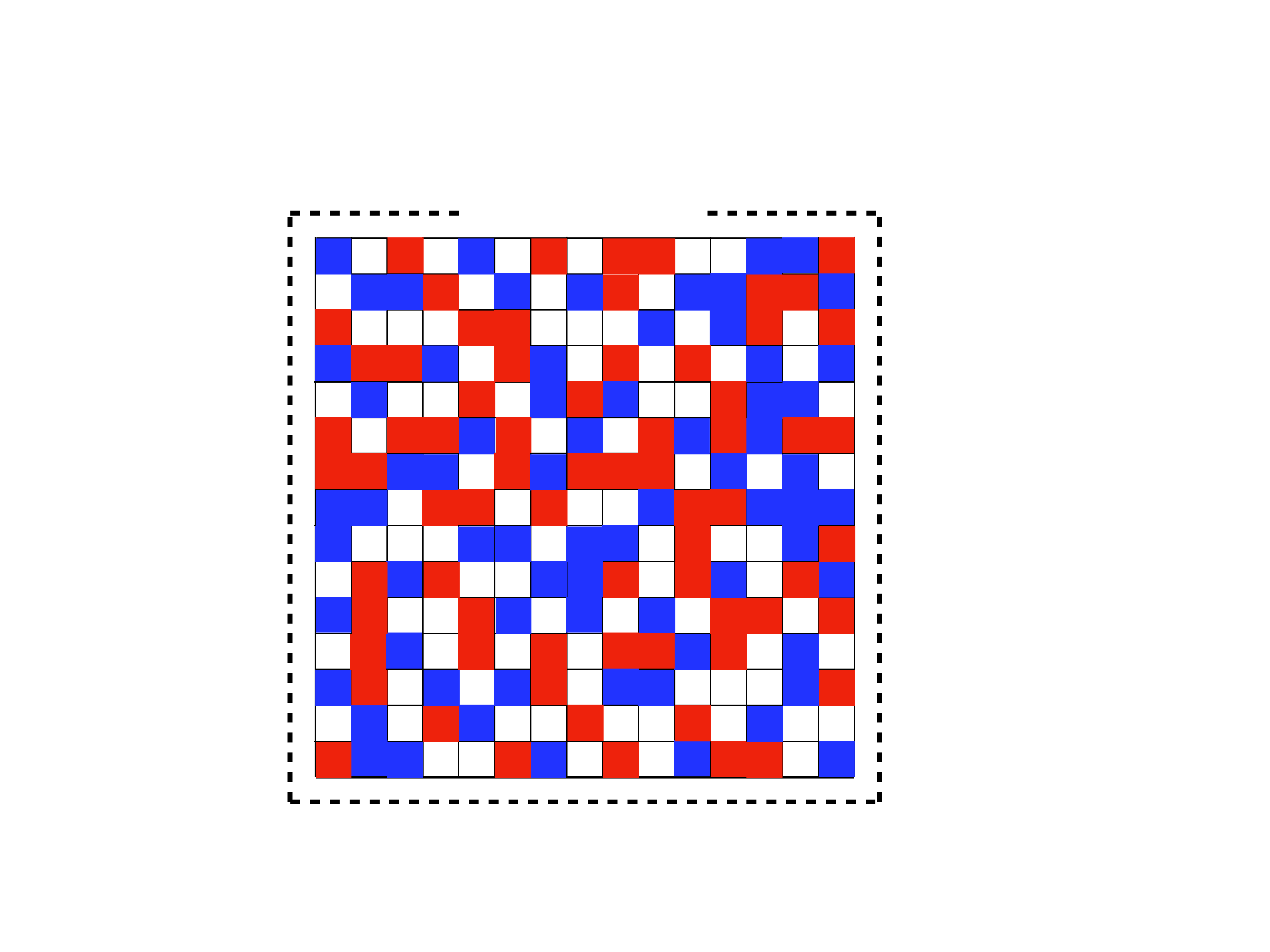}}
    \end{subfigure}
    \caption{Two initial configurations for the lattice gas dynamics. Blue and red pixels represent, respectively, passive and active particles. The thick dashed line surrounding a large fraction of the grid denotes the presence of reflecting boundary conditions. The exit door is located in presence of the missing segment of dashed line. In (a) only $N_{\text{P}}$ passive particles are present. In (b), the passive particles occupy the same initial positions as in (a), and $N_{\text{A}}$ active particles are also included (we fix  $N_{\text{A}}=N_{\text{P}}$). We shall compare the evacuation time relative to the two configurations in (a) and (b).}\label{fig:fig0.5}
\end{figure}

All the simulations are done starting the system from the same initial 
configuration chosen once for all by distributing the particle 
at random with uniform probability. More precisely, two initial 
configurations are considered, one for the case 
$N_{\text{P}}=70$ and $N_{\text{A}}=0$  
and one for the case 
$N_{\text{P}}=70$ and $N_{\text{A}}=70$, chosen in such a way that 
in the two cases the initial positions of the passive particle is the same, see Figure~\ref{fig:fig0.5} 
for a schematic illustration.

\begin{figure}
	\centering
	\includegraphics[width=0.45\textwidth]{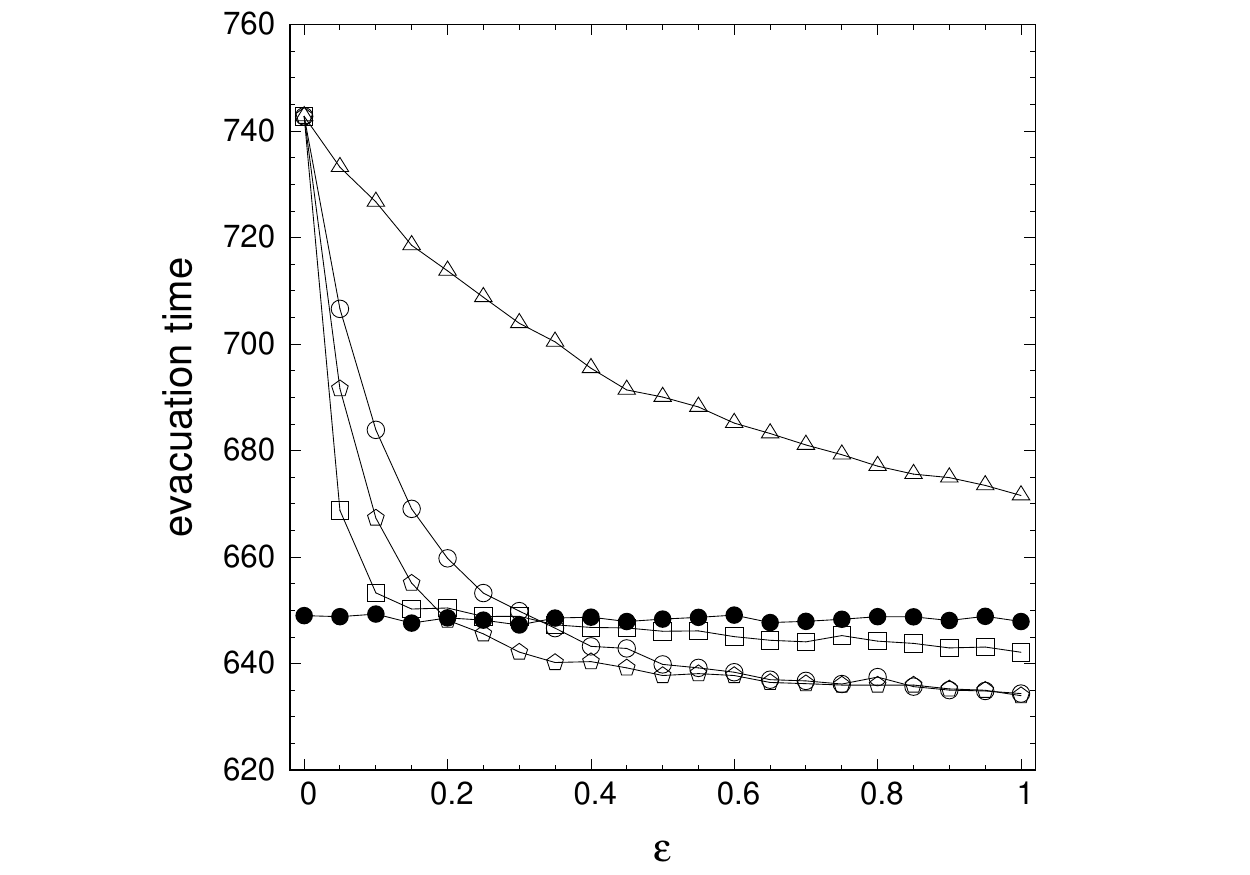}
	\includegraphics[width=0.45\textwidth]{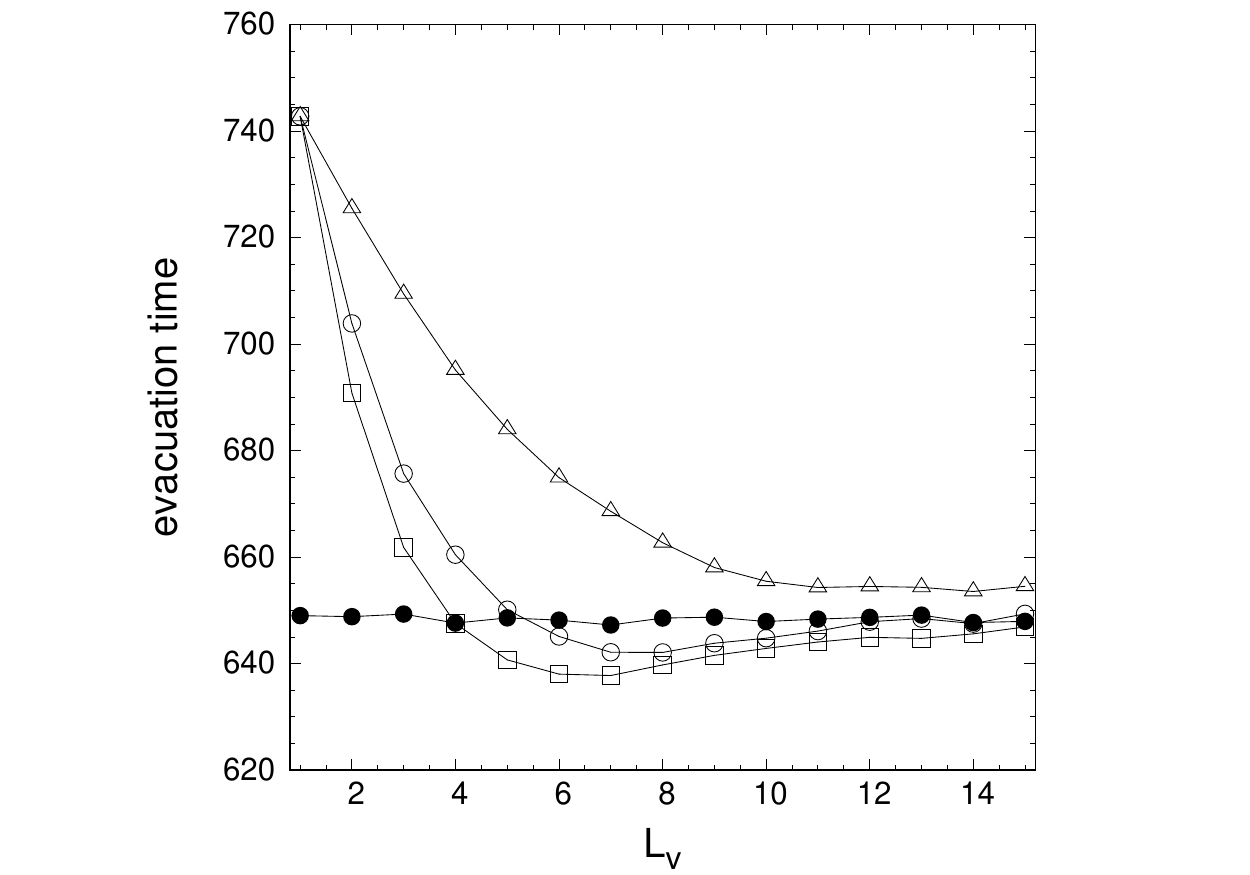}
	\caption{Evacuation time in an empty corridor 
for $L=15$, $w_{\text{ex}}=7$, 
$N_{\text{A}}=0$ and $N_{\text{P}}=70$  (solid disks)
and 
$N_{\text{A}}=N_{\text{P}}=70$ (open symbols). 
Left panel: $L_{\text{v}}=2$ (open triangles), 
$L_{\text{v}}=5$ (open circles),  
$L_{\text{v}}=7$ (open pentagons), 
$L_{\text{v}}=15$ (open squares).
Right panel: $\varepsilon = 0.1$ (open triangles), 
$\varepsilon=0.3$ (open circles), 
$\varepsilon = 0.5$ (open squares).
	}
	\label{fig:fig1}
\end{figure}

We then compute the time needed to evacuate all the particles 
initially in the systems and, by averaging over 
$10^5$ different realizations 
of the process, we compute a numerical estimate of the evacuation
time \eqref{eva020} for the chosen initial condition. Results 
are reported in Figure~\ref{fig:fig1}.

The main result of our investigation is the following: 
the evacuation time $T_{\eta}$ corresponding to the initial configuration 
with active particles 
(see the illustration (b) in Figure~\ref{fig:fig0.5})
is \textit{less} than that corresponding to the initial configuration with 
sole passive particles
(see the illustration (a) in Figure~\ref{fig:fig0.5}).
This result is 
non--trivial since our lattice dynamics is based on a hard core exclusion principle \cite{Sp91} --  the motion of particles towards the exit is hindered by the presence of nearby particles. In the context of this work, we refer to this phenomenon as drafting, marking this way the analogy with the drafting or aerodynamic drag effect encountered by pelotons of cyclists racing towards the goal; we refer the reader, for instance, to \cite{BvT+2018,BTvT2018} and references cited therein, for wind tunnel and computational evidence on drafting.  It is crucial to note that the presence of active particles is essential for the onset of this phenomenon: if all  active particles in the configuration (b) in Figure~\ref{fig:fig0.5} (represented by the red pixels) were replaced by passive ones (blue pixels), the evacuation time would clearly become larger than with the configuration (a). This is essentially due to the exclusion constraint of the lattice gas dynamics.

In the left panel in Figure~\ref{fig:fig1}, the dependence of the 
evacuation time on the drift $\varepsilon$ is shown.
Open symbols refer to the evacuation time for 
$N_{\text{P}}=70$ and $N_{\text{A}}=70$; for each 
value of $\epsilon$ we repeat the measure of the evacuation 
time also for a system in which only passive particles are present. 
We then obtain the sequence of solid disks reported in the figure 
which is approximatively constant, since the dynamics of 
the passive particles does not depend on $\varepsilon$. We observe that the small fluctuations visible in the data represented by solid disks Figure~\ref{fig:fig1} come from considering averages over a finite number of different realizations of the process (all starting from the given initial configuration).

Since the number particles in the initial configuration in the 
experiments with the presence of 
active particles is double with respect to that considered in the case 
of only passive particles, one would expect a larger evacuation time. 
This is indeed the case for a small visibility depth, i.e. for $L_{\text{v}}=2$.
In such a case, it is worth noting that the evacuation time decreases 
when the drift is increased as it is in fact reasonable since a larger drift 
favors the fast evacuation of active particles, but it remains larger than the 
evacuation time in absence of active particles for all the values of 
$\varepsilon$ that we considered. 
On the other hand, for larger values of the visibility depth, as long as 
the drift is large enough, the evacuation time in the presence of active 
particles becomes smaller than the one measured in the presence of only passive 
particles. 

This effect
is rather surprising. It can interpreted by saying 
that the presence of active particles, which have some information about 
the location of the exit, helps passive particles to evacuate the 
corridor even if no information exchange is allowed, and, mostly, even in presence of the exclusion constraint of the lattice gas dynamics. Indeed, passive 
particles continue their blind symmetric dynamics, nevertheless 
their evacuation time is reduced. The sole interaction among passive 
and active particles is the exclusion rules, hence one possible 
interpretation of this effect is that active particles, while walking 
toward the exit, leave behind a sort of empty sites wake. 
Passive particles, on the other hand, can benefit of such an empty 
path and be thus blindly driven towards the exit. 
A different interpretation is that passive particles, due to the 
exclusion rule, are pushed by active particles towards the exit. 
  
In the right panel of Figure~\ref{fig:fig1}, for the same choices of 
parameters and initial conditions, we show the evacuation time 
as a function of the visibility depth $L_{\text{v}}$ 
for several values of the drift $\varepsilon$.
Data can be discussed similarly as we did for those plotted in 
the left panel of the same figure: for small drift, and for any 
choice of the visibility length, the evacuation time in presence
of active particles is larger than the one measured with sole 
passive particles. But, if the drift is increased, for 
a sufficiently large visibility depth, the evacuation time 
in presence of active particles becomes smaller than the one 
for sole passive particles though the total number of 
particles to be evacuated is doubled.  
As before we interpret these data as an evidence of the presence 
of the drafting effect. 

Remarkably, for sufficiently large drift $\varepsilon$, the evacuation 
time is not monotonic with respect to the visibility depth.
In other words, there is an optimal value of $L_{\text{v}}$ which 
minimizes the evacuation time. The fact that for 
$L_{\text{v}}$ too large, i.e., comparable with the side length 
of the corridor, the evacuation time increases with $L_{\text{v}}$ 
can be explained remarking that if active particles exit the system 
too quickly then passive particles, left alone in the corridor, 
evacuate it with their standard time. Hence, the drafting effect is visible as long as the parameters $\varepsilon$ and $L_{\text{v}}$ make the motion of the active particles towards the exit enough faster than that of passive particles, but not too fast. Indeed, if the active particles move too slowly, they behave as passive ones: this would make the evacuation time larger, due to the standard exclusion constraint of our lattice gas dynamics. On the other hand, if the active particles move too fast, passive particles remain soon alone on the lattice, therefore the evacuation time relative to the configuration of type (b) in Figure~\ref{fig:fig0.5} reduces to that relative to the configuration of type (a).

\begin{figure}
	\centering
	\includegraphics[width=0.45\textwidth]{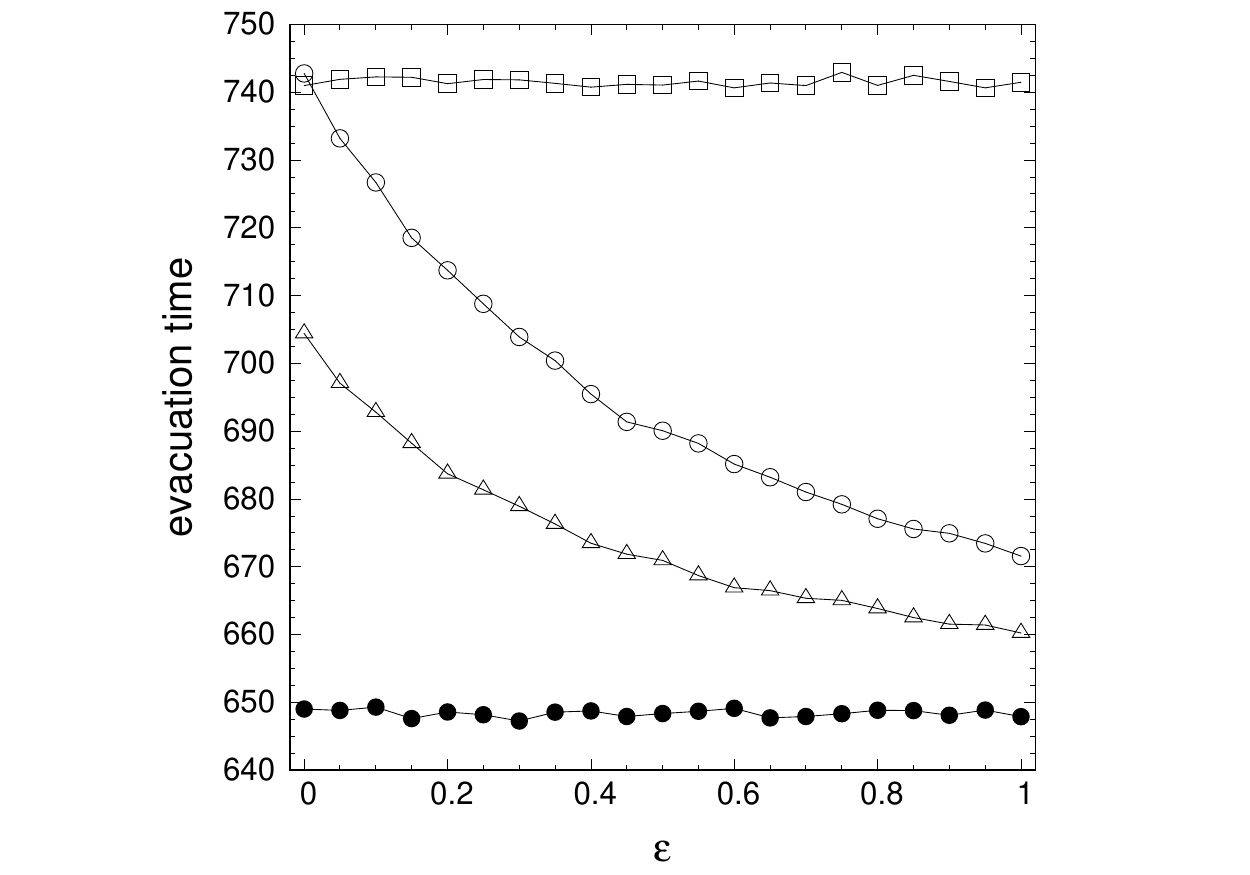}
	\includegraphics[width=0.45\textwidth]{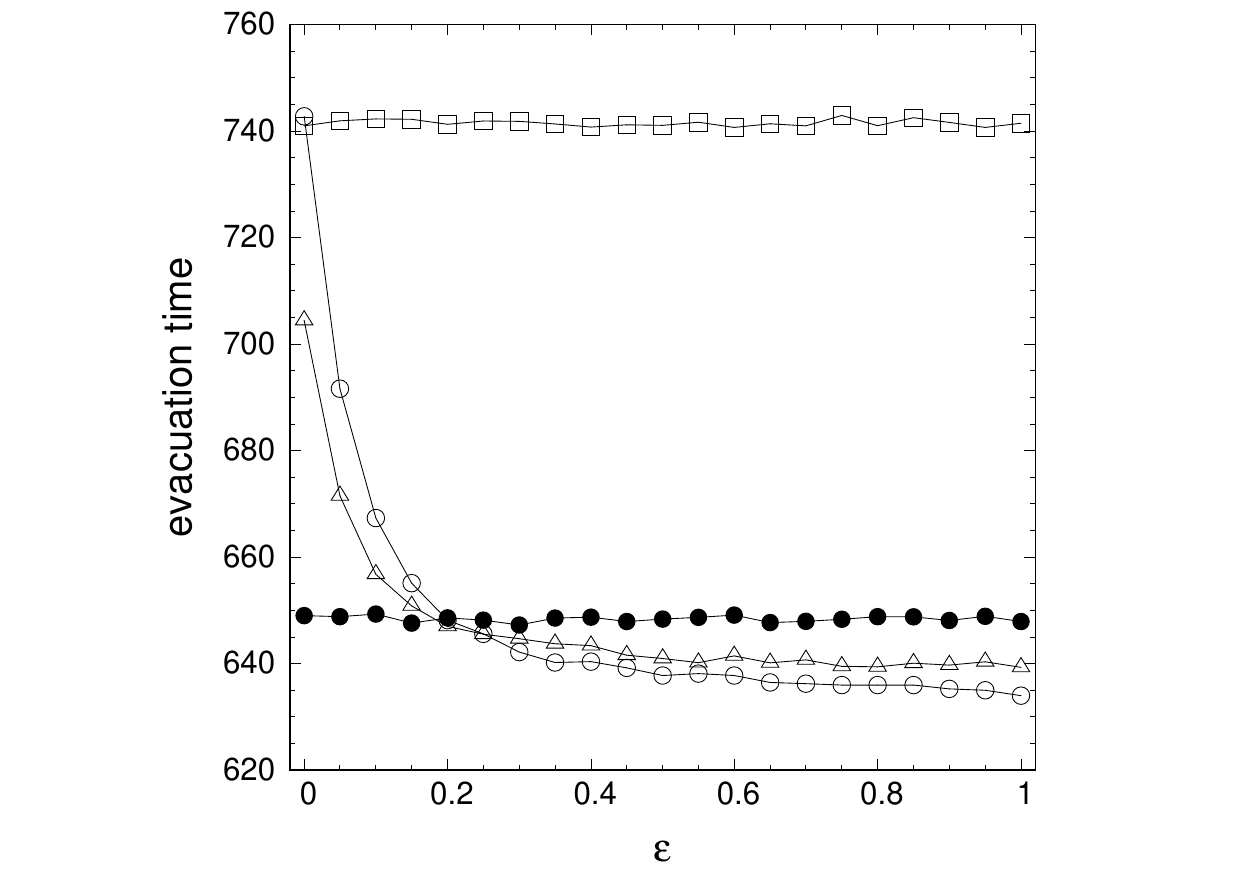}
	\caption{Evacuation time in an empty corridor 
for $L=15$, $w_{\text{ex}}=7$, 
$N_{\text{A}}=0$ and $N_{\text{P}}=70$  (solid disks),
$N_{\text{A}}=35$ and $N_{\text{P}}=70$ (open triangles), $N_{\text{A}}=70$ and $N_{\text{P}}=70$ (open circles) and $N_{\text{A}}=0$ and $N_{\text{P}}=140$ (open squares).
Left panel: $L_{\text{v}}=2$.  
Right panel: $L_{\text{v}}=7$.
	}
	\label{fig:placeholder1}
\end{figure}

Before concluding this Section, we shall also highlight the effect of varying the relative amount of active and passive particles in the initial configuration. In Figure~\ref{fig:placeholder1} we also present the average evacuation times for the cases with $N_{\text{P}}=70$, $N_{\text{A}}=35$  
and $N_{\text{P}}=140$, $N_{\text{A}}=0$, for two different values of $L_{\text{v}}$. 
One readily notices that when the number of passive particles is doubled
(case $N_{\text{P}}=140$ and $N_{\text{A}}=0$)
the evacuation time increases and it obviously results to be independent of 
$\varepsilon$ and $L_{\text{v}}$.
For small visibility depth ($L_{\text{v}}=2$ in the 
left panel in Figure~\ref{fig:placeholder1})
the evacuation time for the case 
$N_{\text{A}}=35$ and $N_{\text{P}}=70$ is smaller than the one 
mesured in the case 
$N_{\text{A}}=70$ and $N_{\text{P}}=70$ 
for any $\varepsilon$
and shows a monotonic decrease. 
The results are more interesting for larger visibility depth 
($L_{\text{v}}=7$ in the 
right panel in Figure~\ref{fig:placeholder1}):
if the drift $\varepsilon$ is large enough, namely, larger than about $0.2$, 
the total evacuation time in presence of active particles becomes 
smaller than the one measured for sole passive particles
both for the case 
$N_{\text{A}}=35$ and $N_{\text{P}}=70$ and 
$N_{\text{A}}=70$ and $N_{\text{P}}=70$, that is to say, 
in both cases the drafting effect shows up. 
More interestingly, the evacuation time is smaller in the 
case in which more active particles are present (open circles in 
the picture): this is a sort of signature of the drafting effect.

\subsection{The corridor with an obstacle}
\label{s:obs}
\begin{figure}
	\centering
	\includegraphics[width=0.45\textwidth]{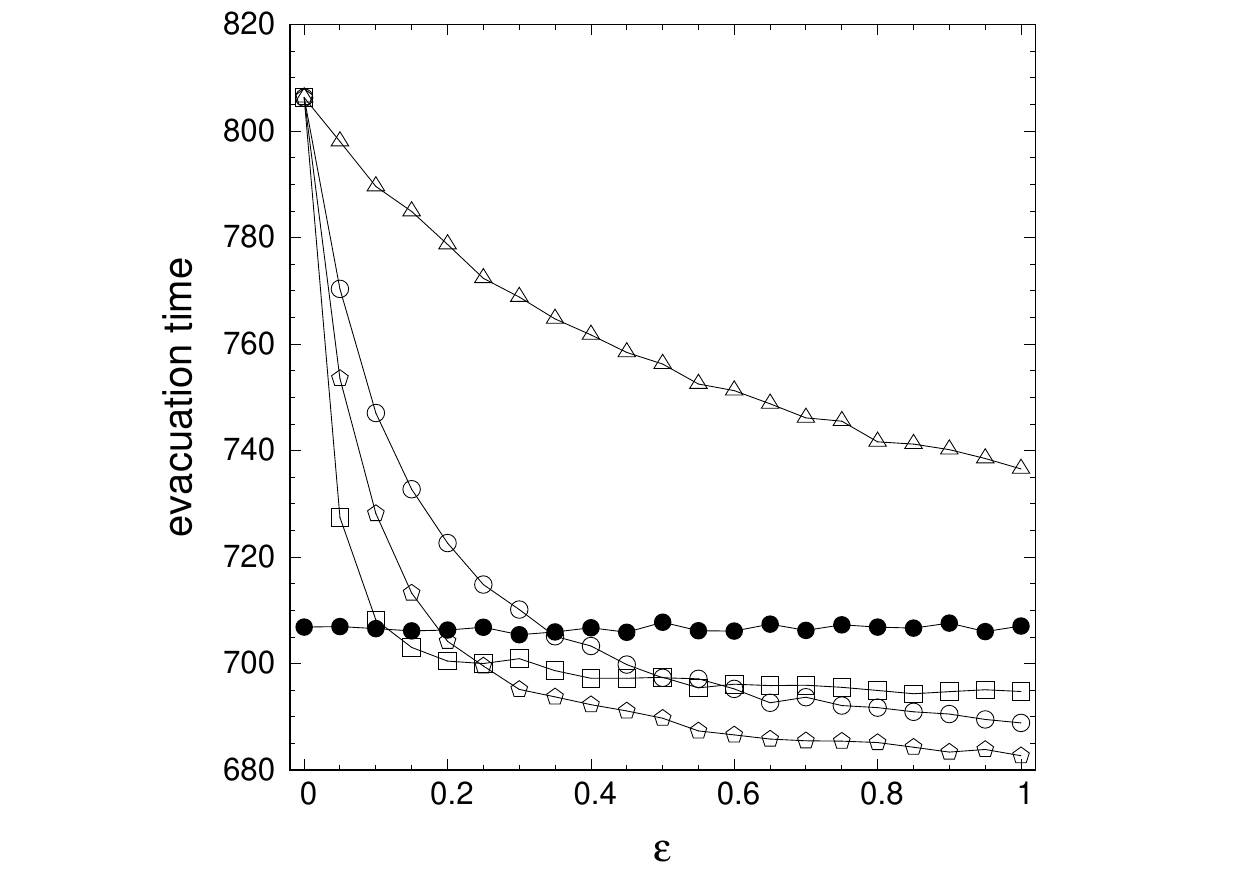}
	\includegraphics[width=0.45\textwidth]{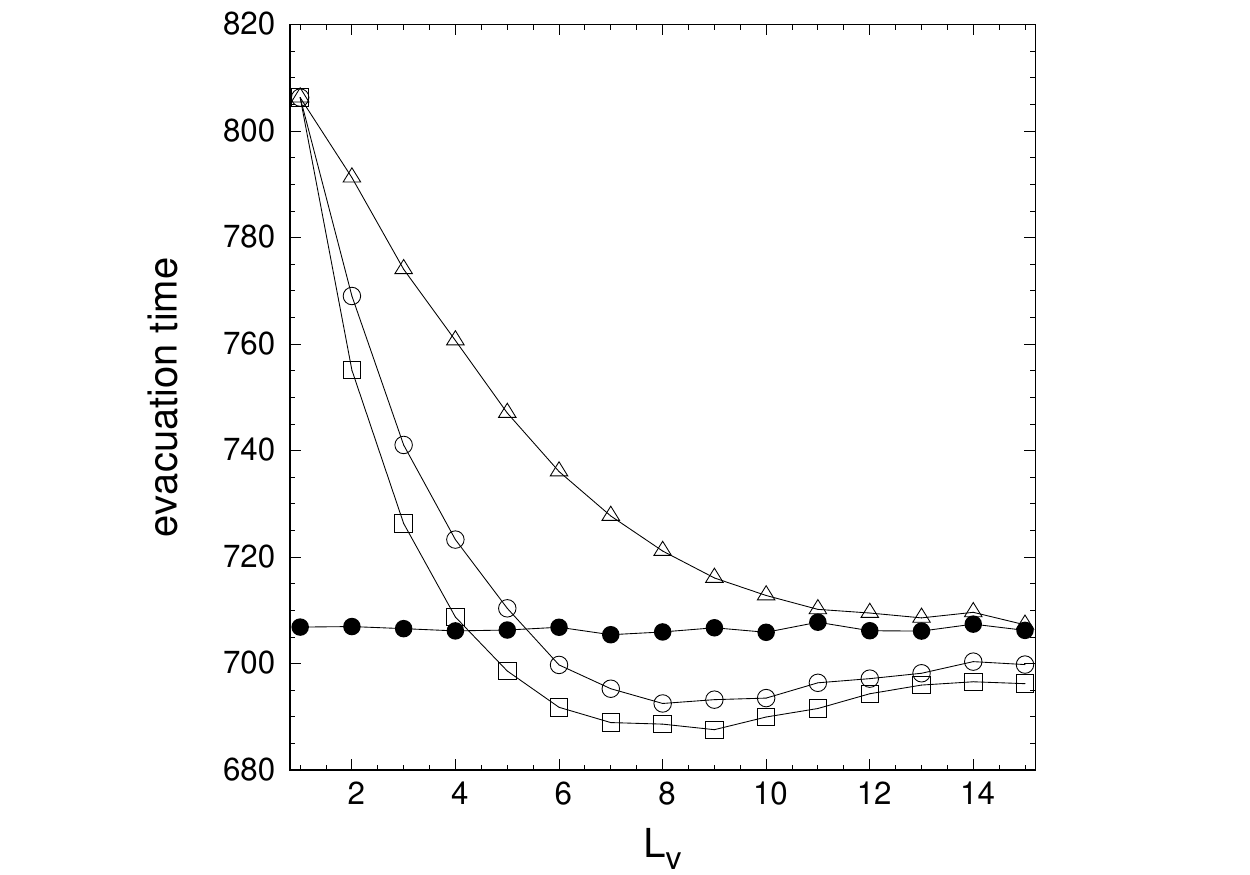}
	\caption{Evacuation time in a corridor 
with a $5\times5$ squared centered obstacle 
for $L=15$, $w_{\text{ex}}=7$, 
$N_{\text{A}}=0$ and $N_{\text{P}}=70$  (solid disks)
and 
$N_{\text{A}}=N_{\text{P}}=70$ (open symbols). 
Left panel: $L_{\text{v}}=2$ (open triangles), 
$L_{\text{v}}=5$ (open circles),  
$L_{\text{v}}=7$ (open pentagons), 
$L_{\text{v}}=15$ (open squares).
Right panel: $\varepsilon = 0.1$ (open triangles), 
$\varepsilon=0.3$ (open circles), 
$\varepsilon = 0.5$ (open squares).
	}
	\label{fig:fig3}
\end{figure}

Simulations similar with those described in Section~\ref{s:emp} have been 
run in presence of a centered square obstacle made of $5\times5$ 
sites of the corridor not accessible by both active and passive 
particles. As before, we have computed the evacuation time 
in such a case and results are reported in Figure~\ref{fig:fig3}. 

It is immediate to remark that plots in Figure~\ref{fig:fig3} are 
very similar to those shown in Figure~\ref{fig:fig1}. Our 
interpretation of the results is then the same. 
We just mention that the vertical scale is slightly different and 
we notice that, in presence of an obstacle, the drafting effect 
is slightly increased. 
The fact that the presence of an obstacle with suitable 
geometry can favor the evacuation of a corridor is a fact already 
established in the literature, see, e.g., 
\cite{CCCM2018,CC2018,CCS2018,CKMSpre2016,CKMSSpA2016,CP2017}
and references therein. 

\section{The stationary flux}
\label{s:flu}
To detect non--trivial behaviors as time elapses beyond  a characteristic walking timescale,  
we consider now a modified version of the model proposed in Section~\ref{s:mod}. Essentially, the current situation is as follows: 
Particles exiting the system are introduced back in one site, randomly chosen among the 
empty sites of the corridor so that the total number of active 
and passive particles is approximatively kept constant during the evolution.
This way, the system reaches a final stationary state and in such 
a state we shall measure the flux of exiting active and passive 
particles. 

The main idea is to add a \emph{reservoir} in which particles exiting the 
corridor are collected. 
Particles in the reservoir are then introduced inside $\Lambda$ with 
rates depending on the number of 
particles in such a reservoir and on the number of empty sites in the corridor.

More precisely, recall the definition of the number of active and passive 
particles $n_A(\eta)$ and $n_P(\eta)$ in the configuration $\eta$ 
and fix two non--negative integer numbers $N_{\text{A}}$ and 
$N_{\text{P}}$. Consider the Markov process defined in 
Section~\ref{s:mod} with an initial configuration with total 
number of active and passive particles respectively equal to 
$N_{\text{A}}$ and $N_{\text{P}}$
and rates
$c(\eta,\eta')$, for $\eta,\eta'\in\Omega$,
defined as in Section~\ref{s:mod} with the following modification:
\begin{itemize}
\item[--]
if $\eta'$ can be obtained by $\eta$ by 
adding a $+1$ at an empty site $x$ 
then 
$c(\eta,\eta')=[N_{\text{A}}-n_{\text{A}}(\eta)]
               /(L^2-n_{\text{A}}(\eta)-n_{\text{P}}(\eta))$
(moving an active particle from the reservoir to an empty site in the 
corridor);
\item[--]
if $\eta'$ can be obtained by $\eta$ by 
adding a $-1$ at an empty site $x$ 
then 
$c(\eta,\eta')=[N_{\text{P}}-n_{\text{P}}(\eta)]
               /(L^2-n_{\text{A}}(\eta)-n_{\text{P}}(\eta))$
(moving a passive particle from the reservoir to an empty site in the 
corridor).
\end{itemize}

At time $t$, the quantities 
$N_{\text{A}}-n_{\text{A}}(\eta(t))$
and 
$N_{\text{P}}-n_{\text{P}}(\eta(t))$
 represent the number of active, and respectively, passive 
particles in the reservoir at time $t$, whereas $L^2-n_{\text{A}}(\eta(t))-n_{\text{P}}(\eta(t))$
is the number of empty sites of the corridor at time $t$. 

With these changes in the definition of the rate, the total number 
of particles in the system (considering the corridor and the 
reservoir) is conserved. The number of particles in the corridor, on the 
other hand, will fluctuate due to the fact that particles can 
accumulate in the reservoir. 

In the study of this dynamics, the main quantity of interest is the 
stationary \emph{outgoing flux} or \emph{current} 
which is the value approached in the infinite time limit 
by the ratio between the total number of particle that in the interval $(0,t)$
jumped from the exit to the reservoir and the time $t$.

\subsection{The empty corridor case}
\label{s:fnob}
We consider the system defined in Section~\ref{s:flu} 
for $L=15$ (side length of the corridor), 
$w_{\text{ex}}=7$ (exit width), 
$N_{\text{P}}=70$  (number of passive particles)
$N_{\text{A}}=0,70$  (number of active particles)
$L_{\text{v}}=2,5,7,15$ (visibility depth), 
and 
$\varepsilon = 0.1,0.3,0.5$ (drift).
More details on the selected parameters regimes are provided in the figure captions. 

As in Section~\ref{s:emp},
all the simulations share the same initial 
configuration obtained by distributing the particle 
at random with a uniform probability. More precisely, two initial 
configurations are considered, one for the case 
$N_{\text{P}}=70$ and $N_{\text{A}}=0$  
and one for the case 
$N_{\text{P}}=70$ and $N_{\text{A}}=70$, chosen in such a way that 
in the two cases the initial positions of the passive particle is the same, see Figure~\ref{fig:fig0.5}.

\begin{figure}
	\centering
	\includegraphics[width=0.45\textwidth]{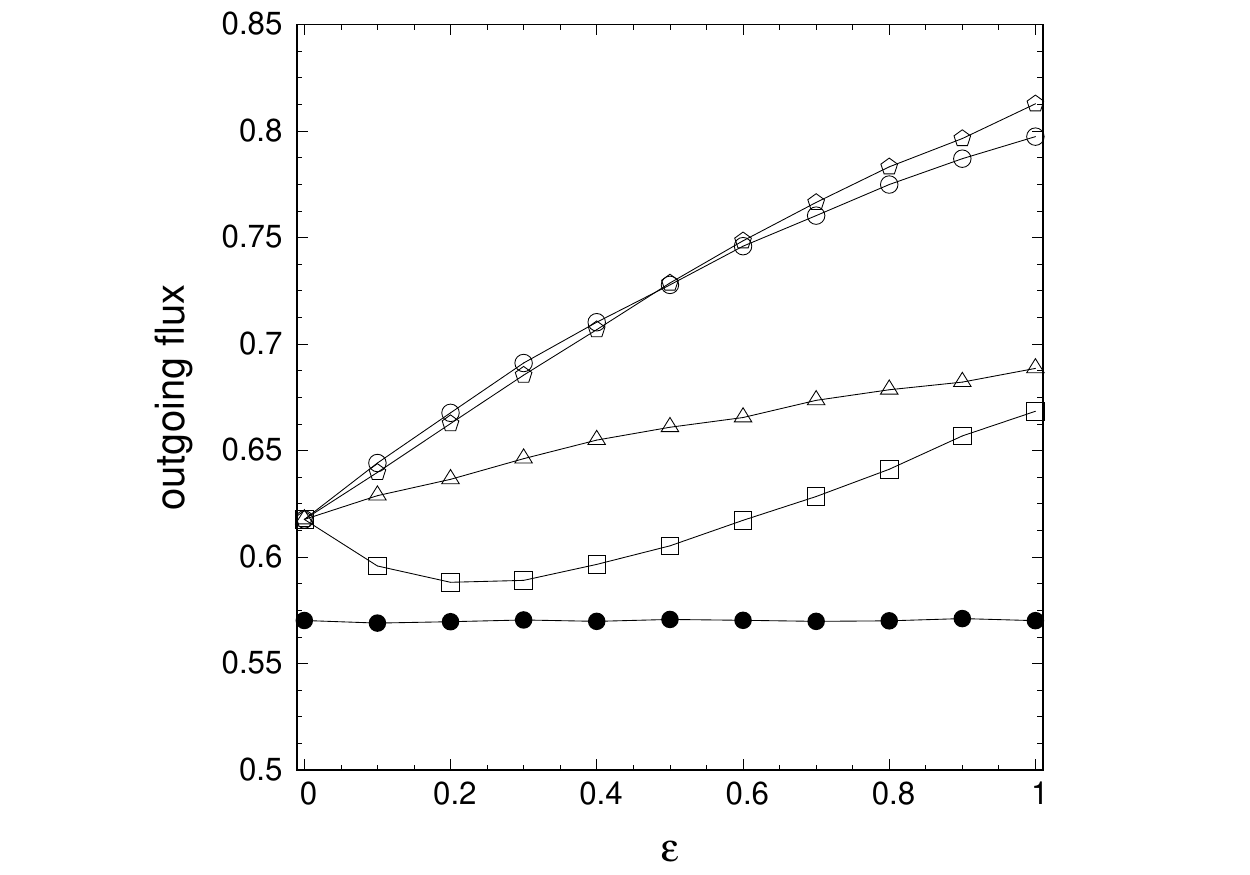}
	\includegraphics[width=0.45\textwidth]{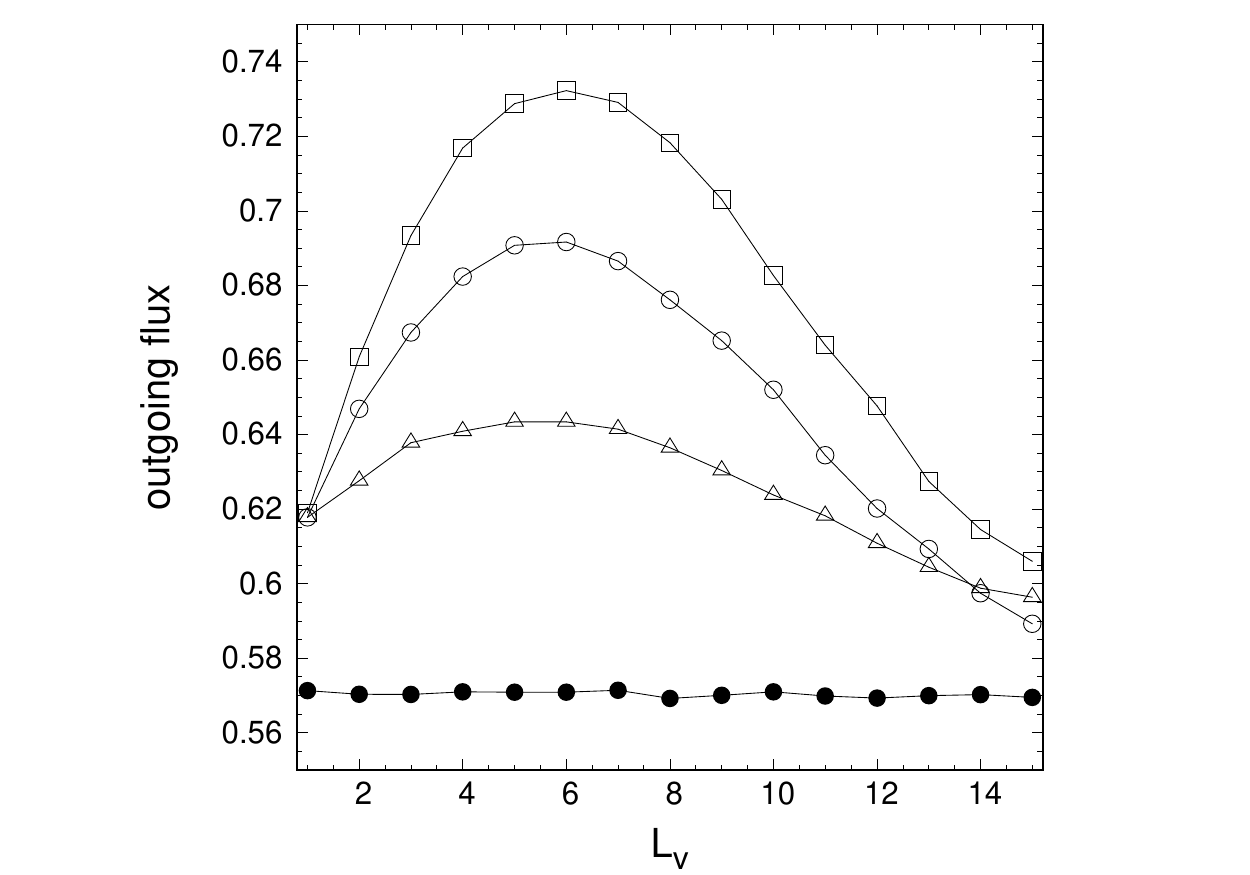}
	\caption{Stationary flux of passive particles in an empty corridor 
for $L=15$, $w_{\text{ex}}=7$, 
$N_{\text{A}}=0$ and $N_{\text{P}}=70$  (solid disks)
and 
$N_{\text{A}}=N_{\text{P}}=70$ (open symbols). 
Left panel: $L_{\text{v}}=2$ (open triangles), 
$L_{\text{v}}=5$ (open circles),  
$L_{\text{v}}=7$ (open pentagons), 
$L_{\text{v}}=15$ (open squares).
Right panel: $\varepsilon = 0.1$ (open triangles), 
$\varepsilon=0.3$ (open circles), 
$\varepsilon = 0.5$ (open squares).
	}
	\label{fig:fig5-1}
\end{figure}

We then let the system evolve and compute the ratio of the number of
particles jumping from the exit to the reservoir to time. We consider, in particular, the flux of passive particles, in \textit{absence} and in \textit{presence} of the active ones. This observable 
fluctuates until it approaches a roughly constant value after about $k=6.36\times 10^7$ MC steps (corresponding, approximately, to time 328342)
is reached.
Our results are reported in Figure~\ref{fig:fig5-1}.

We show  the dependence of the 
stationary flux of passive particles
on the drift $\varepsilon$ in the left panel of Figure~\ref{fig:fig5-1}. 
Open symbols refer to the flux for 
$N_{\text{P}}=70$ and $N_{\text{A}}=70$; for each 
value of $\epsilon$ we repeat the measure of the flux 
time also for a system in which only passive particles are present. 
We then obtain the sequence of solid disks reported in the figure 
which is approximatively constant, since the dynamics of 
the passive particles does not depend on $\varepsilon$. 

\begin{figure}
	\centering
	\includegraphics[width = 0.28\textwidth]{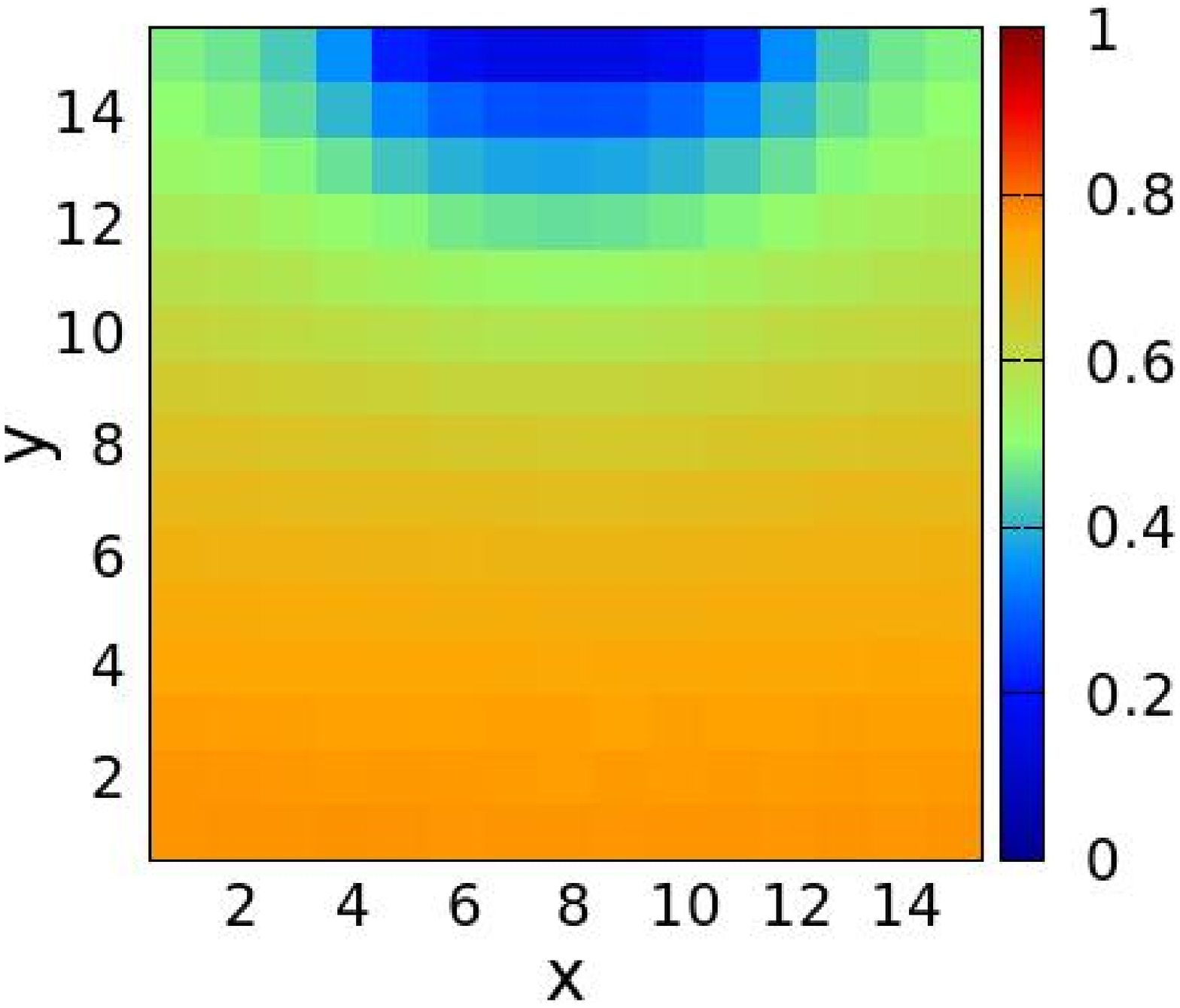} 
	\hspace{-1. cm}
	\includegraphics[width = 0.28\textwidth]{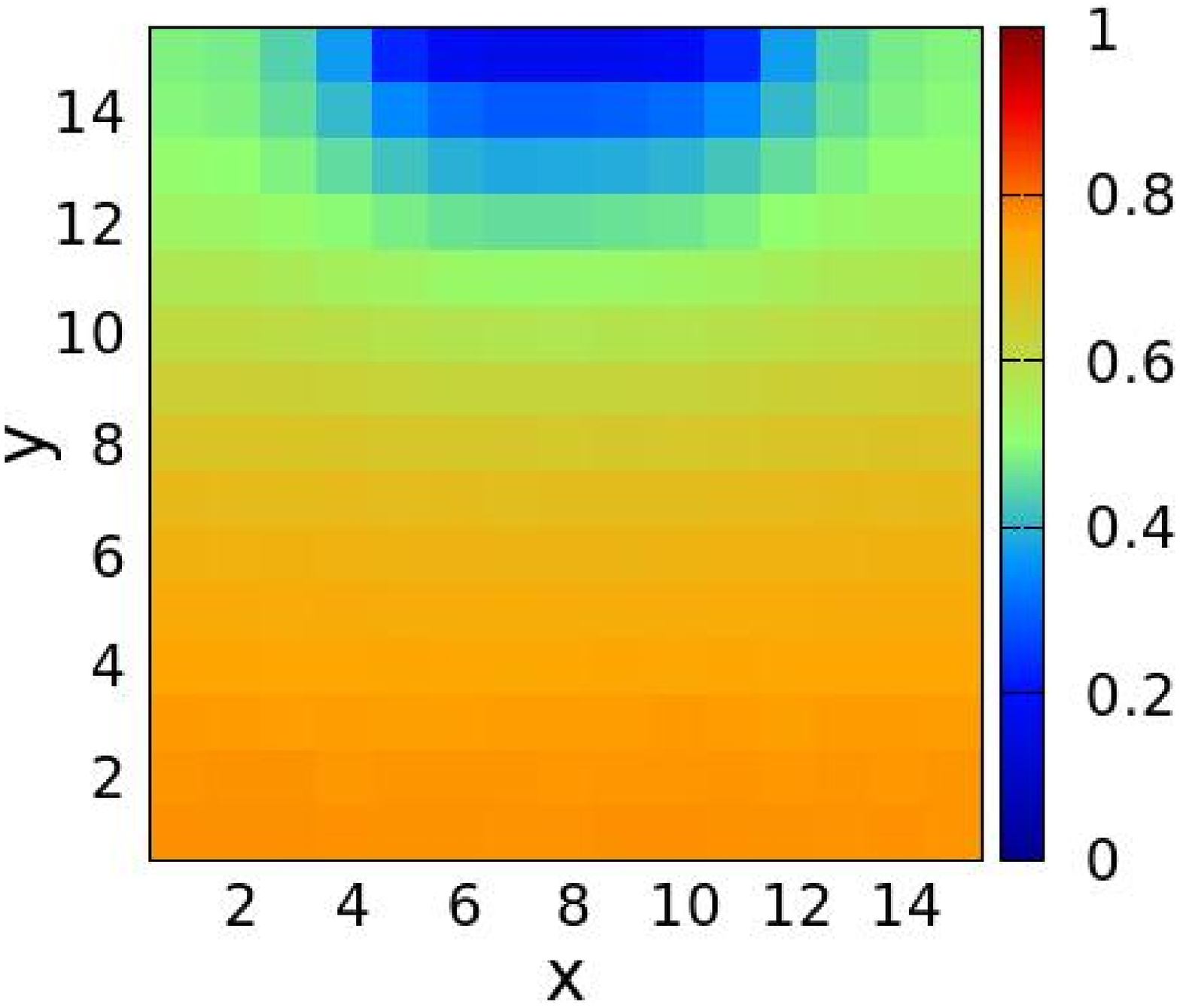} 
	\hspace{-1. cm}
	\includegraphics[width = 0.28\textwidth]{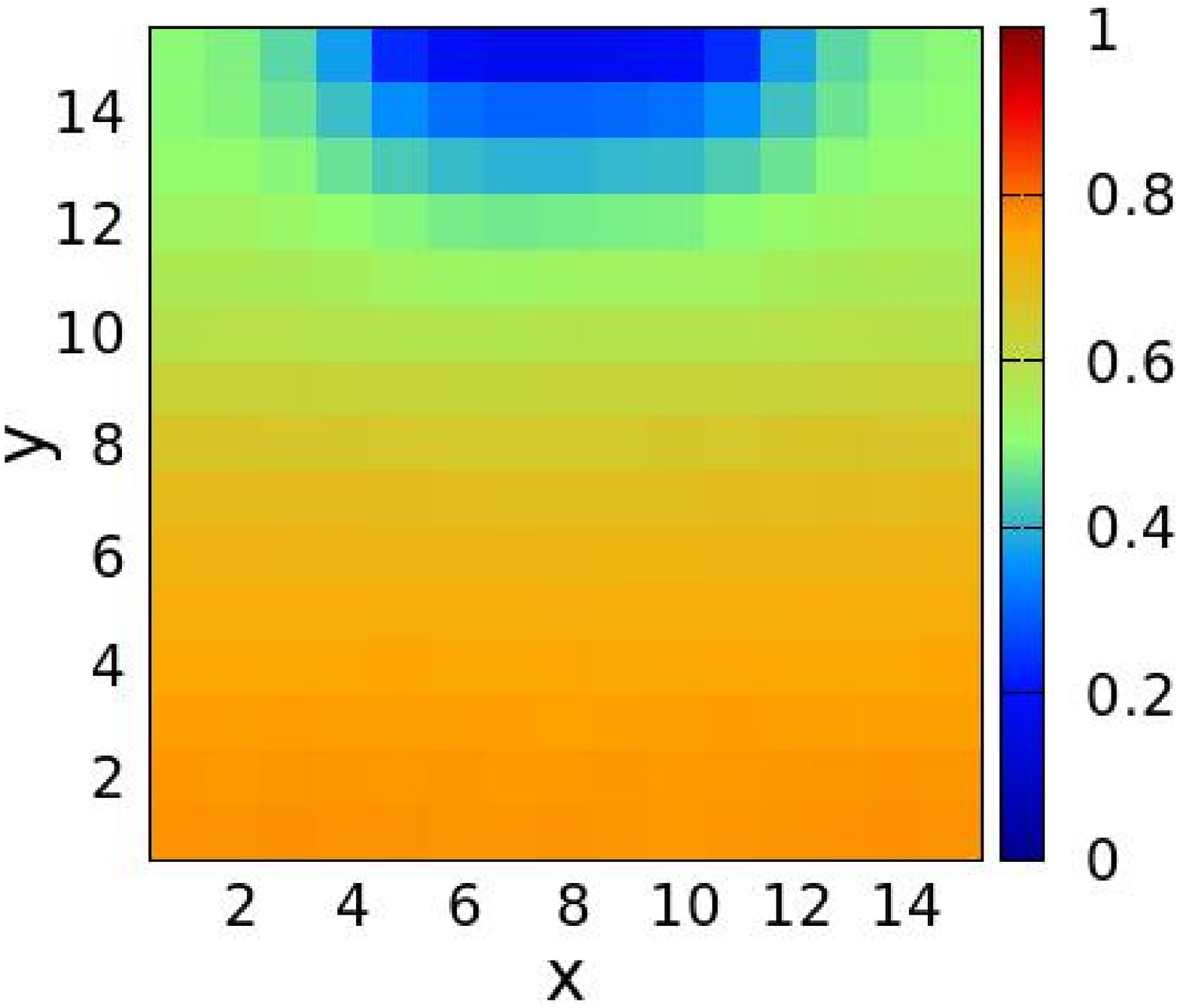} 
	\hspace{-1. cm}
	\includegraphics[width = 0.28\textwidth]{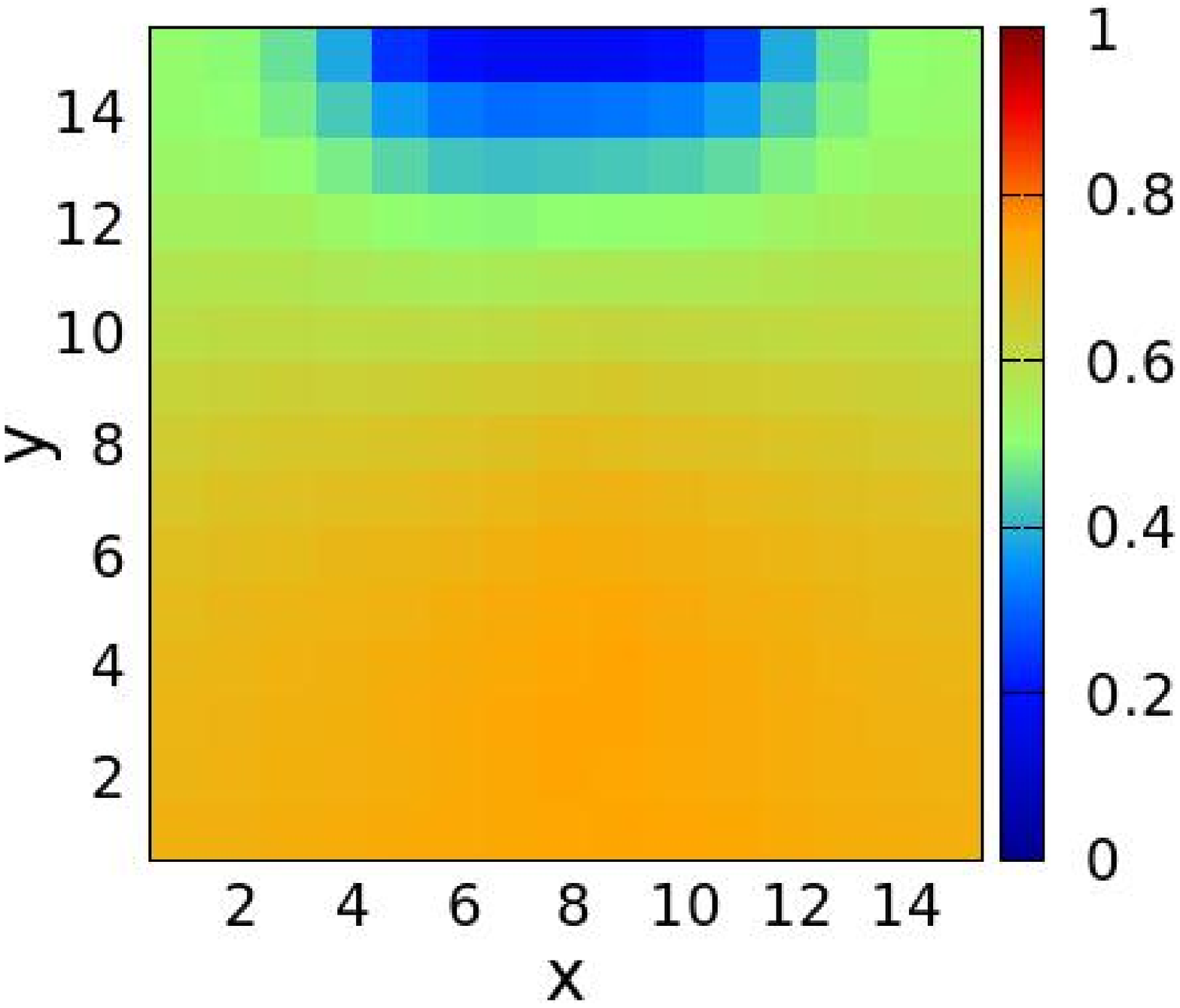} 
        \\
	\includegraphics[width = 0.28\textwidth]{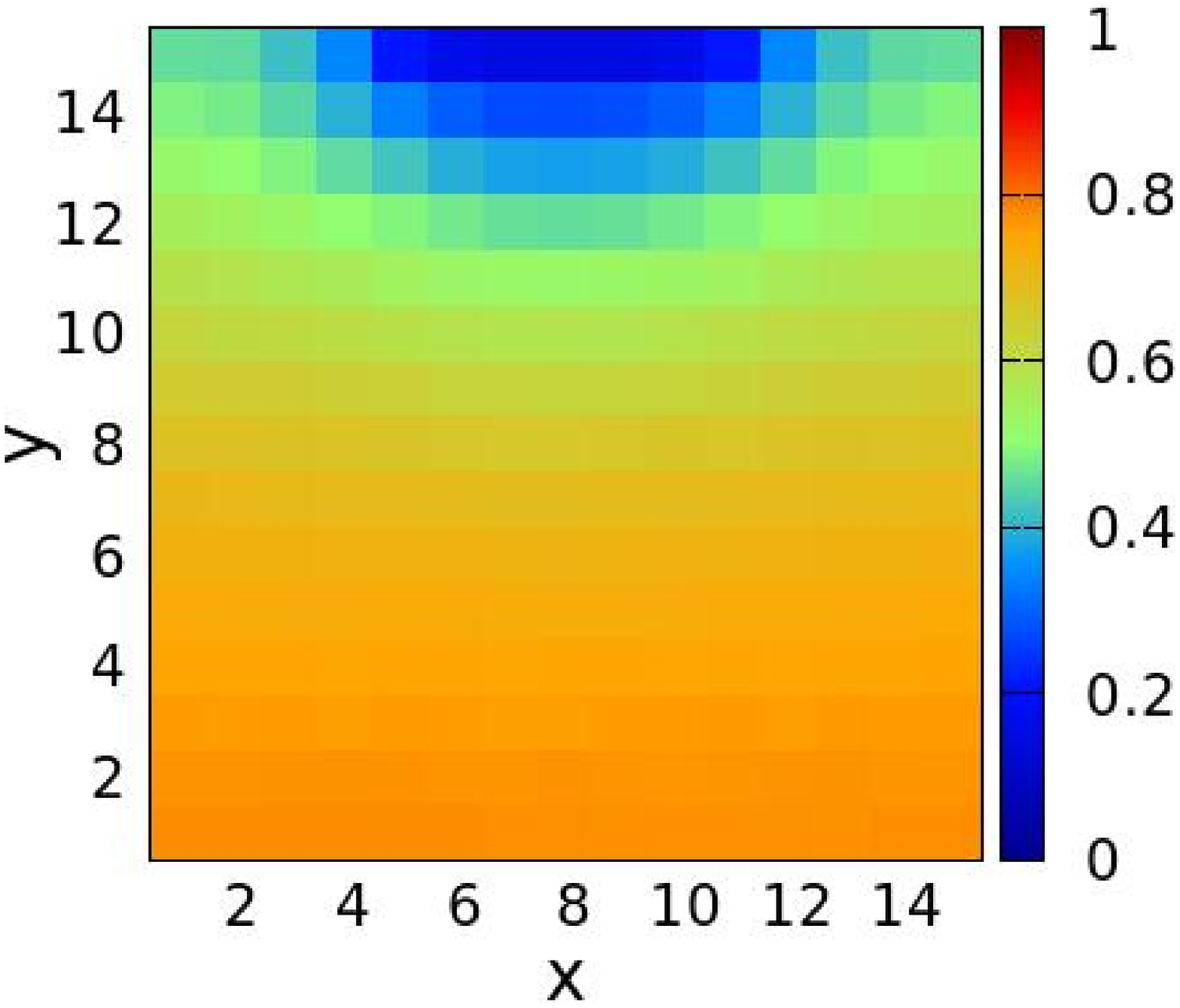}
	\hspace{-1. cm}	
	\includegraphics[width = 0.28\textwidth]{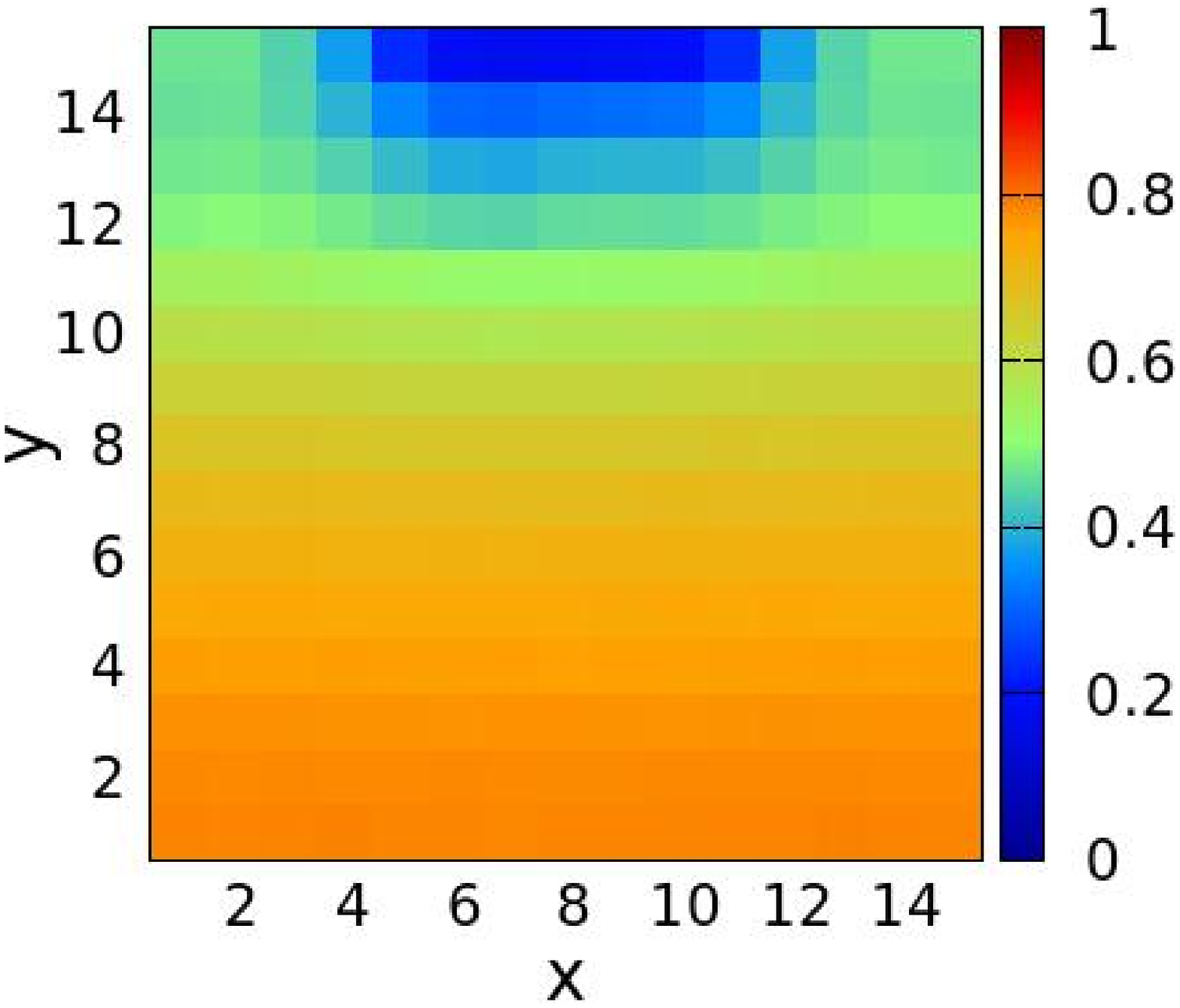}
	\hspace{-1. cm}
	\includegraphics[width = 0.28\textwidth]{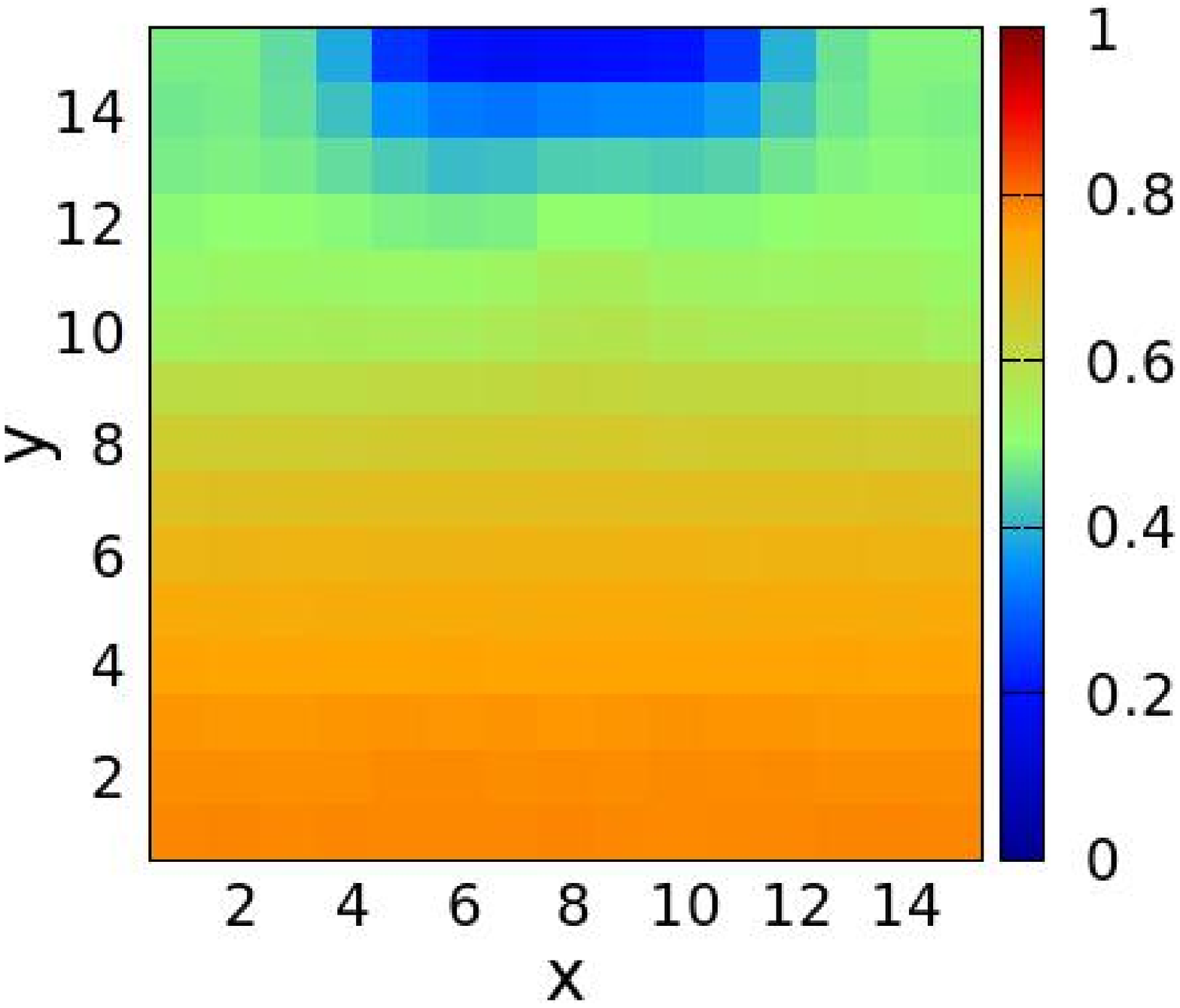}
	\hspace{-1. cm}	
	\includegraphics[width = 0.28\textwidth]{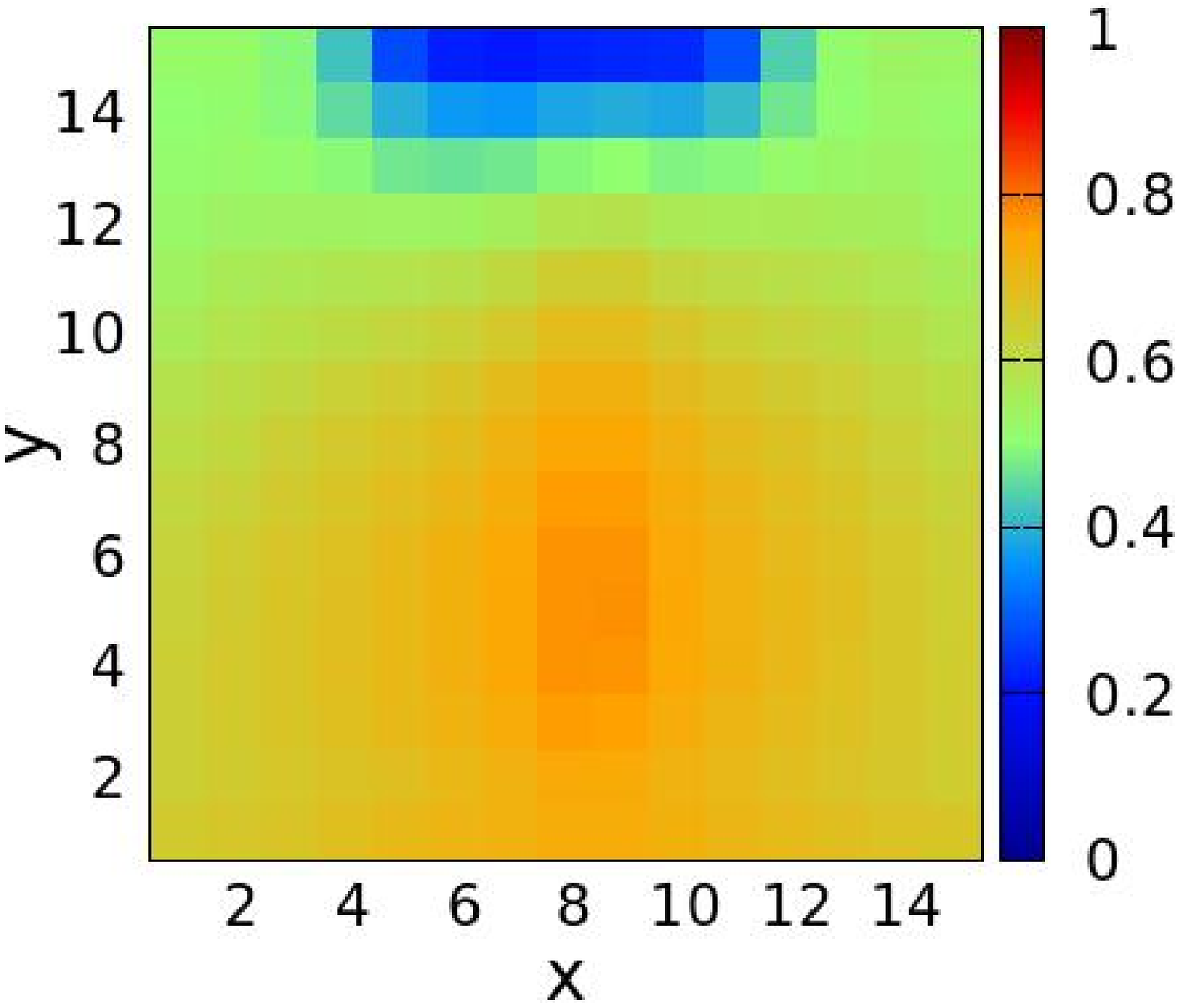}
        \\
	\includegraphics[width = 0.28\textwidth]{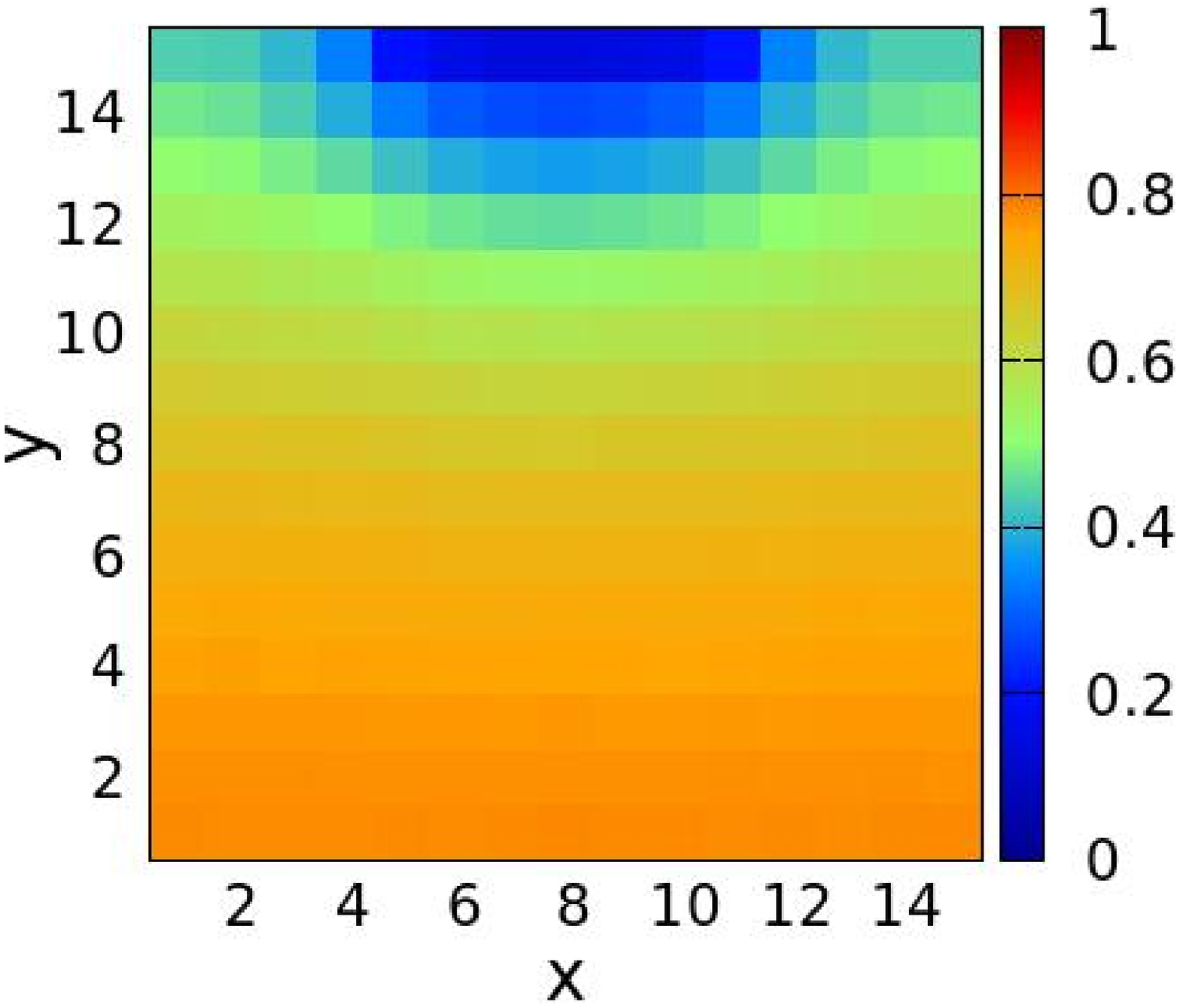} 
	\hspace{-1. cm}	
	\includegraphics[width = 0.28\textwidth]{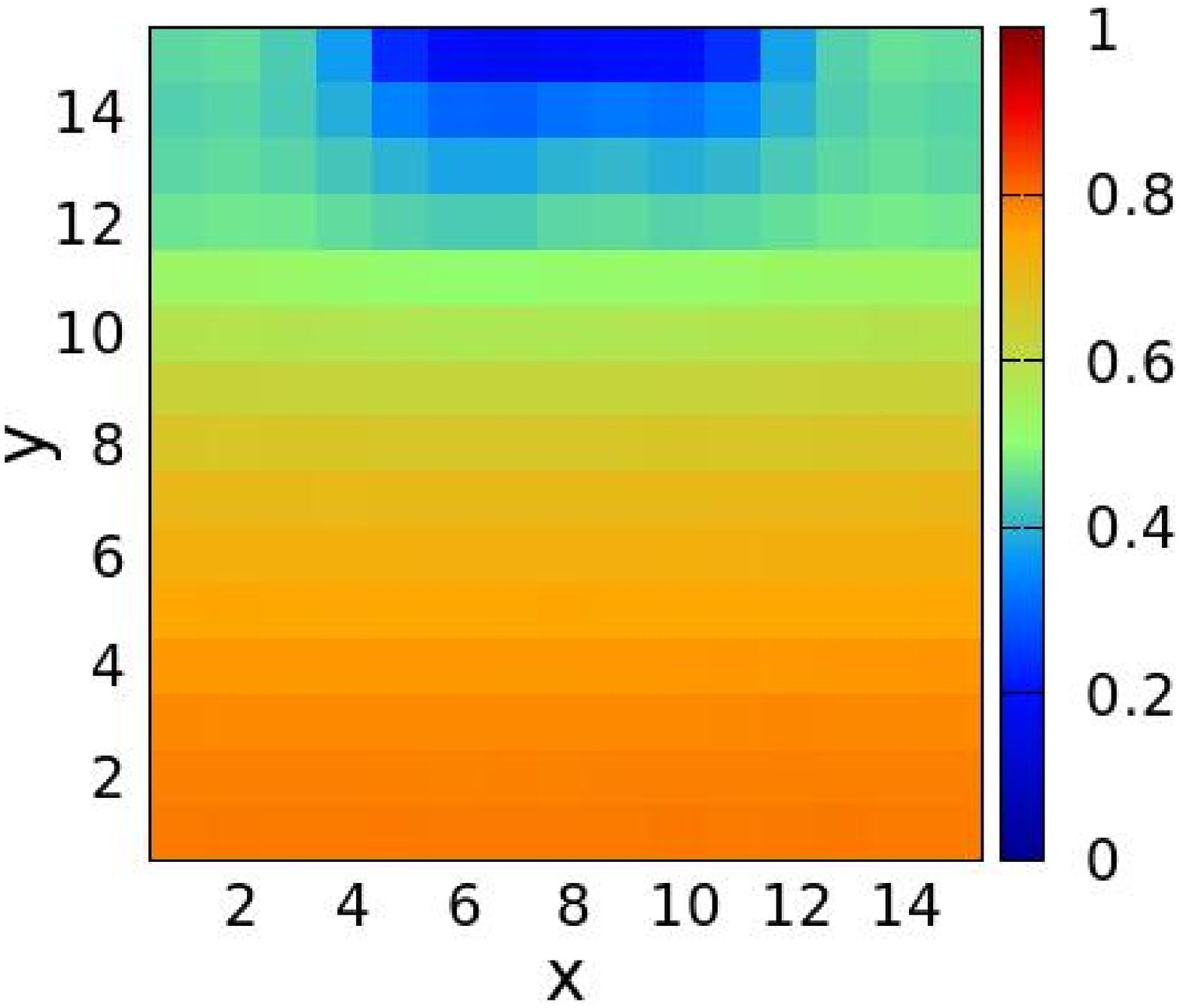} 
	\hspace{-1. cm}
	\includegraphics[width = 0.28\textwidth]{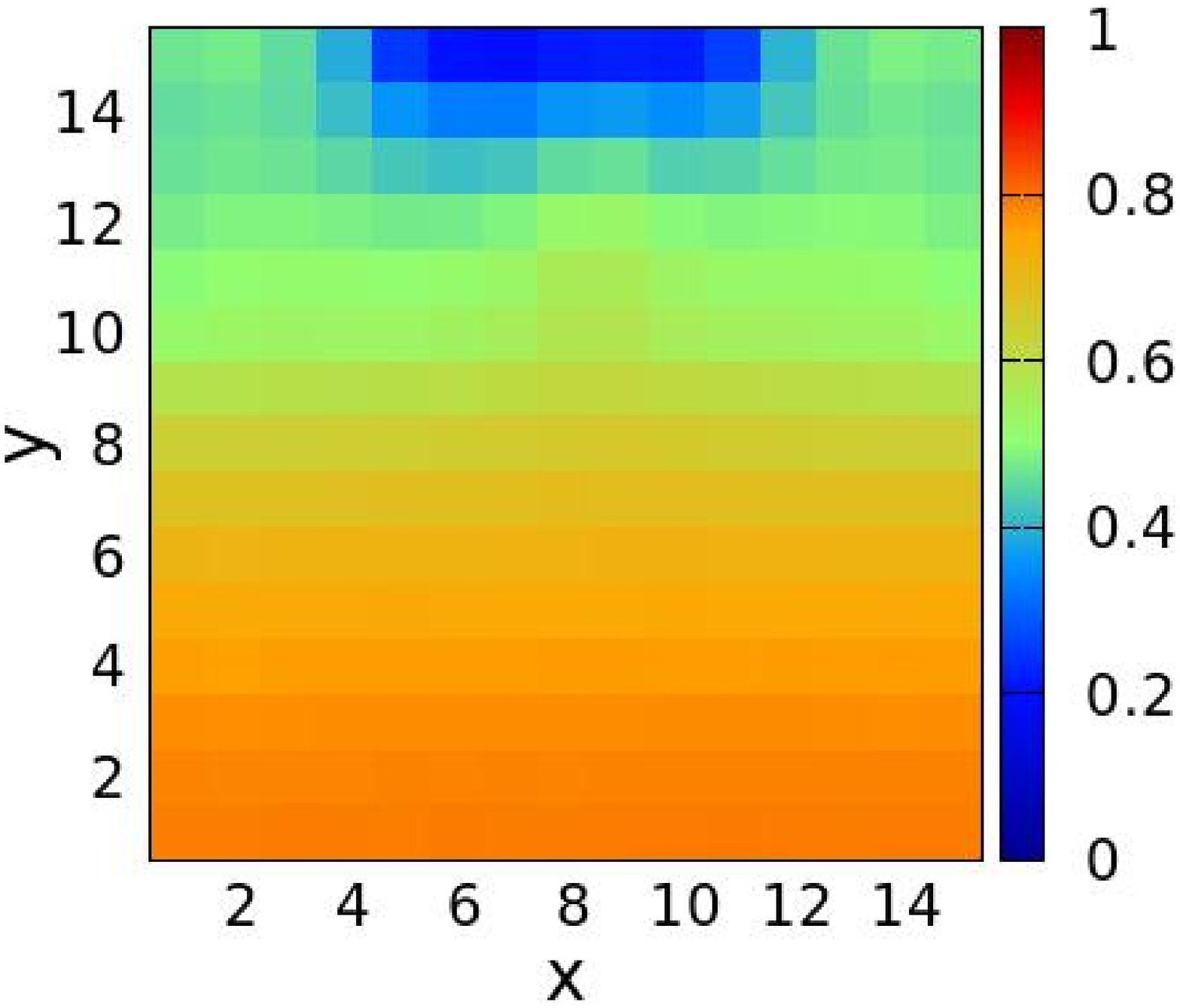} 
	\hspace{-1. cm}	
	\includegraphics[width = 0.28\textwidth]{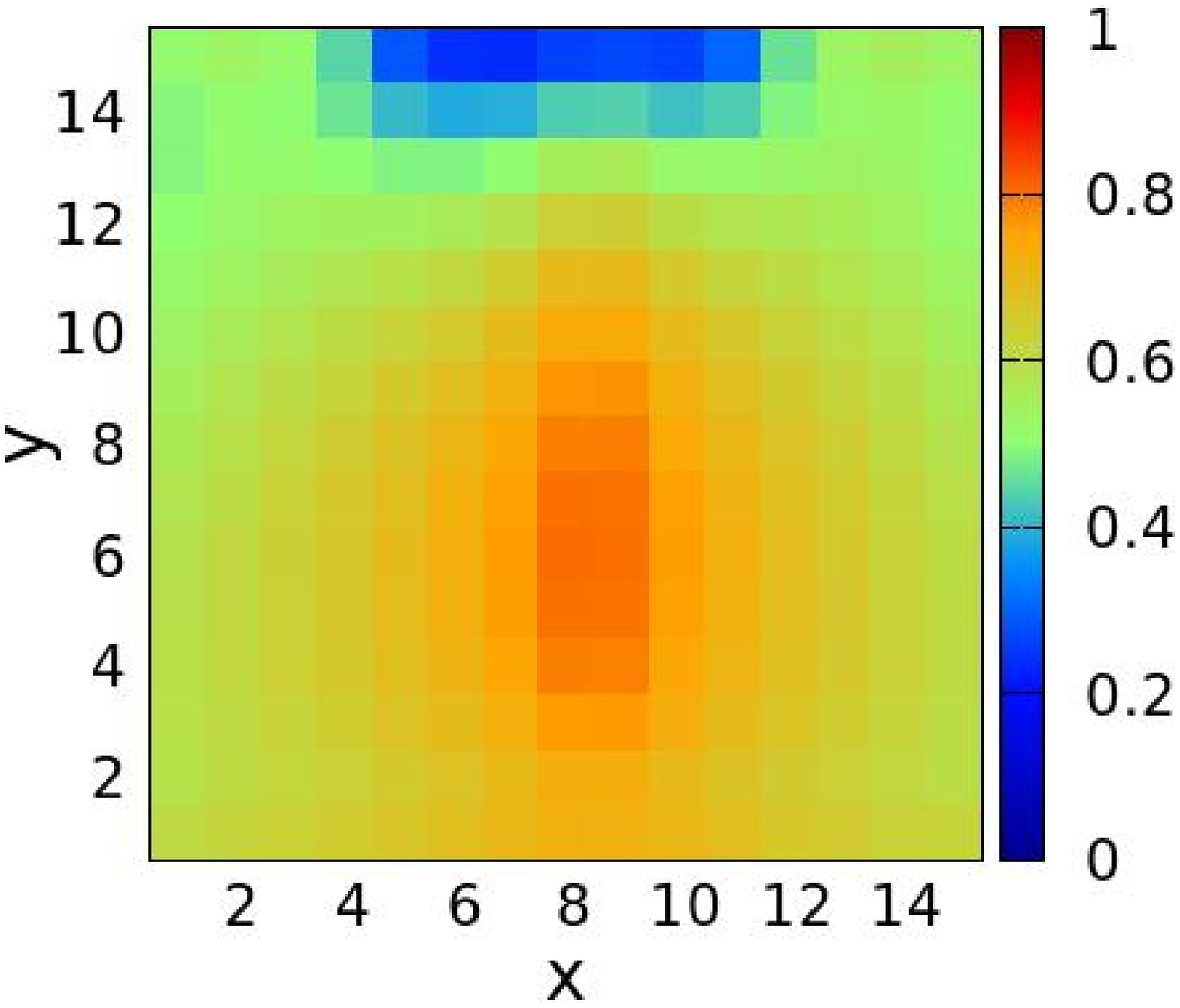} 
	\caption{Occupation number profile at stationarity for
                 $L=15$, 
                 $w_{\text{ex}}=7$, 
                 $x_{\text{ex}}=5$, $N_A=N_P=70$, 
                 $\varepsilon=0.1,0.3,0.5$ (from the top to the bottom),
                 $L_{\text{v}}=2,5,7,15$  (from the left to the right).
	        }
	\label{fig:fighm1}
\end{figure}

In the right panel of Figure~\ref{fig:fig5-1}, for the same choices of 
parameters and initial conditions, we show the stationary flux 
as a function of the visibility depth $L_{\text{v}}$ 
for several values of the drift $\varepsilon$.

Both figures exhibit firstly an increase of the flux in presence of active particles at zero drift or zero visibility depth with respect to the case in which only passive particles are present (filled disks in Figure~\ref{fig:fig5-1}.
This situation can be understood by considering that, despite the exclusion constraint of the lattice gas dynamics, doubling the number of particles can  
justify the increase of the flux at zero drift, if no complete clogging is reached. It is instructive to follow the sequence of empty symbols in Figure~\ref{fig:fig5-1} for increasing values of  $\varepsilon$ or $L_{\text{v}}$. Note that an increase of the drift yields a monotonic increase of the stationary flux as long as the visibility depth is not too large. In fact, the monotonic increase of the flux is not observed with $L_{\text{v}}=15$ (see the 
open squares in the left panel). This  means that in the presence of such a 
large visibility depth the evacuation of the passive particles can be hindered by the presence of active particles if the drift is not large enough.

A quite interesting and a priori unexpected fact is the non--monotonicity of  the stationary flux 
with respect to the visibility depth: this can be seen by looking at the 
different curves in the left panel and it is also put in evidence 
in the plots of the right panel. 

To get a deeper insight in  this interesting effect, 
we show in Figure~\ref{fig:fighm1} the stationary 
occupation number profile. To obtain these results, we have run the dynamics for a sufficiently 
long time (order of $6.36\times 10^7$ MC steps)
so that the system reaches the stationary state. Starting off from that 
time, we have averaged the occupation number $|\eta_x(t)|$ over time 
at each site of the corridor. The resulting function takes 
values between zero and one; see Figure~\ref{fig:fighm1} for an illustration.

The plots indicate that for large drift and large visibility depth that 
clogging along the median vertical line can take place. 
the occurrence of such clogging situations partly explain the not--monotonic behavior of 
the stationary flux with varying the visibility depth.

\subsection{The corridor with an obstacle}
\label{s:fwob}
\begin{figure}
	\centering
	\includegraphics[width=0.45\textwidth]{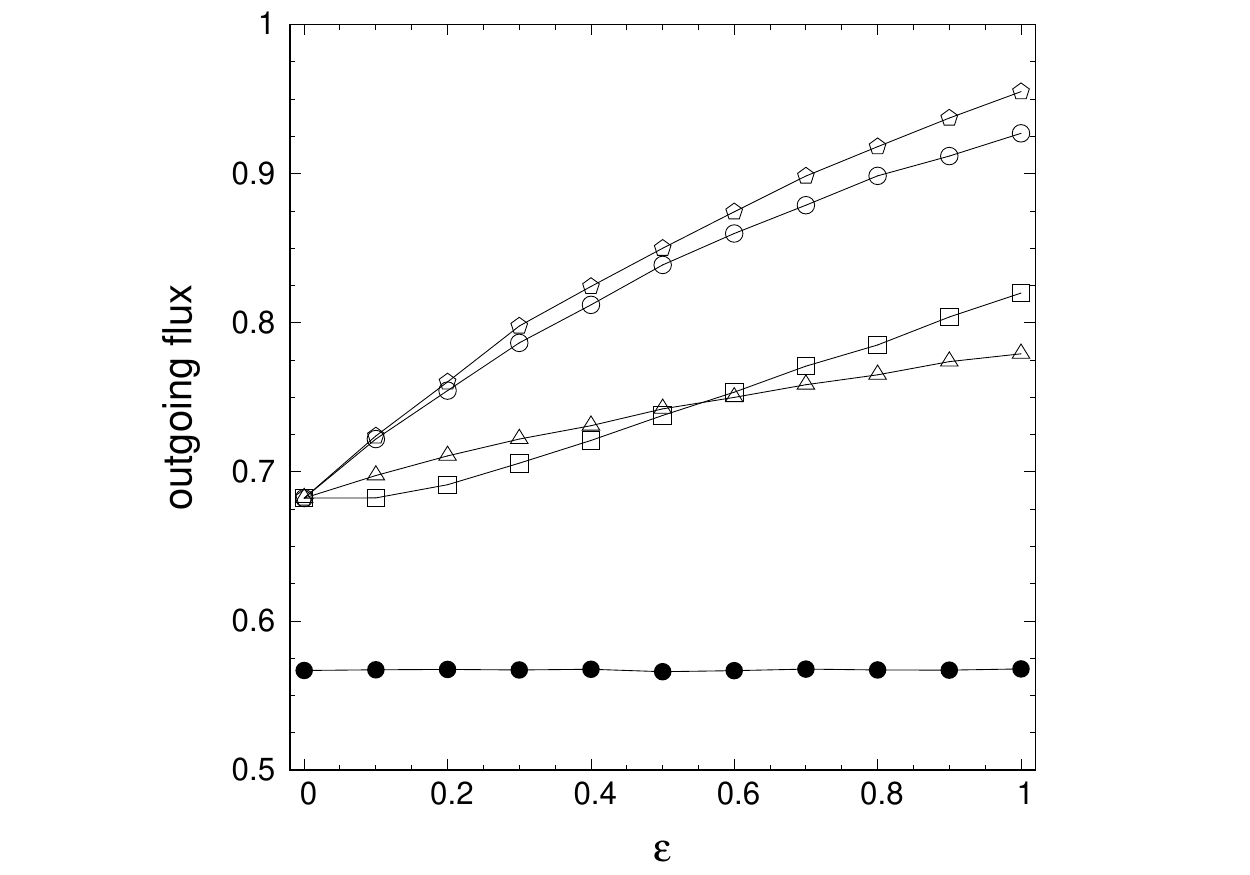}
	\includegraphics[width=0.45\textwidth]{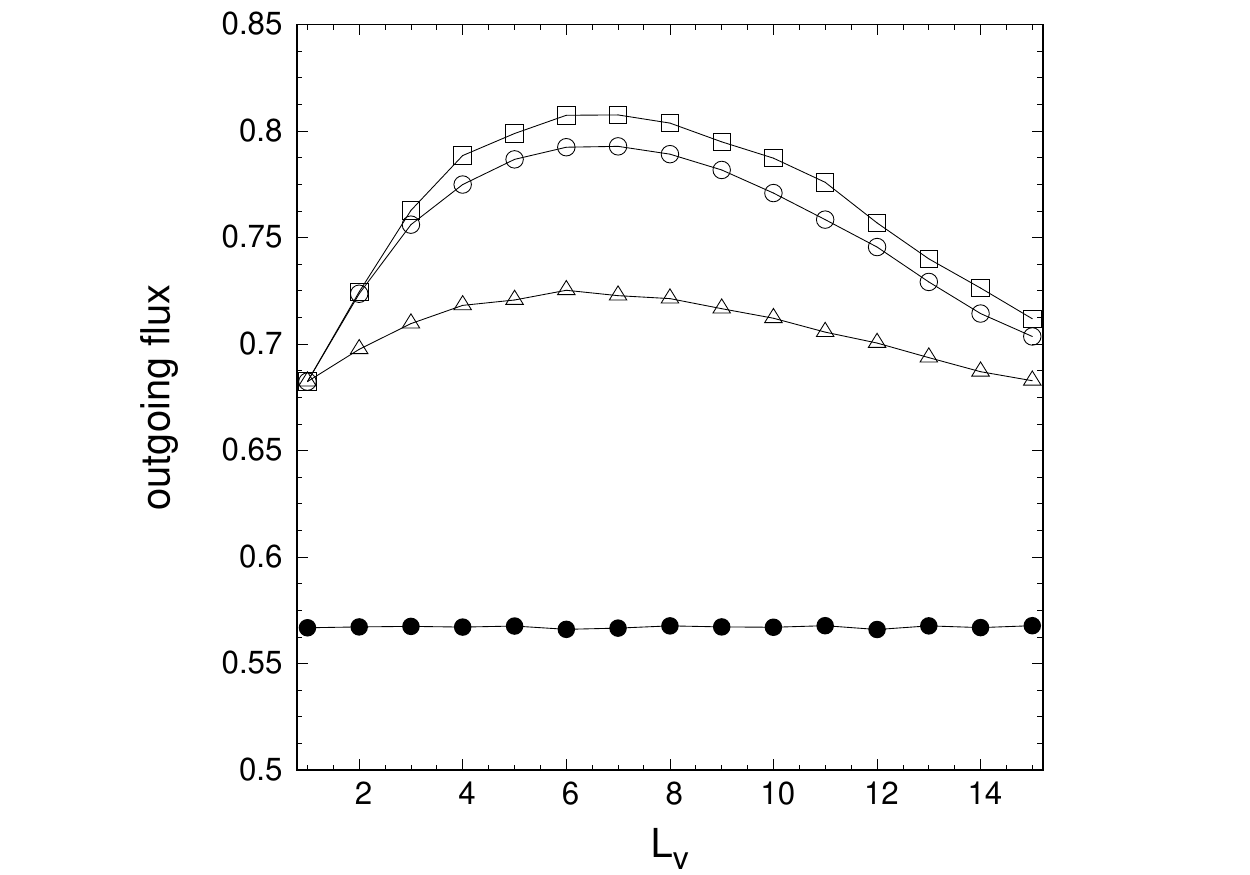}
	\caption{Stationary flux in a corridor 
with a $5\times5$ squared centered obstacle 
for $L=15$, $w_{\text{ex}}=7$, 
$N_{\text{A}}=0$ and $N_{\text{P}}=70$  (solid disks)
and 
$N_{\text{A}}=N_{\text{P}}=70$ (open symbols). 
Left panel: $L_{\text{v}}=2$ (open triangles), 
$L_{\text{v}}=5$ (open circles),  
$L_{\text{v}}=7$ (open pentagons), 
$L_{\text{v}}=15$ (open squares).
Right panel: $\varepsilon = 0.1$ (open triangles), 
$\varepsilon=0.3$ (open circles), 
$\varepsilon = 0.5$ (open squares).
	}
	\label{fig:fig5-2}
\end{figure}

Simulations similar with those described in Section~\ref{s:fnob} have been 
run in presence of a centered square obstacle made of $5\times5$ 
sites of the corridor not accessible by both active and passive 
particles. As before, we have computed the stationary flux and our
results are reported in Figure~\ref{fig:fig5-2}. 

\begin{figure}
	\centering
	\includegraphics[width = 0.28\textwidth]{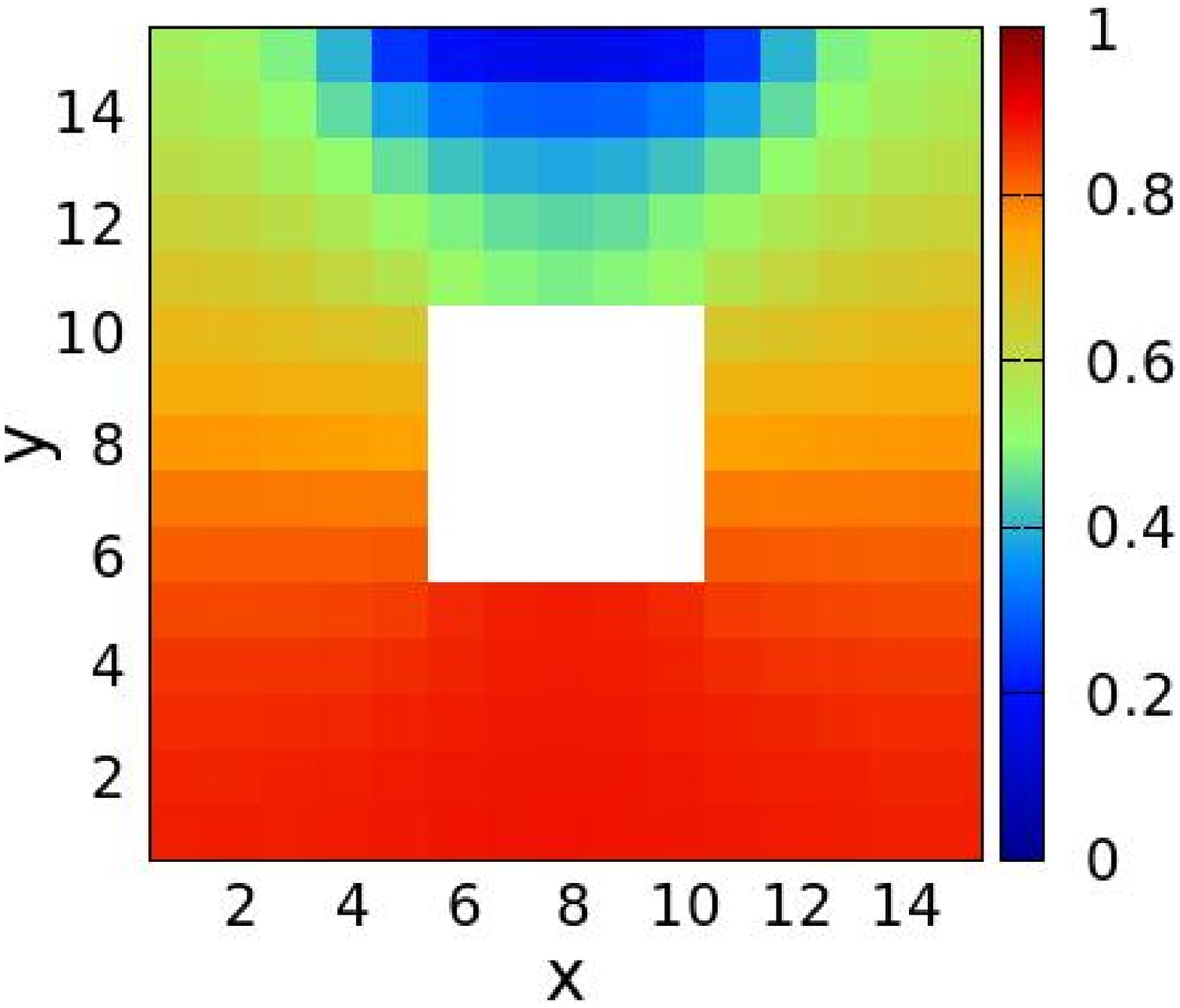} 
	\hspace{-1. cm}
	\includegraphics[width = 0.28\textwidth]{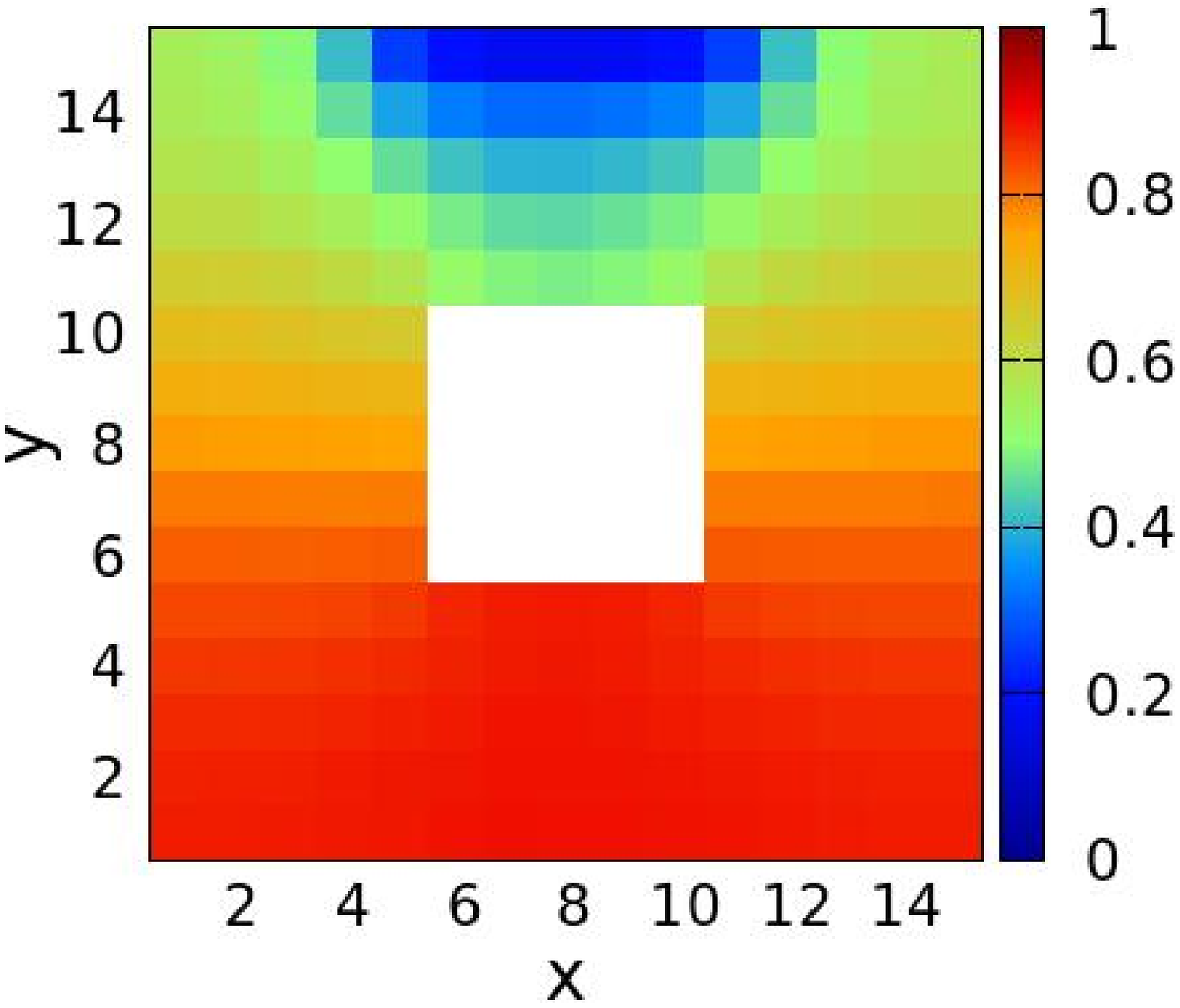} 
	\hspace{-1. cm}
	\includegraphics[width = 0.28\textwidth]{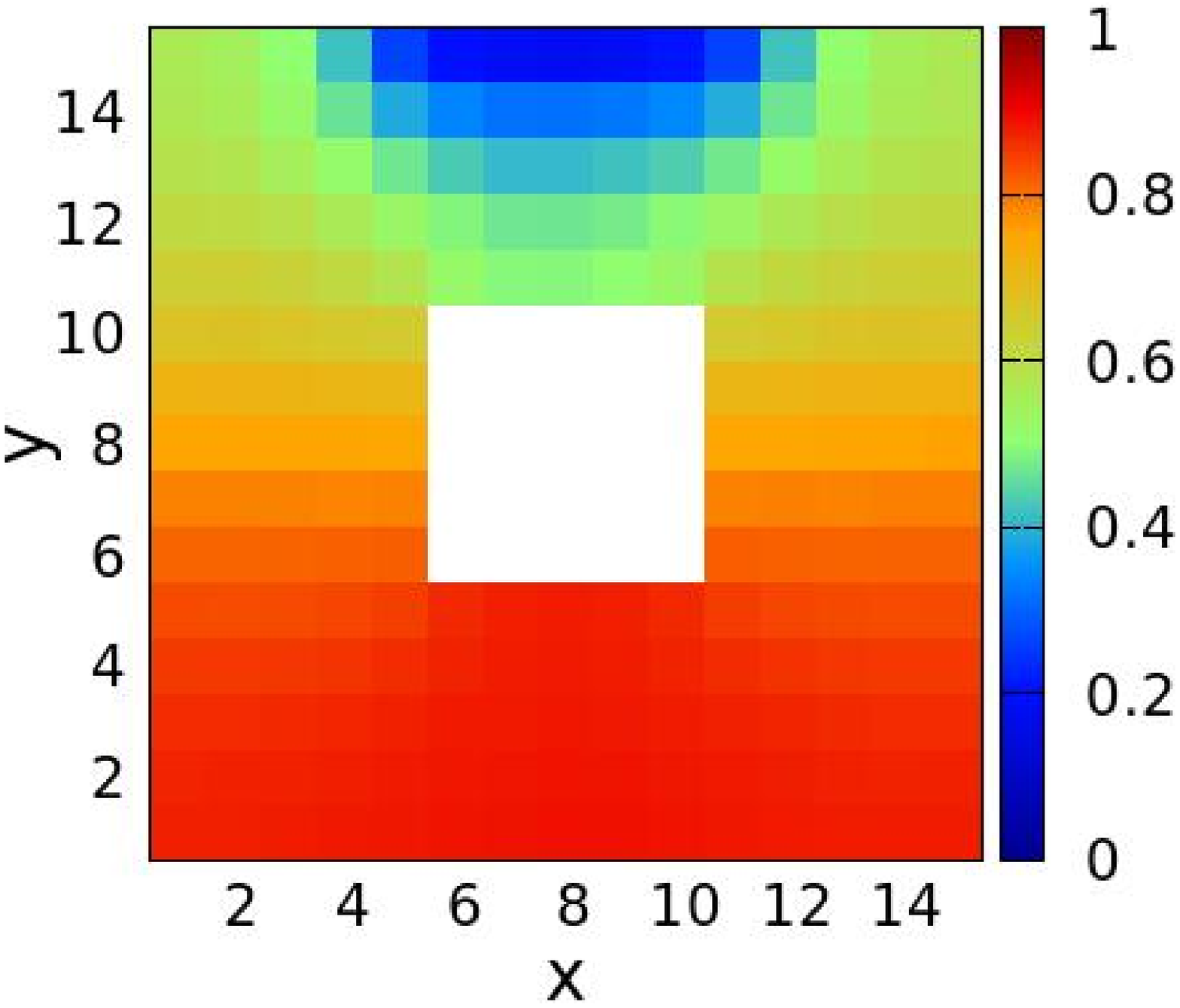} 
	\hspace{-1. cm}
	\includegraphics[width = 0.28\textwidth]{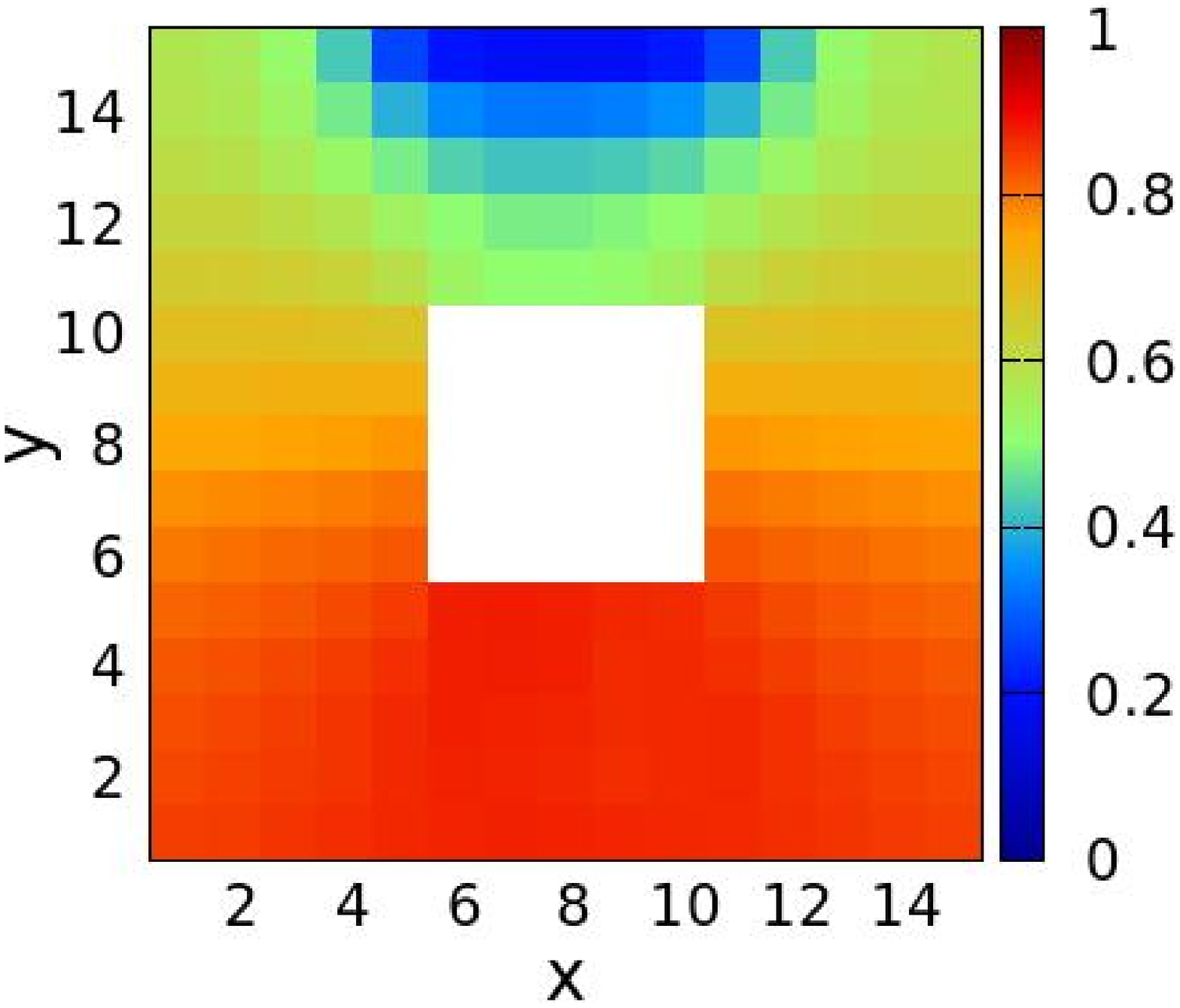} 
        \\
	\includegraphics[width = 0.28\textwidth]{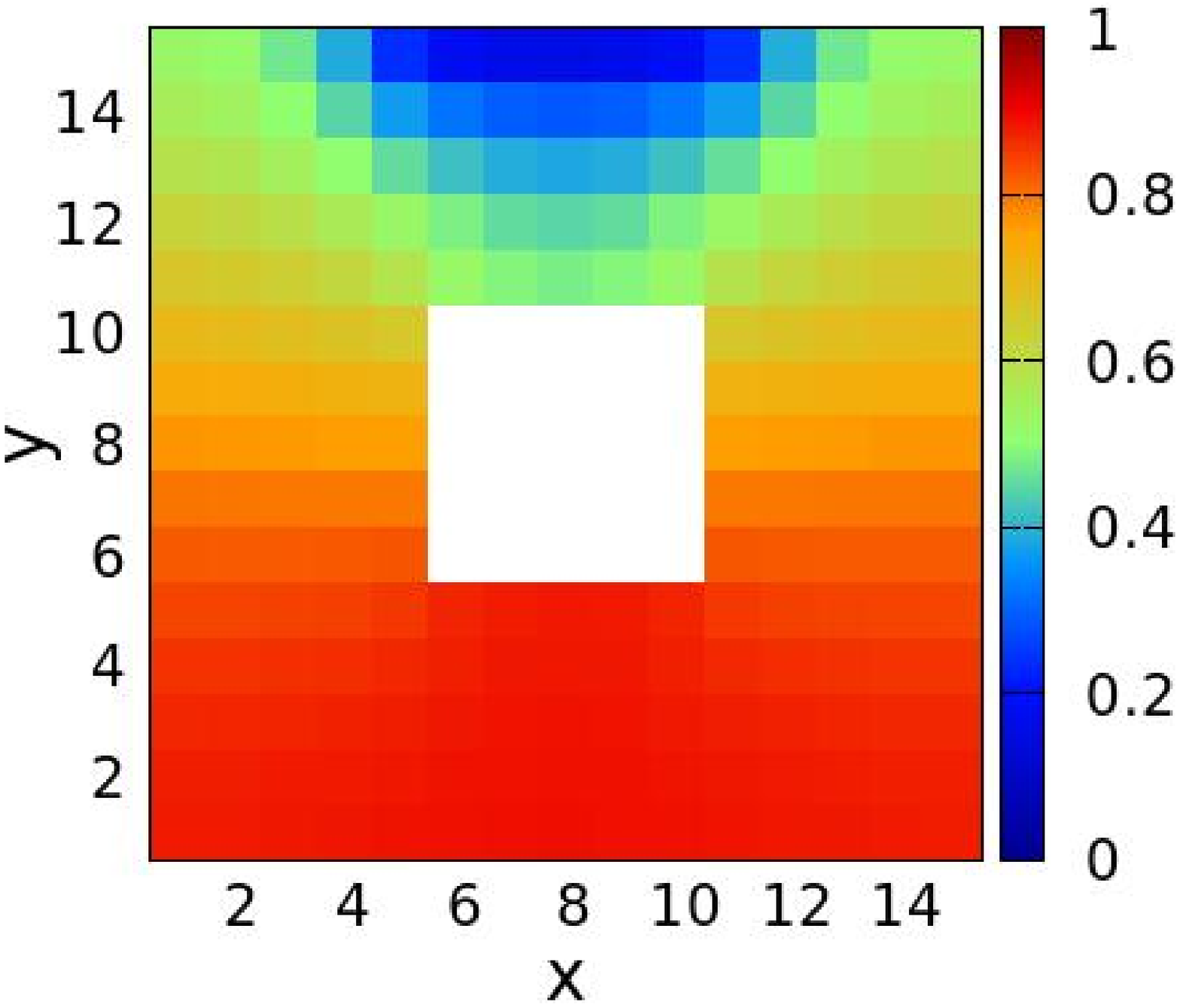}
	\hspace{-1. cm}	
	\includegraphics[width = 0.28\textwidth]{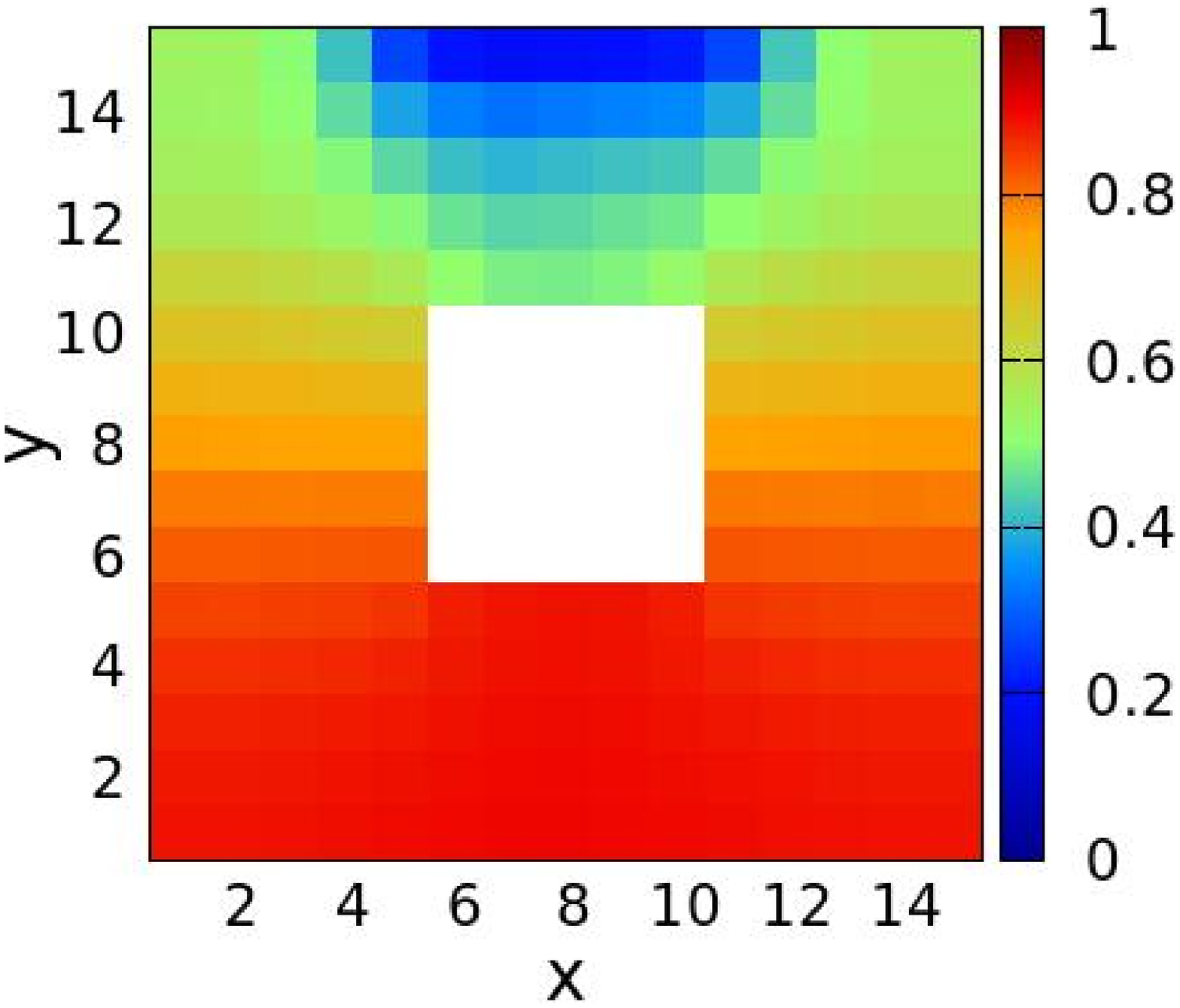}
	\hspace{-1. cm}
	\includegraphics[width = 0.28\textwidth]{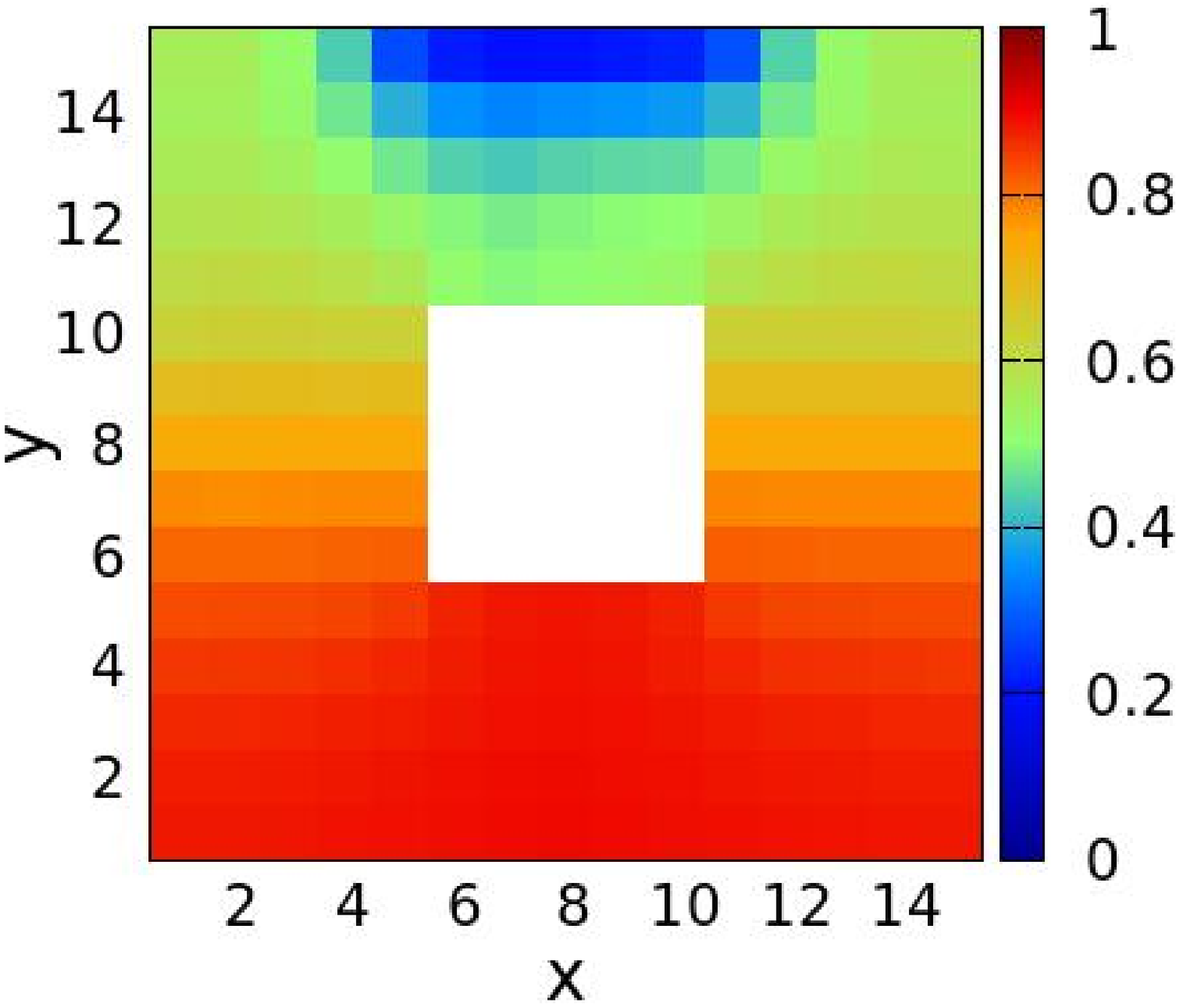}
	\hspace{-1. cm}	
	\includegraphics[width = 0.28\textwidth]{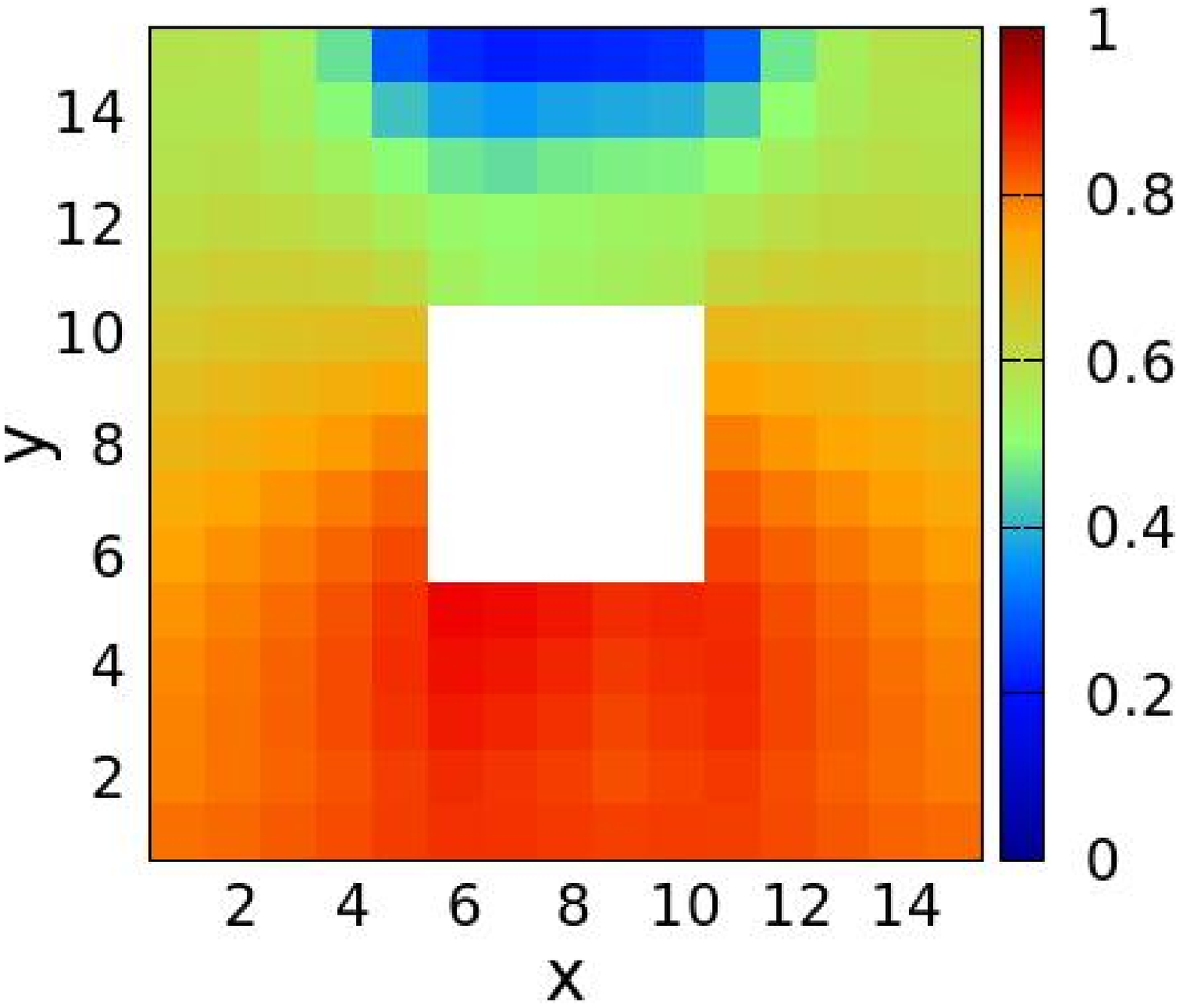}
        \\
	\includegraphics[width = 0.28\textwidth]{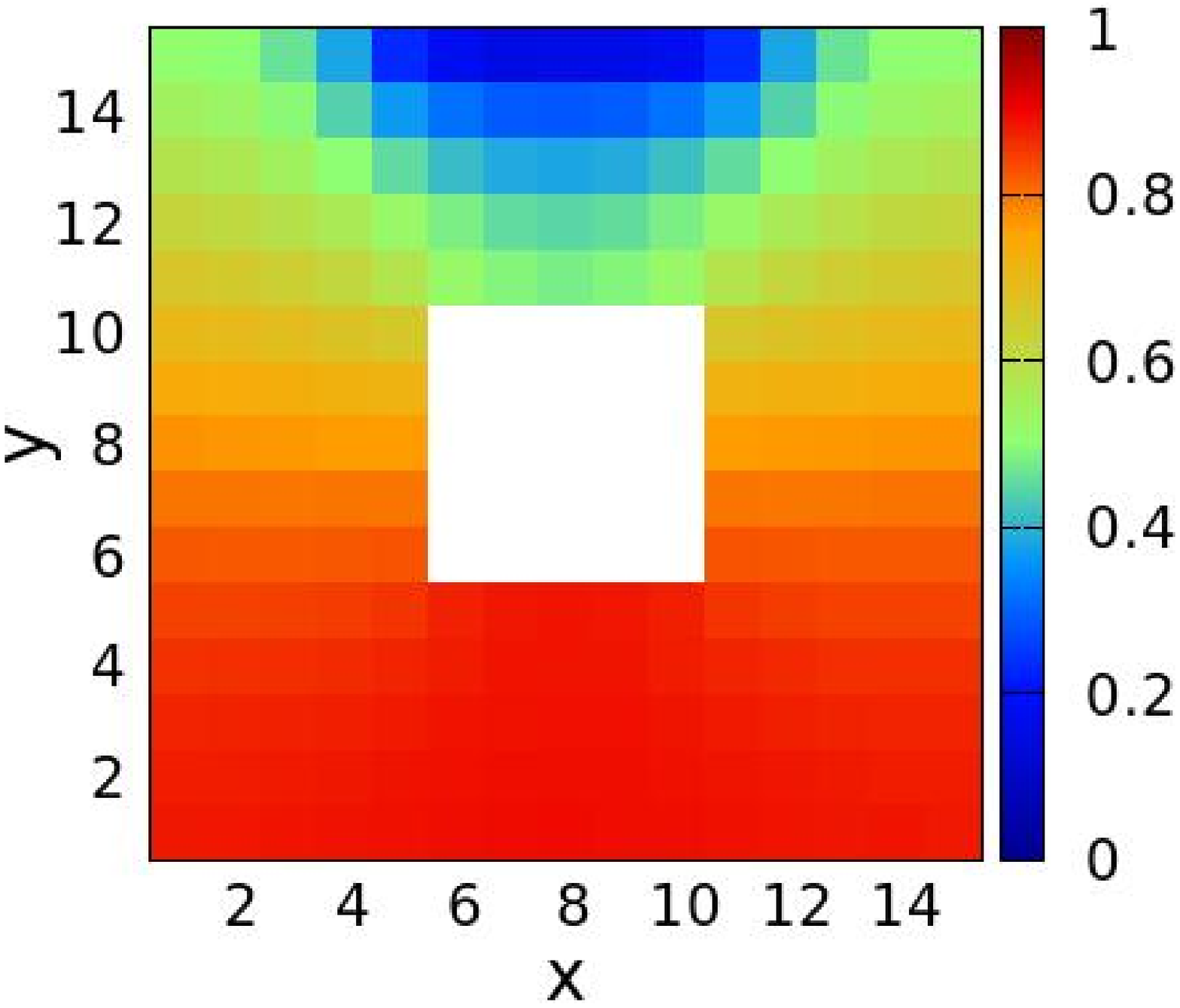} 
	\hspace{-1. cm}	
	\includegraphics[width = 0.28\textwidth]{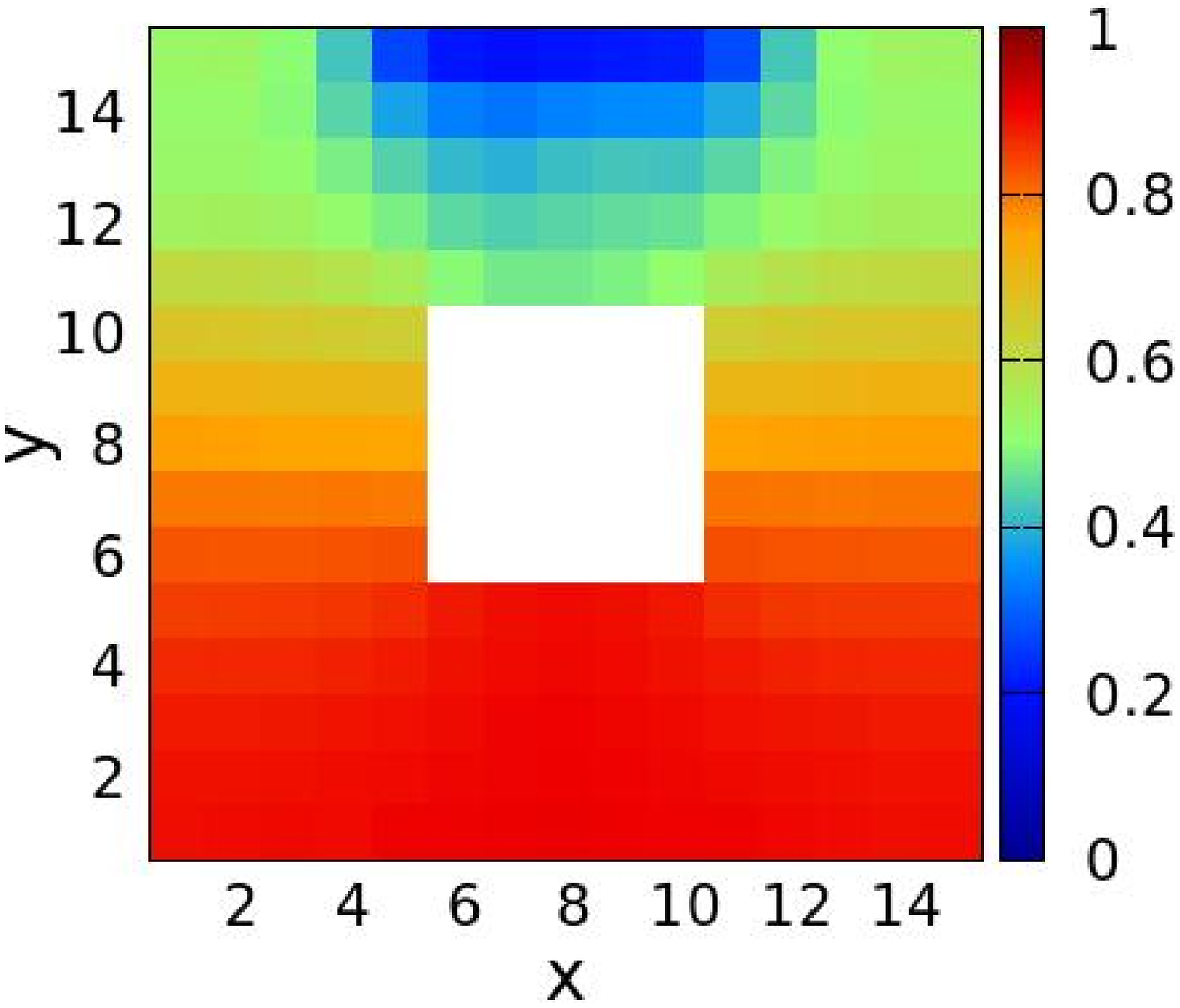} 
	\hspace{-1. cm}
	\includegraphics[width = 0.28\textwidth]{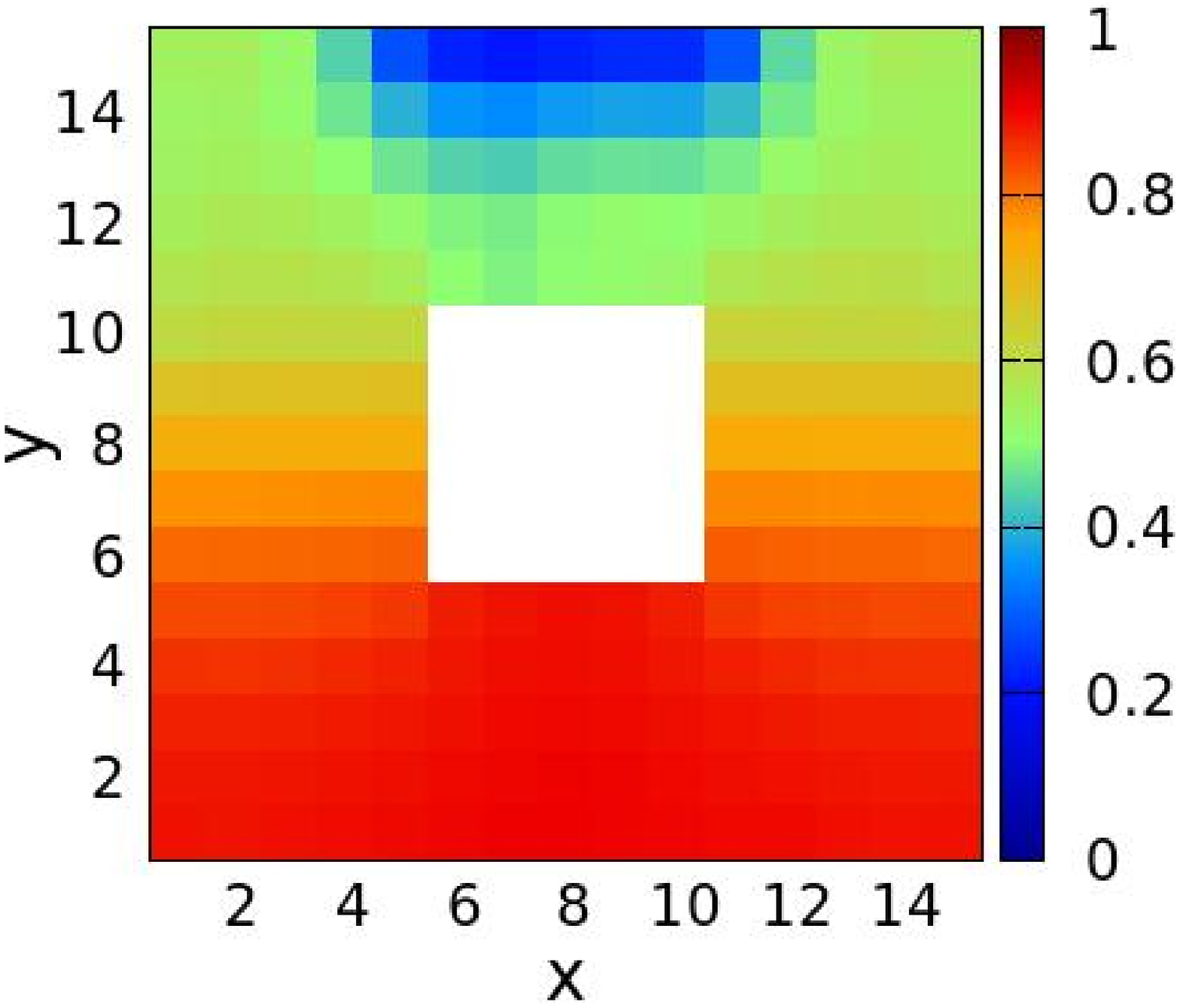} 
	\hspace{-1. cm}	
	\includegraphics[width = 0.28\textwidth]{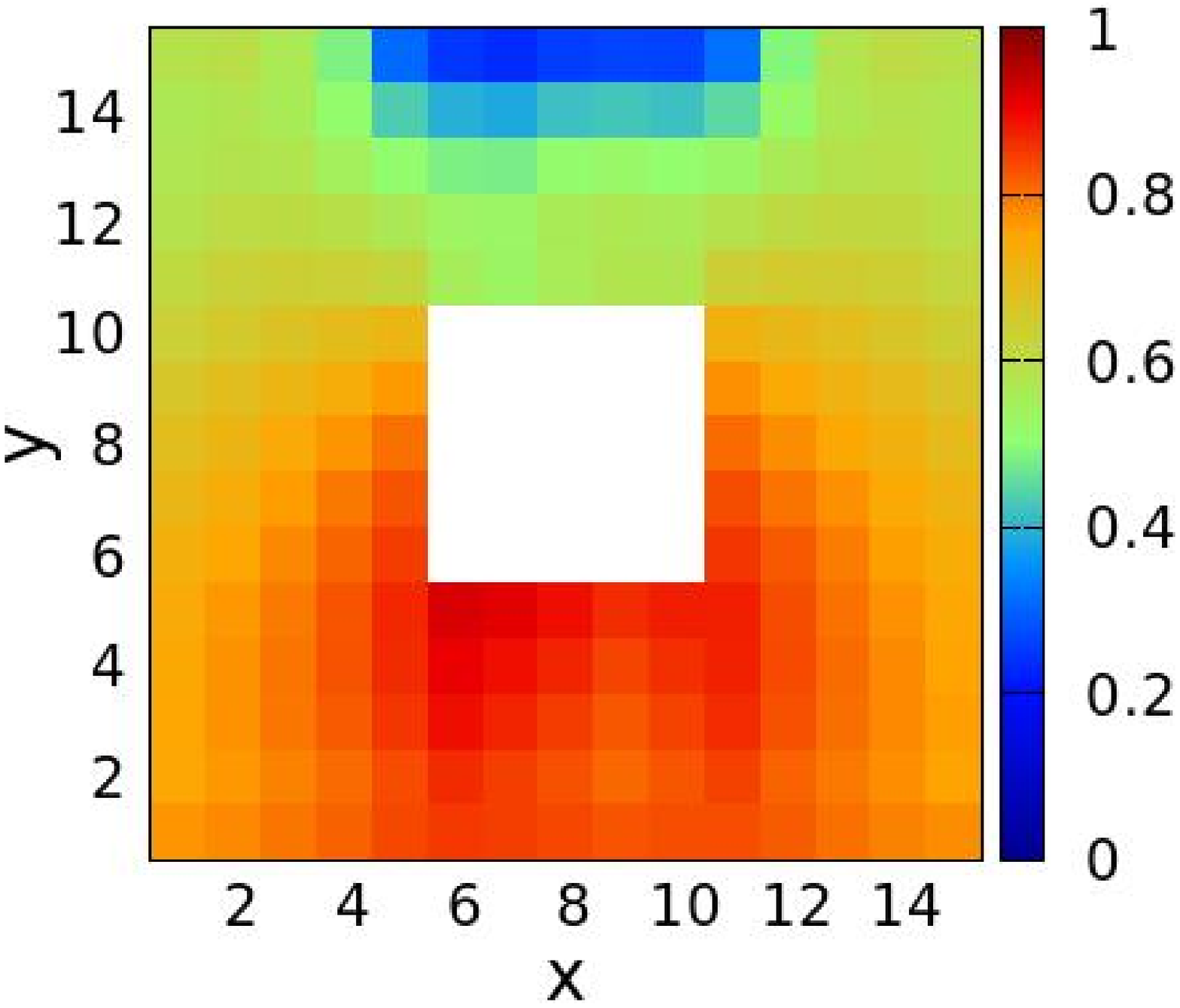} 
	\caption{Occupation number profile at stationarity in presence 
                 of a $5\times 5$ centered obstacle for
                 $L=15$, 
                 $w_{\text{ex}}=7$, 
                 $x_{\text{ex}}=5$, $N_A=N_P=70$, 
                 $\varepsilon=0.1,0.3,0.5$ (from the top to the bottom),
                 $L_{\text{v}}=2,5,7,15$  (from the left to the right).
	        }
	\label{fig:fighm2}
\end{figure}

The results plotted in Figures~\ref{fig:fig5-2} and \ref{fig:fighm2} are 
very similar to those shown in Figure~\ref{fig:fig5-1} and \ref{fig:fighm1}. 
Our 
interpretation of the results is essentially the same. Mind though that, in order to reach the stationary state, we had to run the dynamics for a larger time than in the case of the empty corridor model (i.e., of order of $9.0\times10^7$ MC steps).
Also from the point of view of the computed stationary 
flux measures, our results suggest that the presence of the obstacle 
slightly favors the exit of particles from the corridor.
This was noted in Section~\ref{s:obs} in connection 
with the evacuation time measurements; see also  
\cite{CCCM2018,CC2018,CCS2018,CKMSpre2016,CKMSSpA2016,CP2017}.

Comparing Figures~\ref{fig:fig5-1} and \ref{fig:fig5-2} 
one realizes that the dependence of the stationary flux
on the drift and on the visibility depth is milder. 
This fact can be explained remarking that the 
phenomenon of accumulation of particles along the 
median vertical line of the corridor discussed in Section~\ref{s:fnob}
is less evident. Again, the obstacle seems to be  keeping particles far apart so that clogging is reduced.

\section{Conclusions}
\label{s:con}
We  studied the problem of the evacuation of a large crowd of pedestrians 
from an obscure region. We start from the assumption that the crowd is made  of both active and passive pedestrians. 
 The hazardous motion of pedestrians due 
to lack of light and, possibly, combined also to a high level of stress is modeled via a 
simple random walk with exclusion.  
The active (smart, informed, aware, ...) pedestrians, which are aware of the 
location of the exit, are supposed to be 
subject to a given drift towards the exit, while the passive (unaware, uninformed) pedestrians are performing a random walk within the  walking geometry and eventually evacuate if they accidentally find the exit. The particle system is strongly interacting via the site exclusion principle -- each site can be occupied by only a single particle. 

The main observable is  the evacuation time as a function of the 
parameters caracterizing the motion of the aware pedestrians. 
We have found that the presence of the active pedestrians favors the evacuation of the passive ones. This is rather surprising since we explicitly do not allow for any communication among the pedestrians.  This seems to be due to some sort of drafting effect. A drag seems to arise  
due to the empty spaces left behind by the active pedestrians 
moving towards the exit and naturally filled by the 
completely random moving unaware pedestrians. 

We have also remarked that too smart active pedestrians 
can limit the drafting effect: indeed, if they exit 
the corridor too quickly the unaware pedestrian do not have 
the time to take profit of the wakes of empty side that they 
left during their motion towards the exit. 

A promising research line concerns the investigation of evacuation times when different species of particles are assumed to choose among  different exit doors. Such topic is relevant not only for urban situations but also for  tunnel fires or for forrest fires expanding towards the neighborhood of inhabited regions.  

The main open question in this context is the model validation. A suitable experiment design is needed to make any progress in this sense.  This will be our target in forthcoming  work.

With regards to the building up of the aerodynamic drag, it would also be interesting to verify the onset of the drafting phenomenon in lattice gas models in the presence of 
non--standard transport regimes leading to uphill diffusion of particles; see \cite{CDMP16,CDMP17,CDMP17bis,CGGV18,CGVK19} for the study of such transport mechanisms.

\section*{Acknowledgments}
\par\noindent
ENMC and MC thank FFABR 2017 for financial support. We thank our collaborator Omar Richardson (Karlstad, Sweden) and Errico Presutti (Gran Sasso Science Institute, Italy) for very useful discussions on this and related topics. 

\section*{Conflict of interest}
\par\noindent
There is no conflict of interest.



\end{document}